\documentclass[english]{JHEP3}
\usepackage[T1]{fontenc}
\usepackage[latin9]{inputenc}
\usepackage{float}
\usepackage{amsmath}
\usepackage{graphicx}
\usepackage{amssymb}
\usepackage{esint}

\makeatletter

\newcommand{\lyxdot}{.}

\usepackage{cite}
\usepackage[mathscr]{euscript}

\title{Improving NLO-parton shower matched simulations with higher order matrix elements}

\author{Keith Hamilton \\ INFN, Sezione di Milano-Bicocca, Piazza della Scienza 3, 20126 Milan, Italy. \\ Email: \email{keith.hamilton@mib.infn.it}}

\author{Paolo Nason \\ INFN, Sezione di Milano-Bicocca, Piazza della Scienza 3, 20126 Milan, Italy. \\ Email: \email{paolo.nason@mib.infn.it}}
\preprint{ MCnet/10/05 }

\abstract{

In recent times the algorithms for the simulation of hadronic collisions
have been subject to two substantial improvements: the inclusion, within parton
showering, of exact higher order tree level matrix elements
(\textsc{Meps}) and, separately, next-to-leading order corrections
(\textsc{Nlops}). In this work we examine the key criteria to be met
in merging the two approaches in such a way that the accuracy of both
is preserved, in the framework of the \textsc{Powheg} approach to
\textsc{Nlops}.  We then ask to what extent these requirements may be
fulfilled using existing simulations, without modifications.  The
result of this study is a pragmatic proposal for merging \textsc{Meps}
and \textsc{Nlops} events to yield much improved \textsc{Menlops}
event samples. We apply this method to W boson and top quark pair
production. In both cases results for distributions within the remit
of the \textsc{NLO} calculations exhibit no discernible changes with
respect to the pure \textsc{Nlops} prediction; conversely, those
sensitive to the distribution of multiple hard jets assume, exactly,
the form of the corresponding \textsc{Meps} results.}


\newdimen\hbigcirc
\newdimen\wbigcirc

\makeatother

\usepackage{babel}

\begin{document}

\section{Introduction\label{sec:Introduction}}

In recent years the promise of new and exciting data from the \textsc{\small LHC}
experiments has led to renewed interest and vigor in the research and
development of Monte Carlo event generators. In this time significant
progress has been made and a number of long standing goals have been
achieved. The most significant innovations have been the inclusion
of full next-to-leading order (\textsc{NLO}) corrections within parton
shower simulations and, separately, the merging of event generators
based on tree level matrix elements together with parton showers.

The central challenge to be addressed for both of these advances was
that of overcounting, since the parton shower dynamics encodes the
collinear limits of the relevant higher order, real and virtual, matrix
elements. In the case of matching with \textsc{NLO} calculations a
further complication lay in how to arrange the \textsc{NLO} formulae
such that they lend themselves to a physical, probabilistic, interpretation
and hence the formulation of a practical Monte Carlo algorithm.

Currently there are two proven methods, \textsc{MC@NLO} and \textsc{Powheg}
\cite{Frixione:2002ik,Nason:2004rx,Frixione:2007vw}, for including
\textsc{NLO} corrections within parton shower algorithms. Simulations
based on these approaches provide predictions for infrared safe observables
with full \textsc{NLO} accuracy. The hardest emitted parton in each
event is distributed according to the exact real, single emission,
matrix elements, and \textsc{NLO} virtual corrections are consistently
included. Moreover, effects of further, higher order, soft and collinear
emissions to all orders in the leading log approximation are also
included. Both methods may be considered to be mature, having been
applied to several processes, and the subject of a number
of comparative studies \cite{Frixione:2003ei,Nason:2006hfa,Frixione:2007nw,Alioli:2008gx,Hamilton:2008pd,Alioli:2008tz,Hamilton:2009za,Frixione:2008yi,Alioli:2009je,Nason:2009ai,Torrielli:2010aw,Mangano:2006rw}.

Despite the clear advantages associated with promoting parton shower
event generators to \textsc{NLO} accuracy (\textsc{Nlops}), this class
of simulations are not, by themselves, sufficiently versatile to model
all features of the data in adequate detail. In particular, since
the only exact real corrections included are those of the next-to-leading
order calculation, the parton shower approximation is used to describe
all but one of the emissions. As such these event generators do not
offer a satisfactory description of particle production in association
with multiple hard jets, as is widely anticipated to occur in the
case of new physics signals and backgrounds \cite{Mangano:2006rw,Alwall:2008qv}.

The other leading advancement in this line of research greatly improves
the ordinary parton shower description of a given final state in association
with additional QCD radiation, by making use of higher order tree
level matrix elements. These matrix element-parton shower (\textsc{Meps})
merging procedures \cite{Catani:2001cc,Krauss:2002up,Lonnblad:2001iq,MLM} take parton level events, of assorted multiplicity,
from tree level event generators and carefully dress them with parton
showers, vetoing and/or weighting them in such a way as to yield inclusive
event samples smoothly populating all of phase space, while not overcounting
any regions. The introduction of exact high multiplicity matrix elements
acts to correct the radiation pattern with respect to the standalone
parton shower description. The distribution of the hard emissions,
and hence the distribution of the jets, follows from the real matrix
elements while only the internal structure of jets is determined by
the parton shower, where previously the former was given erroneously
by the multiple soft-collinear limits of those same matrix elements.

As with the\textsc{ Nlops} case, the\textsc{ Meps} methods are not
without their shortcomings. Since the underlying physics ingredients
from which the \textsc{Meps} simulations are constructed are tree
level matrix elements and leading-log resummation, they are subject
to many of the same theoretical uncertainties as leading order calculations.
Predictions for observables based on these calculations exhibit an
acute sensitivity to the renormalization and factorization scales,
predominantly affecting the overall normalization but also, to a lesser
extent, the shapes of distributions. These limitations are in contrast
to the \textsc{Nlops} output for which the complete set of $\mathcal{O}\left(\alpha_{\mathrm{S}}\right)$
corrections greatly reduces such ambiguities.

Plainly there is a high degree of complementarity between the \textsc{Nlops}
and \textsc{Meps} schemes in regards to their strengths and weaknesses.
Moreover, for both classes of event generators there now exists a significant
and rapidly growing number of phenomenologically important simulations.
In fact, the \textsc{Meps} method has been largely automated within
tree level event generators \cite{Mangano:2002ea,Alwall:2007st,Gleisberg:2008ta}
and further progress has been made toward automated production of
\textsc{Nlops} simulations \cite{Alioli:2010xd}. It is therefore
natural to look for a means of combining the two approaches, preserving
their virtues and forgoing their weaknesses. This problem may be approached
in several different ways. For example, if we rely upon the \textsc{Powheg}
method for our \textsc{Nlops} approach, what we would need is a \textsc{Meps}
procedure that can start from a given process with a given kinematics,
and builds higher multiplicity events on top of it. In other words,
we would like a \textsc{Meps} that behaves as any standard shower
Monte Carlo program. However, current \textsc{Meps} methods are not
designed to work in this way. They generate full event samples with
no constraints on their kinematics. So, if we want to follow this
direction, we may expect that a lot of work would be needed to reformulate
the whole \textsc{Meps} approach.

A theoretical formalism aiming at such a merging, albeit in the same vein as
the \textsc{MC@NLO} approach to \textsc{Nlops} matching, has been proposed in
Ref.\,\cite{Giele:2007di} and an implementation realised for the process
$\mathrm{H}\rightarrow\mathrm{gg}$ at \textsc{NLO}, including real emission
corrections for the first radiated gluon only. A further, more ambitious
endeavour, by Lavesson and L{\"{o}}nnblad \cite{Lavesson:2008ah}, aims instead
to augment the lowest order parton shower simulation with \emph{both} higher
multiplicity real \emph{and} virtual corrections, in the spirit of
\cite{Nagy:2007ty}. For the time being, the theoretical construction and
implementation is limited to the simulation of $\mathrm{e}^{+}\mathrm{e}^{-}$
collisions. We shall briefly return to discuss how this last work compares
to the one we shall propose at the end of
Section~\ref{sec:Combining-Powheg-and-Meps}.

In this paper we approach the problem of merging an \textsc{Meps} simulation
with an \textsc{Nlops} simulation in a radically different way. We compare
the \textsc{Meps} and \textsc{Nlops} approaches, identifying and quantifying
their best features, with a view to defining what is required to obtain
an exact theoretical solution of the merging problem. Motivated by
the presence of the large body of validated, trusted \textsc{Meps}
and \textsc{Nlops} simulations available today, we then seek to address
the question of how close one may get to achieving a theoretically
exact merging, simply by manipulating their event samples. We go on
to show that in this way one may obtain a \textsc{Menlops}
merging that is, in practice, very satisfactory. We will describe
the application of this merging method to $\mathrm{W}$ boson
production and to $\mathrm{t}\bar{\mathrm{t}}$ production in
hadronic collisions.

Although the method that we propose proves to be very much adequate for
practical applications, we emphasize that it is not an exact solution
to the matching problem. We have however achieved two goals: firstly,
we have found a practical method to merge \textsc{Meps} and \textsc{Nlops}
simulations that can be immediately applied to processes for which such
simulations already exist; second, we have clarified what is needed in
order to achieve a full theoretical solution of the merging problem.

We begin in Section~\ref{sec:Hardest-emission-xsecs} by briefly
reviewing key features of the \textsc{Powheg} formalism, in particular
the hardest emission cross section. We do not provide a summary
of the methods used to implement \textsc{Meps} algorithms. We will
simply assume that we have at hand an \textsc{Meps} simulation that
is capable of predicting small angle radiation with the same accuracy
as a shower Monte Carlo program and, at the same time, also has the
ability to describe high multiplicity jet production with leading
order accuracy. Based upon these assumptions, we derive a cross section
differential in the phase space variables associated with the hardest
emitted parton in the \textsc{Meps} approach. On this basis, we formulate
an exact theoretical solution to the matching problem. In
Section~\ref{sec:Combining-Powheg-and-Meps} we formulate our practical
matching prescription and, based upon the findings in
Section~\ref{sec:Hardest-emission-xsecs}, we derive its range of
applicability. In Section~\ref{sec:Results} we demonstrate the efficacy
of our approach using W boson and top-quark pair production
as case studies, elaborating on the conditions necessary for its success.
Our findings and conclusions are summarized in Section~\ref{sec:Conclusions}.

\section{Hardest emission cross section in P\textsc{\large owheg} and M\textsc{\large eps}{\large{}
}simulations\label{sec:Hardest-emission-xsecs}}

Since we shall frequently use the term \emph{NLO accuracy} in
the course of this work, and since the way in which this is manifest in
\textsc{Nlops} simulations generalises that of, more familiar, fixed order
calculations, before we begin, we wish to take a moment to clarify what
we mean by it. As an instructive example consider the case of $\mathrm{W}$
production in hadronic collisions. At leading order in perturbation theory,
the cross section is of zeroth order in the strong coupling constant and
the transverse momentum of the $\mathrm{W}$ is zero. When 
$\cal{O}(\alpha_{\mathrm{S}})$ corrections to the \textsc{LO} process are
included, inclusive observables that do not vanish at the Born level are
predicted with \textsc{NLO} accuracy, while those that do vanish are only
known with \textsc{LO} accuracy. Thus, for example, the prediction for the
$\mathrm{W}$ production cross section with the constraint
$p_T^{\mathrm{W}}<p_T^{\mathrm{cut}}$ is comprised of contributions of order
$1$ and order $\alpha_{\mathrm{S}}$, it is thus known with \textsc{NLO}
accuracy. On the other hand, the cross section with the cut
$p_T^{\mathrm{W}}>p_T^{\mathrm{cut}}$ vanishes at the Born level, thus the
\textsc{NLO} calculation of $\mathrm{W}$ production only yields a leading
order accurate prediction for this observable. Here, our use
of the term \textsc{NLO} accuracy is restricted to inclusive observables
that do not vanish at the Born level.

We further remark that, in \textsc{Nlops} simulations, in contrast to fixed
order calculations, the distinction made above is slightly more subtle.
While in the fixed order \textsc{NLO} calculation the virtual contribution
to the cross section in our example sits at the point $p_T=0$, in an
\textsc{Nlops} simulation it is spread out over the whole Sudakov region of
the $p_T$ distribution. \textsc{NLO} accuracy is thus also spread out in this
region in a physically consistent way.

The basic ideas behind the \textsc{Powheg} method are most readily introduced 
in the context of a simple example wherein the leading order process is comprised
of a single colour dipole with a massless parton. One can consider, for definiteness, the top quark decay process
$\mathrm{t}\rightarrow\mathrm{bW}$, neglecting the $\mathrm{b}$ quark mass and taking the
$\mathrm{W}$ boson to be stable.
In general we denote the leading order differential cross section $B\left(\Phi_{B}\right)$,
corresponding, in our example, to that of the $\mathrm{t}\rightarrow\mathrm{bW}$ process, 
parametrized by the so-called Born phase space variables $\Phi_{B}$. The differential
cross section for the real emission process is similarly denoted $R\left(\Phi_{R}\right)$,
where the phase space variables, $\Phi_{R}$, determine the kinematics of the relevant
processes; they are routinely defined in terms of the Born phase space variables $\Phi_{B}$
together with additional radiative phase variables $\Phi_{\mathrm{rad}}$ \emph{i.e.}
$\Phi_{R}=\Phi_{R}\left(\Phi_{B},\Phi_{\mathrm{rad}}\right)$. 
In the context of our heuristic example, the Born phase space is characterized by the direction of the
$\mathrm{W}$, while the radiation phase space may be described
by the angle of the emitted gluon with respect to the $\mathrm{W}$
direction, its azimuth and its energy. 

In the \textsc{Powheg} approach~\cite{Nason:2004rx}, the simulation
process starts with the generation of a two or three
body final state according to the distribution \begin{equation}
\mathrm{d}\sigma_{\mathrm{PW}}^{\mathrm{HE}}=\overline{B}\left(\Phi_{B}\right)\,\mathrm{d}\Phi_{B}\,\left[\Delta_{R}\left(p_{T}^{\min}\right)+\frac{R\left(\Phi_{R}\right)}{B\left(\Phi_{B}\right)}\,\Delta_{R}\left(k_{T}\left(\Phi_{R}\right)\right)\,\mathrm{d}\Phi_{\mathrm{rad}}\right],\label{eq:sec2_powheg_hard_emission_xsec}\end{equation}
 that represents the cross section for the hardest radiated particle in the inclusive process.\footnote{The superscript $\rm HE$ stands here for ``hardest emission''.} The function $\overline{B}\left(\Phi_{B}\right)$
is defined as \begin{equation}
\overline{B}\left(\Phi_{B}\right)=B\left(\Phi_{B}\right)+\left[V\left(\Phi_{B}\right)+\int\mathrm{d}\Phi_{\mathrm{rad}}\, R\left(\Phi_{R}\right)\right],\label{eq:sec2_bbar}\end{equation}
 where $B\left(\Phi_{B}\right)$ is the leading order contribution.
The virtual term $V\left(\Phi_{B}\right)$ has soft and collinear
divergences that cancel against the integral of the real term over
the radiation variables. We thus assume that, within the square bracket,
some regularization procedure (like dimensional regularization) is
adopted. The technicalities concerning how this formula is realised
in \textsc{Powheg} are highly complex; however, they are not directly relevant to the present
discussion (such details can be found in \emph{e.g.}
Refs.\,\cite{Frixione:2007vw,Alioli:2010xd}). The modified
Sudakov form factor is defined as \begin{equation}
\Delta_{R}\left(p_{T}\right)=\exp\left[-\int\mathrm{d}\Phi_{\mathrm{rad}}\,\frac{R\left(\Phi_{R}\right)}{B\left(\Phi_{B}\right)}\,\theta\left(k_{T}\left(\Phi_{R}\right)-p_{T}\right)\right],\label{eq:sec2_powheg_sudakov}\end{equation}
 where $k_{T}\left(\Phi_{R}\right)$ is equal to the transverse momentum
of the extra parton in the collinear and soft limits. We implicitly assume, as in a conventional
parton shower simulation, that $k_{T}$ has always an implicit lower cut-off $p_{T}^{\min}$
in Eq.\,\ref{eq:sec2_powheg_hard_emission_xsec}. Note also that the explicit dependence
of $\Delta_{R}$ on $\Phi_{B}$ has been suppressed for ease of notation.

We now briefly recount the key features of the \textsc{Powheg} formula through which
 \textsc{NLO} accuracy is achieved. First of all, in the large transverse momentum
region, the hardest emission cross section Eq.\,\ref{eq:sec2_powheg_hard_emission_xsec}
becomes, up to terms of higher order, equal to $R\left(\Phi_{R}\right)$. In fact,
for large transverse momenta only the second term in the square bracket of 
Eq.\,\ref{eq:sec2_powheg_hard_emission_xsec} contributes. Furthermore, the 
associated Sudakov form factor tends to one. Hence, for these kinematics, 
neglecting terms beyond NLO accuracy, we have
\begin{equation}
\mathrm{d}\sigma_{\mathrm{PW}}^{\mathrm{HE}}=\overline{B}\left(\Phi_{B}\right)\,\frac{R\left(\Phi_{R}\right)}{B\left(\Phi_{B}\right)}\,\mathrm{d}\Phi_{B}\,\mathrm{d}\Phi_{\mathrm{rad}}\approx R\left(\Phi_{R}\right)\,\mathrm{d}\Phi_{R}.\label{eq:sec2_powheg_accuracy}\end{equation}
Of equal importance in achieving \textsc{NLO} accuracy is the requirement that the
integral of the \textsc{Powheg} hardest emission cross section with respect to the
radiative phase space variables should be identical to that of the exact NLO cross
section, \emph{i.e.} equal to Eq.\,\ref{eq:sec2_bbar}. This property is guaranteed
by the form of the \textsc{Powheg} Sudakov form factor which, by construction, satisfies
the following identity \begin{equation}
\frac{\mathrm{d}\Delta_{R}\left(p_{T}\right)}{\mathrm{d}p_{T}}=\Delta_{R}\left(p_{T}\right)\,\int \frac{R\left(\Phi_{R}\right)}{B\left(\Phi_{B}\right)}\,\delta\left(k_{T}\left(\Phi_{R}\right)-p_{T}\right)\,\mathrm{d}\Phi_{\mathrm{rad}}.\label{eq:sec2_exact_differential}\end{equation}
Using this relation it is trivial to show that the term in square brackets in Eq.\,\ref{eq:sec2_powheg_hard_emission_xsec} integrates to one for all $\Phi_{B}$.

Note that taking the Sudakov form factor exactly as laid out in the original
\textsc{Powheg} proposal~\cite{Nason:2004rx} is not a strict requirement for attaining
\textsc{NLO} accuracy, merely the most convenient one. A modified
\textsc{Powheg} formula \begin{equation}
\mathrm{d}\sigma_{\mathrm{PW}}^{\mathrm{HE}}=\overline{B}\left(\Phi_{B}\right)\,\mathrm{d}\Phi_{B}\left[\frac{\Delta_{S}\left(p_{T}^{\min}\right)+\phantom{\int}\mathrm{d}\Phi_{\mathrm{rad}}\,\frac{R\left(\Phi_{R}\right)}{B\left(\Phi_{B}\right)}\,\Delta_{S}\left(k_{T}\left(\Phi_{R}\right)\right)}{\Delta_{S}\left(p_{T}^{\min}\right)+\int\mathrm{d}\Phi_{\mathrm{rad}}\,\frac{R\left(\Phi_{R}\right)}{B\left(\Phi_{B}\right)}\,\Delta_{S}\left(k_{T}\left(\Phi_{R}\right)\right)}\right],\label{eq:sec2_modified_powheg_hard_emission}\end{equation}
where $\Delta_{S}\left(p_{T}\right)$ is an alternative Sudakov form factor given by
\begin{equation}
\Delta_{S}\left(p_{T}\right)=\exp\left[-\int\mathrm{d}\Phi_{\mathrm{rad}}\,\frac{S\left(\Phi_{R}\right)}{B\left(\Phi_{B}\right)}\,\theta\left(k_{T}\left(\Phi_{R}\right)-p_{T}\right)\right],\label{eq:sec2_modified_sudakov}
\end{equation}
would satisfy the same properties as Eq.\,\ref{eq:sec2_powheg_hard_emission_xsec}, provided
that $S\left(\Phi_{R}\right)$ coincides with $R\left(\Phi_{R}\right)$ in the regions of phase
space corresponding to soft and collinear emissions. In
other words, to achieve \textsc{NLO} accuracy it is mandatory that the term in square
bracket is unitary, \emph{locally} in the Born phase space. However, there is some
flexibility regarding precisely how that unitarity is achieved.

In the \textsc{Powheg} framework the parton shower simulation is promoted to
\textsc{NLO} accuracy by distributing non-radiative events according to the
first term in Eq.\,\ref{eq:sec2_powheg_hard_emission_xsec} and the hardest
(highest $p_{T}$) emission according to the second term. Whereas in a
conventional parton shower simulation, an \emph{N}-body configuration is
generated according to $B\left(\Phi_{B}\right)$ and then showered using a
conventional, process-independent, Sudakov form factor, the \textsc{Powheg}
technique requires that the \emph{N}-body configuration is generated instead
according to $\overline{B}\left(\Phi_{B}\right)$ and showered with the
process-dependent modified Sudakov form factor in Eq.\,\ref{eq:sec2_powheg_sudakov}.
Inclusive observables computed using events generated from this distribution
have full \textsc{NLO} accuracy, in contrast to the corresponding predictions
from a conventional parton shower simulation, which are only \textsc{LO} accurate.

With these points in mind we move to express, in similar terms, the analogous
cross section in \textsc{Meps} based simulations. In this way we aim to clarify 
the differences between the \textsc{Meps} and \textsc{Powheg} methods, in order 
to guide us in attempting to consistently combine the two. To this end, we need
only assume that event generators which utilise these merging methods, are capable
of correctly describing widely separated jets of any multiplicity according to the
leading order matrix elements, and small angle radiation according to the leading
logarithmic approximation. In simpler words, we will assume that \textsc{Meps}
algorithms have the same accuracy as a shower Monte Carlo in the small angle limit,
and leading order QCD accuracy for jet cross sections, even for widely separated jets.

Momentarily, putting aside the fact that a very small fraction of events contain
no radiation at all, given an otherwise arbitrary \textsc{Meps} event, we may cluster
it according to a $k_{\perp}$ jet algorithm, until it is resolved as a 1-jet event.
The kinematics of this 1-jet structure may be parametrized in terms of the real emission
phase space $\Phi_{R}$, which may in turn be expressed in terms of the Born and
radiative phase space variables $\Phi_{B}$ and $\Phi_{\mathrm{rad}}$ introduced
earlier. Effectively the
jet algorithm defines a pair of \emph{unique} mappings $\widehat{\Phi}_{B}\left(\Phi\right)$
and $\widehat{\Phi}_{\mathrm{rad}}\left(\Phi\right)$ from the phase space of arbitrary
multiplicity \textsc{Meps} events, generically denoted $\Phi$, to the Born and radiative
phase spaces respectively. 

Having specified the Born and radiative phase space projections, the \textsc{Meps}
hardest emission cross section, for radiative events, follows as
\begin{eqnarray}
 &  & \int\mathrm{d}\Phi\,\frac{\mathrm{d}\sigma_{\mathrm{ME}}}{\mathrm{d}\Phi}\,\delta\left(\Phi_{B}-\widehat{\Phi}_{B}\left(\Phi\right)\right)\,\delta\left(\Phi_{\mathrm{rad}}-\widehat{\Phi}_{\mathrm{rad}}\left(\Phi\right)\right)\nonumber \\
 & = & \widehat{R}\left(\Phi_{R}\right)\Delta_{\widehat{R}}\left(k_{T}\left(\Phi_{R}\right)\right)\,,\label{eq:sec2_C_function}\end{eqnarray}
where $\widehat{R}(\Phi_{R})$ differs from $R(\Phi_{R})$ by a factor $1+\cal{O}\left(\alpha_{\mathrm{S}}\right)$.
$\Delta_{\widehat{R}}\left(k_{T}\left(\Phi_{R}\right)\right)$
is an effective Sudakov form factor, of equivalent logarithmic accuracy to those
used in \textsc{Powheg} and conventional parton shower simulations.
Equation (\ref{eq:sec2_C_function}) follows directly from our assertions regarding
the \textsc{Meps} algorithms, namely, that widely separated jets are described with
leading order accuracy (this is why we recover $R\left(\Phi_{R}\right)$ up to terms
formally of higher order in $\alpha_{S}$), while radiation in the soft and collinear
regions must be described as accurately as in a parton shower Monte Carlo (this is
why we recover the $\Delta_{\widehat{R}}\left(k_{T}\left(\Phi_{R}\right)\right)$ factor).
The difference between the real emission cross section used in \textsc{NLO} calculations,
$R\left(\Phi_{R}\right)$, and the \textsc{Meps} approximation to it,
$\widehat{R}\left(\Phi_{R}\right)$, arises from spurious higher order terms, of
NNLO significance, that will in general be present in an \textsc{Meps} algorithm.

We can now express the hardest jet cross section in an \textsc{Meps}
simulation in a similar form to that of the \textsc{Powheg} hardest
emission cross section \emph{viz}
\begin{equation}
\mathrm{d}\sigma_{\mathrm{ME}}^{\mathrm{HE}}=B\left(\Phi_{B}\right)\,\mathrm{d}\Phi_{B}\,\left[\Delta_{\widehat{R}}\left(p_{T}^{\min}\right)+\frac{\widehat{R}\left(\Phi_{R}\right)}{B\left(\Phi_{B}\right)}\,\Delta_{\widehat{R}}\left(k_{T}\left(\Phi_{R}\right)\right)\,\mathrm{d}\Phi_{\mathrm{rad}}\right].\label{eq:sec2_ME_hard_emission_xsec}\end{equation}
Following the manipulations leading to Eq.\,\ref{eq:sec2_modified_powheg_hard_emission}
we proceed to rewrite the \textsc{Meps} cross section in such a way that the piece
corresponding to the generation of the radiation is manifestly unitary: \begin{equation}
\mathrm{d}\sigma_{\mathrm{ME}}^{\mathrm{HE}}=\overline{B}_{\mathrm{ME}}\left(\Phi_{B}\right)\,\mathrm{d}\Phi_{B}\,\frac{\Delta_{\widehat{R}}\left(p_{T}^{\min}\right)+\phantom{\int}\mathrm{d}\Phi_{\mathrm{rad}}\,\frac{\widehat{R}\left(\Phi_{R}\right)}{B\left(\Phi_{B}\right)}\,\Delta_{\widehat{R}}\left(k_{T}\left(\Phi_{R}\right)\right)}{\Delta_{\widehat{R}}\left(p_{T}^{\min}\right)+\int\mathrm{d}\Phi_{\mathrm{rad}}\,\frac{\widehat{R}\left(\Phi_{R}\right)}{B\left(\Phi_{B}\right)}\,\Delta_{\widehat{R}}\left(k_{T}\left(\Phi_{R}\right)\right)}\,,\label{eq:sec2_modified_ME_hard_emission}\end{equation}
 where \begin{equation}
\overline{B}_{\mathrm{ME}}\left(\Phi_{B}\right)=B\left(\Phi_{B}\right)\left[\Delta_{\widehat{R}}\left(p_{T}^{\min}\right)+\int\mathrm{d}\Phi_{\mathrm{rad}}\,\frac{\widehat{R}\left(\Phi_{R}\right)}{B\left(\Phi_{B}\right)}\,\Delta_{\widehat{R}}\left(k_{T}\left(\Phi_{R}\right)\right)\right].\end{equation}
 It is also clear that \begin{equation}
\overline{B}_{\mathrm{ME}}\left(\Phi_{B}\right)=\Delta_{\widehat{R}}\left(p_{T}^{\min}\right)+\int\,\mathrm{d}\Phi\,\frac{\mathrm{d}\sigma_{\mathrm{ME}}}{\mathrm{d}\Phi}\,\delta\left(\Phi_{B}-\widehat{\Phi}_{B}\left(\Phi\right)\right).\text{ }\label{eq:sec2_ME_bbar}\end{equation}
 We thus state the following simple result: in order to achieve \textsc{NLO}
accuracy in a \textsc{Meps} simulation it is sufficient to reweight the
events with a factor \begin{equation}
\frac{\overline{B}\left(\widehat{\Phi}_{B}\left(\Phi\right)\right)}{\overline{B}_{\mathrm{ME}}\left(\widehat{\Phi}_{B}\left(\Phi\right)\right)}\,,\label{eq:sec2_Powheg_to_ME_bbar}\end{equation}
 where, as before, $\Phi$ represents the kinematics of an arbitrary multiplicity
\textsc{Meps} event and $\overline{B}_{\mathrm{ME}}$ is given by Eq.\,\ref{eq:sec2_ME_bbar}.

Although easy to state, Eq.~\ref{eq:sec2_Powheg_to_ME_bbar}
may turn out in practice to be very difficult to use. In fact, only for
simple processes one may be able to compute the ratio $\bar{B}/\bar{B}^{\rm ME}$
and store it in a sufficiently dense grid of points in the Born phase space,
such that given any Born phase space configuration the value of the ratio may
be interpolated with enough precision.

\section{Combining P\textsc{\large owheg} and M\textsc{\large eps} samples\label{sec:Combining-Powheg-and-Meps}}

In this section we shall first discuss the relative merits of jet cross
sections and their constituent events in \textsc{Nlops} and \textsc{Meps}
simulations. Based on the exact merging outlined in
Section~\ref{sec:Hardest-emission-xsecs} and the following jet cross section
analysis, we propose an approximate \textsc{Menlops} scheme requiring
no modifications to existing codes. As we shall see later, for some LHC
processes the exact approach we advocated earlier can be totally obviated
by this simplified scheme.

\subsection{Jet cross sections\label{sec:Jet-cross-sections}}

We will now examine the \textsc{Powheg} and \textsc{Meps} samples
by clustering their events according to a given jet resolution scale
$y_{0}$. We assume that the clustering parameter is related to the
transverse momentum \emph{i.e.} that it is of the Durham variety.
The events will be thus characterized by the number of jets at the
given $y_{0}$ value. We will still stick to our example, where only
a single massless coloured parton is present in the external leg of
our basic process. We begin by comparing the differential cross sections
for the production of events in which no additional radiated jets
are present. In \textsc{Powheg}, this is given by \begin{equation}
\mathrm{d}\sigma_{\mathrm{PW}}\left(0\right)=\overline{B}\left(\Phi_{B}\right)\,\mathrm{d}\Phi_{B}\,\left[\Delta_{R}\left(p_{T}^{\min}\right)+\int\mathrm{d}\Phi_{\mathrm{rad}}\,\frac{R\left(\Phi_{R}\right)}{B\left(\Phi_{B}\right)}\,\Delta_{R}\left(k_{T}\left(\Phi_{R}\right)\right)\,\theta\left(y_{0}-y\left(\Phi_{R}\right)\right)\right],\label{eq:sec3_powheg_0_jet_a}\end{equation}
depending upon the Born variables alone. Equation~\ref{eq:sec3_powheg_0_jet_a}
is obtained by assuming that clustering showered \textsc{Powheg} events
to the point where only one radiated jet is resolved recovers the basic
\textsc{Powheg} cross section Eq.\,\ref{eq:sec2_powheg_hard_emission_xsec}.
This is certainly the case for the non-emission term, and also for
the radiation term, since when \textsc{Powheg} is interfaced to a parton
shower Monte Carlo it is forbidden to generate radiation harder than the
\textsc{Powheg} generated one. Formula~\ref{eq:sec3_powheg_0_jet_a} is clearly
leading-log accurate when $y_{0}$ is very small. Appealing to unitarity
it can be rewritten in the following way \begin{equation}
\mathrm{d}\sigma_{\mathrm{PW}}\left(0\right)=\overline{B}\left(\Phi_{B}\right)\,\mathrm{d}\Phi_{B}\,\left[1-\int\mathrm{d}\Phi_{\mathrm{rad}}\,\frac{R\left(\Phi_{R}\right)}{B\left(\Phi_{B}\right)}\,\Delta_{R}\left(k_{T}\left(\Phi_{R}\right)\right)\,\theta\left(y\left(\Phi_{R}\right)-y_{0}\right)\right].\label{eq:sec3_powheg_0_jet_b}\end{equation}
When $y_{0}$ is not small, the factor in square bracket differs from
one by terms of order $\alpha_{s}$, and the full formula is accurate at
\textsc{NLO}. 

The corresponding cross section in the \textsc{Meps} simulation is given by an
expression of the form 
\begin{equation}
\mathrm{d}\sigma_{\mathrm{ME}}\left(0\right)=\overline{B}_{\mathrm{ME}}\left(\Phi_{B}\right)\,\mathrm{d}\Phi_{B}\,\left[1-\int\mathrm{d}\Phi_{\mathrm{rad}}\,\,\frac{\widehat{R}\left(\Phi_{R}\right)}{\overline{B}_{\mathrm{ME}}\left(\Phi_{B}\right)}\,\Delta_{\widehat{R}}\left(k_{T}\left(\Phi_{R}\right)\right)\,\theta\left(y\left(\Phi_{R}\right)-y_{0}\right)\right],\label{eq:sec3_meps_0_jet}\end{equation}
 where $\Delta_{\widehat{R}}$ is an effective Sudakov form factor
accounting for the combination of the \textsc{ME} and \textsc{PS} Sudakov form
factors in the \textsc{Meps} algorithm and, for brevity,
$\Phi_{B}=\widehat{\Phi}_{B}\left(\Phi\right)$.
Thus, from the point of view of \textsc{NLO} accuracy, the \textsc{Meps} result
differs from the \textsc{Powheg} one by the weight factor in
Eq.~\ref{eq:sec2_Powheg_to_ME_bbar}, with remaining differences, due to the terms
in the square brackets, being of relative order $\alpha_{S}^{2}$. Moreover, we
point out that this weight factor is in fact the \emph{only} difference between
the \textsc{Meps} result and that which would be obtained using the exact merging
procedure outlined in Section~\ref{sec:Hardest-emission-xsecs}. It is therefore
clear that the \textsc{Powheg} prediction for this quantity is always better
than the \textsc{Meps} one.

We now examine the cross section for radiating a single additional jet.
In \textsc{Powheg} it is given by \begin{equation}
\mathrm{d}\sigma_{\mathrm{PW}}\left(1\right)=\overline{B}\left(\Phi_{B}\right)\,\mathrm{d}\Phi_{B}\,\left[\frac{R\left(\Phi\right)}{B\left(\Phi_{B}\right)}\,\Delta_{R}\left(k_{T}\left(\Phi\right)\right)\,\theta\left(y\left(\Phi\right)-y_{0}\right)\mathrm{d}\Phi_{\mathrm{rad}}\right]\Delta_{\mathrm{MC}}\left(y_{0}\right).\label{eq:sec3_powheg_1_jet}\end{equation}
 This is equivalent to the cross section for the first radiation to
be above the clustering scale, times and extra factor $\Delta_{\mathrm{MC}}\left(y_{0}\right)$,
that represents the probability that the subsequent shower does not
generate more jets. It is often stated that, as far as the radiation
cross section is concerned, \textsc{Meps} and \textsc{Powheg} are equivalent.
This is certainly the case if the clustering scale $y_0$ is large enough.
In this limit $\Delta_{\rm MC}(y_0)$ differs from 1 by terms of higher order,
and the one jet cross section itself becomes of order $\alpha_{\mathrm{S}}$.
However, as the
clustering scale becomes smaller, and the fraction of one jet events becomes
a sizeable fraction of the total cross section, we should recall that an NLO
$K$-factor becomes visible in the \textsc{Powheg} cross section that is
not present in the \textsc{Meps} one. Thus, there is at least one limiting case
in which the  \textsc{Powheg} cross section is better than the
\textsc{Meps} one, and so \textsc{Powheg} should be preferred for this quantity.

As we go to higher jet multiplicity, however, the \textsc{Meps} becomes
more accurate than the \textsc{Powheg} approach. In fact, the cross
section for more than one jet is determined in the \textsc{Powheg}
approach in part by the generation of the hardest jet, and in part
by the shower Monte Carlo, that will generate the second jet. The
whole cross section will thus be accurate only in the kinematic region
where the second jet is either collinear to the first jet, or (depending
upon the Shower Monte Carlo ability to predict correctly soft emissions)
when it is soft. On the other hand, in the \textsc{Meps} approach,
this cross section will be correctly predicted at all angles. 

\subsection{A simple \textsc{\large Menlops} merging procedure\label{sec:Menlops-Recipe}}

Based on our deductions in Section.~\ref{sec:Jet-cross-sections}, regarding
the accuracy of the jet cross sections and the description of the events which
comprise them, we propose that the exact reweighting method outlined in
Section~\ref{sec:Hardest-emission-xsecs} may be very well approximated by simply
mixing the 0- and 1-jet events output from a \textsc{Powheg} simulation, together
with events including at least two jets output from an \textsc{Meps} simulation.
If the fraction of events with more than one jet in the final \textsc{Menlops} 
sample is at least as small as $\alpha_{s}$ relative to the total, \textsc{NLO}
accuracy for shape variables will be clearly preserved, and the LO description
of high multiplicity jet samples will also be retained. 


The proportions in which the various contributions should be mixed are non-trivial,
they are chosen so as to respect our assertions regarding the accuracy of the
various jet cross sections in the different approaches. 
Let us denote the total
\textsc{Powheg} cross section for $j$ jets by $\sigma_{\mathrm{PW}}\left(j\right)$
and the corresponding \textsc{Meps} cross section by $\sigma_{\mathrm{ME}}\left(j\right)$.
By analogy we label the fully differential cross section for \textsc{Powheg} events
containing $j$ jets by $\mathrm{d}\sigma_{\mathrm{PW}}\left(j\right)$, and that of
their \textsc{Meps} counterparts by $\mathrm{d}\sigma_{\mathrm{ME}}\left(j\right)$.
Note that, by fully differential, we mean differential in the momenta of \emph{all}
produced particles, after showering has taken place. In other words, \begin{equation}
\mathrm{d}\sigma_{\mathrm{PW}} = \sum_{j}\mathrm{d}\sigma_{\mathrm{PW}}\left(j\right)\label{eq:sec3_powheg_differential_in_particles}\end{equation}
 represents the differential cross section for multi-particle production
as simulated by the \textsc{Powheg} algorithm. We also use the
notation $(\geqslant j)$ to indicate the total or differential
cross section for a number of jets greater than or equal to $j$.
We build a
sample by combining \textsc{Meps} and \textsc{Powheg} event samples
according to their jet multiplicities in the following proportions
\begin{equation}
\mathrm{d}\sigma = \mathrm{d}\sigma_{\mathrm{PW}}\left(0\right)+\frac{\sigma_{\mathrm{ME}}\left(1\right)}{\sigma_{\mathrm{ME}}\left(\geqslant1\right)}\,\frac{\sigma_{\mathrm{PW}}\left(\geqslant1\right)}{\sigma_{\mathrm{PW}}\left(1\right)}\,\,\mathrm{d}\sigma_{\mathrm{PW}}\left(1\right)+\frac{\sigma_{\mathrm{PW}}\left(\geqslant1\right)}{\sigma_{\mathrm{ME}}\left(\geqslant1\right)}\,\,\mathrm{d}\sigma_{\mathrm{ME}}\left(\geqslant2\right).\label{eq:sec3_menlops_master_formula}\end{equation}
Notice that the total cross section for the combined sample equals that of
\textsc{Powheg}.
Plainly the 0-jet cross section is as generated by \textsc{Powheg}
alone, as is the 1-jet cross section, except for the overall factor
\begin{equation}
\frac{\sigma_{\mathrm{ME}}\left(1\right)}{\sigma_{\mathrm{ME}}\left(\geqslant1\right)}\,\frac{\sigma_{\mathrm{PW}}(\geqslant1)}{\sigma_{\mathrm{PW}}(1)}.\label{eq:sec3_menlops_1_jet_Kfactor}\end{equation}
 In other words, the total 1-jet fraction is corrected, as if it was
assumed that the ratio of the 1-jet fraction to the $\geqslant1$-jet
fraction is better determined by the \textsc{Meps} program. This is
in fact the case if $y_{0}$ is not too small. The cross section for
two or more jets is instead given by \begin{equation}
\mathrm{d}\sigma\left(\geqslant2\right) = \frac{\sigma_{\mathrm{PW}}\left(\geqslant1\right)}{\sigma_{\mathrm{ME}}\left(\geqslant1\right)}\,\mathrm{d}\sigma_{\mathrm{ME}}\left(\geqslant2\right),\label{eq:sec3_menlops_2_jet_Kfactor}\end{equation}
 \emph{i.e.} it carries an extra overall $K$-factor with respect to the
bare \textsc{Meps} result, given precisely by the \textsc{NLO} $K$-factor
for the $\geqslant1$-jet cross section.

We now discuss to what extent the procedure outlined above retains
the best features of the \textsc{Powheg} and \textsc{Meps} approaches.
There are two questions to answer. The first one is to what extent
the proposed procedure yields the correct \textsc{NLO} cross
sections for inclusive quantities. The second one is to what extent
jet cross sections for widely separated jets are generated according
to the exact leading order matrix elements. As far as \textsc{NLO}
accuracy is concerned, it is clear that a problem may arise from the
$(\geqslant2)$ sample. The contribution of this sample to inclusive
quantities does not include the \textsc{NLO} corrections with their
full dependence on the underlying Born kinematics. It thus violates
the \textsc{NLO} accuracy of the calculation. However, if $y_{0}$
is not too small, the $\geqslant2$ jets contribution to the cross
section is, relatively, an effect of order $\alpha_{s}^{2}$.
It is enough for us to choose $y_{0}$ such that this fraction is not
larger than $\alpha_{s}$ to maintain \textsc{NLO} accuracy of the full
sample. Observe also that the presence of the $K$ factor in
Eq.\,\ref{eq:sec3_menlops_2_jet_Kfactor} improves the situation.
In other words, even if we are not capable to correct the $\geqslant2$
jets sample with the factor of Eq.\,\ref{eq:sec2_Powheg_to_ME_bbar},
we can at least correct the overall rate in such a way that if the
factor in Eq.\,\ref{eq:sec2_Powheg_to_ME_bbar} is constant the correction
becomes exact. Conversely, the $\geqslant2$-jet sample is certainly
accurate for jet production at large angles and with large multiplicity.
The 0- and 1-jet samples, however, can have jet substructures at relatively
large angles, that thus violate ME accuracy. This is certainly the
case if $y_{0}$ is too large. In practice, we must thus require $y_{0}$
to have a value which is small enough to be acceptable for \textsc{Meps}
matching, but large enough so that the $\geqslant2$ jet sample comprises
a relative fraction no greater than ${\cal O}\left(\alpha_{s}\right)$.
Notice that also in this case we apply a constant correction factor to
the 1-jet fraction, such that the ratio of 1 to $\geqslant1$ jet is equal
to the \textsc{Meps} prediction.

The tension in the choice of $y_{0}$ (neither too large, nor too
small) is what prevents this method from being an exact solution to
the \textsc{Meps}-\textsc{Nlops} merging problem.
As in typical \textsc{Meps} matching methods, one expects
that making the clustering parameter small should yield at some point
the correct answer. This is not the case in the present method. By
making $y_{0}$ too small the fraction of $\geqslant2$ jet events
becomes substantial, yielding a contribution that does not
correctly include the \textsc{NLO} corrections (in order for it to be correct
at \textsc{NLO}, we would need to include the factor in
Eq.\,\ref{eq:sec2_Powheg_to_ME_bbar}, whereas we only include
a constant $K$-factor).

We now briefly comment on the method of Ref.\,\cite{Lavesson:2008ah}, which
differs markedly from our approach. \textsc{NLO} accuracy is achieved there
by computing jet distributions at \textsc{NLO} to begin with, using a clustering
scale called $y_{MS}$. In order for the method to work, this clustering scale
has to be set large enough so that most of the Sudakov region is already included
by it; in the example of $\mathrm{W}$ production, this means that the 0-jet cross
section should already include most of the total cross section. This is more
restrictive than in our method, for which we only require that most of the total
cross section be confined to the 0- \emph{and} 1-jet samples, thereby allowing
for a lower merging scale.

In finishing, we wish to emphasize that, as far as
the most simple processes are concerned, a complete solution of the
merging problem straightforwardly follows from the discussion presented
so far.
For example, in case of Higgs production,
the underlying Born kinematics depends upon a single parameter. In
the FKS implementation of \textsc{Powheg} for Higgs production, such
parameter is the rapidity of the Higgs. One can easily compute and
parametrize the factor in Eq.\,\ref{eq:sec2_Powheg_to_ME_bbar},
and use it to reweight the ME sample. For more complex processes,
we have at least clarified what corrections are needed to obtain a
full solution to the matching problem. However, from the studies reported
in the following pages, we also stress that, depending on the process,
the practical gain of
an exact implementation with respect to our proposed method,
should be carefully assessed, since it may
only be marginal, and may not be worth the effort.

\section{Results\label{sec:Results}}

\noindent In order to assess our proposal we have applied it to two
processes, $\mathrm{W}$ production and top quark pair production.
Besides being of considerable phenomenological significance in their
own right, these processes represent a reasonably wide range testing
ground; on the one hand $\mathrm{W}$
production is a quark anti-quark annihilation
process with a relatively low mass final-state, while on the other,
$\mathrm{t}\bar{\mathrm{t}}$ pair production is predominantly gluon
initiated with a high invariant mass final-state. Here our intention
is to demonstrate the effectiveness of the \textsc{Menlops} approach
and to give a more quantitative understanding of its domain of applicability.
The analyses presented in this section are therefore carried out at
the parton level, after parton showering, without applying acceptance
cuts. 

\noindent \smallskip{}

\noindent The following conventions are adopted throughout for the
histograms:
\begin{itemize}
\item Dashes (red) - the pure \textsc{Nlops} result
\item Dots (green) - the pure \textsc{Meps} result \emph{rescaled} by a global
\emph{K}-factor: $\sigma_{\mathrm{PW}}\left(\ge0\right)/\sigma_{\mathrm{ME}}\left(\ge0\right)$
\item Solid (blue) - the \textsc{Menlops} sample
\item Dashes with ``$\times$'' symbols (red) - the \textsc{Nlops} component of the \textsc{Menlops} result
\item Dots with ``$+$'' symbols (green) - the \textsc{Meps} component of the \textsc{Menlops} result
\end{itemize}
Unless stated otherwise, in each plot the jet resolution scale is
the same as the merging scale used to create the corresponding \textsc{Menlops}
sample. 

\noindent \smallskip{}

\subsection{P\textsc{owheg} and M\textsc{eps} simulations}

For both processes under study \textsc{Meps} merged samples were generated
using the \textsc{Madgraph} package \cite{Alwall:2007st}. This program
employs the \textsc{MLM} merging scheme,\emph{ }with minor differences
in the form of the cuts used for the generation of the tree level
events, and in the use of the $k_{\perp}$-jet measure 
\cite{Ellis:1993tq,Catani:1993hr},
as opposed to a cone jet measure, to perform the parton-jet matching.
The P{\footnotesize YTHIA }\cite{Sjostrand:2006za} virtuality ordered
parton shower is used to simulate radiation from the external legs
of the events generated according to tree-level matrix elements. This
implementation is referred to as the $k_{\perp}$-jet \textsc{MLM}
scheme in the documentation \cite{Madgraph:2007}, a full account
of which is given in Ref.\,\cite{Alwall:2007fs}. 

\textsc{Nlops} $\mathrm{W}$ and top-quark pair production events
were simulated using the \textsc{Powheg}-w \cite{Alioli:2008gx} and
\textsc{Powheg}-hvq \cite{Frixione:2007nw,Frixione:2007nu} codes
respectively, showering the output Les Houches event files with the
same P{\footnotesize YTHIA } library included in \textsc{Madgraph}. In
order to be completely faithful to the \textsc{Powheg} formalism,
in showering the events from the Les Houches files we have opted to
use the transverse momentum ordered P{\footnotesize YTHIA }shower
algorithm, setting the starting scale for each event to the value
given in the Les Houches file, i.e. the transverse momentum of
the hardest emission. Other types of shower algorithms, not ordered
in transverse momentum, may be used, but care should then be taken
to veto emissions which have a transverse momentum greater than that
of the hardest in the input \textsc{Powheg} event. 

Although the \textsc{Nlops} samples used in obtaining the final results
were generated in complete adherence to the \textsc{Powheg} formalism,
we have also experimented with various combinations of shower orderings
and starting scales. In all cases we found only small, inconsequential
differences. In particular, we have repeated the following analyses
using the P{\footnotesize YTHIA} virtuality ordered shower algorithm,
with the so-called \emph{Herwig} \emph{scale} as the initial condition,
when showering \textsc{Powheg} events. This scale is given by the
invariant mass of the least massive pair of colour connected particles
in the parton level event.
Hence, this amounts to using the same \textsc{Meps}
showering apparatus for the \textsc{Nlops} events. The differences
seen with respect to the results shown here were marginal and of no
interest. This is perhaps not surprising given that the pair of colour
connected partons with the lowest invariant mass is generally that
with the smallest relative transverse momentum, which in the \textsc{Powheg}
case is essentially always given by that of the radiated parton. 

In choosing the tree level event generation parameters and the \textsc{Meps}
merging scale, we have closely followed the settings recommended in
Refs.\,\cite{Madgraph:2007,Alwall:2007fs,Alwall:2008qv} for applying
the $k_{\perp}$-jet \textsc{MLM} scheme to $\mathrm{W}$ boson and
top quark pair production at LHC. In the event that an alternative
value of a parameter was not advised, the default value in the \textsc{Madgraph}
program was used.

Excepting the differences in the choice of shower algorithm used for
the \textsc{Nlops} and \textsc{Meps} samples, we aimed to have the
remaining inputs as consistent as possible in generating the two.
In particular, we have used the same parton density functions in the
\textsc{Powheg}, \textsc{Madgraph} and P{\footnotesize YTHIA} programs,
\textsc{MRST 2002 NLO} \cite{Martin:2002aw}, provided in all cases
through the \textsc{LHAPDF} interface \cite{Whalley:2005nh}. In all
programs the top quark and $\mathrm{W}$ boson masses have been duly
set to 174.3 and 80.419 GeV; similarly, a value of 2.124 GeV was used
for the width of the $\mathrm{W}$ boson. The default \textsc{Madgraph}
input was also adjusted in order to include the contributions of $\mathrm{b}$-quarks
in the initial- and final-state, as in the \textsc{Powheg} simulations.
Finally, we have generated our event samples assuming the nominal
LHC hadronic centre-of-mass energy, $\sqrt{s}=14\,\mathrm{TeV}$.

The authors of Ref.\,\cite{Alwall:2008qv} observe that the emission
spectrum from the transverse momentum ordered shower tends to be
markedly harder than that of the virtuality ordered shower, modulo
differences in the starting scales. This enables the former to
populate additional, higher $p_{T}$, regions of phase space which the
latter fails to reach. It is argued on these grounds that this shower
approximation therefore has a greater range of validity, justifying a
higher value of the \textsc{Meps} merging scale. We prefer to
interpret this observation more cautiously. Naturally, by emitting
radiation where previously there was none, the transverse momentum
ordered shower will produce results in seemingly better agreement with
exact tree level, resummed, \textsc{Meps} predictions. This is
nevertheless an approximation, merely a less conservative one than
that obtained with the virtuality ordered shower, as evidenced by the
uncertainties surrounding the starting scales \cite{Alwall:2008qv}.
Because of this reason, and since using higher scales for merging would
result in an advantage for the application of our method, we opt to
generate the \textsc{Meps} sample with the virtuality ordered shower
and the 30 GeV \textsc{Meps} merging scale in the present work,
in order not to diminish its value as proof of concept of our method.

\subsection{M\textsc{enlops} implementation\label{sub:MENLOPS-implementation}}

We have realised the merging algorithm described in
Section~\ref{sec:Hardest-emission-xsecs} by dividing each
\textsc{Nlops} and \textsc{Meps} sample into three sub-samples: a
sample containing events with no additional radiated jets, a sample
containing events with one radiated jet and a third sample comprised
of events with greater than one jet. In the \textsc{Nlops} case the
latter sample is discarded, while in the \textsc{Meps} case it is kept
and instead the other two are deleted. The fraction of events
in the final Menlops sample with no additional jets is given by the
same fraction found in the \textsc{Nlops} sample. The fraction of
events with one additional jet in the \textsc{Menlops} sample is given
by one minus this 0-jet fraction, multiplied by the ratio of the
number of 1-jet events to the number with one or more jets in the
\textsc{Meps} sample.

The exact choice of jet measure used to divide up the samples is not
particularly important beyond the requirement that it be infrared
safe, to avoid degrading, or losing altogether, \textsc{NLO} accuracy.
To obtain the results in this section we used the $k_{\perp}$-jet
measure \cite{Ellis:1993tq,Catani:1993hr}, as implemented in the
FastJet jet finder package \cite{Cacciari:2005hq}. Specifically,
the $k_{\perp}$ separation between between two final state particles
$i$ and $j$ is defined to be\begin{equation}
d_{ij}=\min\left(k_{\perp i}^{2},\, k_{\perp j}^{2}\right)\,\Delta R_{ij}^{2}/R\,,\label{eq:results_k_perp_jet_measure}\end{equation}
with $\Delta R_{ij}^{2}=\left(Y_{i}-Y_{j}\right)^{2}+\left(\phi_{i}-\phi_{j}\right)^{2}$,
where $k_{\perp i}$, $Y_{i}$ and $\phi_{i}$ are the transverse
momentum, rapidity and azimuth of particle $i$. The factor $R$ is
a jet-radius parameter which has been set equal to one in our analysis.
In addition, in this scheme the \emph{beam distance} is defined as
$d_{iB}=k_{\perp i}^{2}$. 

The \textsc{Menlops} merging scale according to which the \textsc{Meps}
and \textsc{Nlops} are sorted into 0-, 1- and greater than 1-jet
samples is defined as a cut in the $k_{\perp}$ separation measure,
that is $y_{0}=\sqrt{d_{\mathrm{cut}}}$ (see Sect.\,\ref{sec:Combining-Powheg-and-Meps}).
In the following analyses we shall also present distributions of the
differential jets rates, where we query each event to establish the
threshold in $y_{nm}=\sqrt{d_{nm}}$ at which an $n$-jet event is
resolved as an $m=n+1$-jet event.

\subsection{W boson production\label{sub:W-boson-production}}

We now turn to discuss the results obtained using \textsc{Menlops}
samples for the case of $\mathrm{W}^{-}$ boson production. We consider
the case wherein the $\mathrm{W}$ decays to an electron and neutrino.
Naturally, these leptons are excluded from the jet finding procedure.
Hence, when no additional radiation
occurs, the jet finding algorithm returns 0 jets.

To generate the \textsc{Meps} sample with \textsc{Madgraph} we have
used a $k_{\perp}$ jet measure cut on the tree level event generation
of 15 GeV and taken the \textsc{Meps} merging scale to be 20 GeV.
These are the values advocated for the generation of inclusive $\mathrm{W}$
production samples in Refs.\,\cite{Alwall:2007fs,Madgraph:2007}. 

The default \textsc{Menlops} sample used to produce the results in
this subsection was constructed by combining the \textsc{Nlops} and
\textsc{Meps} samples with a merging scale of 25 GeV, only 5 GeV above
the \textsc{Meps} merging scale. The \textsc{Menlops} cross section
is equal to that of the \textsc{Nlops} event generation, 8150 pb,
and the fraction of \textsc{Meps} events in the total sample is 5\%,
safely within the recommended maximum fraction $\alpha_{\mathrm{S}}$.
We also show
some results obtained using a \textsc{Menlops} sample generated with
a merging scale of 40 GeV, for comparison. For this higher merging
scale the number of \textsc{Meps} events in the \textsc{Menlops} sample
is 2\%.

\subsubsection{Jet multiplicities}

\begin{figure}[H]
\begin{centering}
\includegraphics[scale=0.37,angle=90]{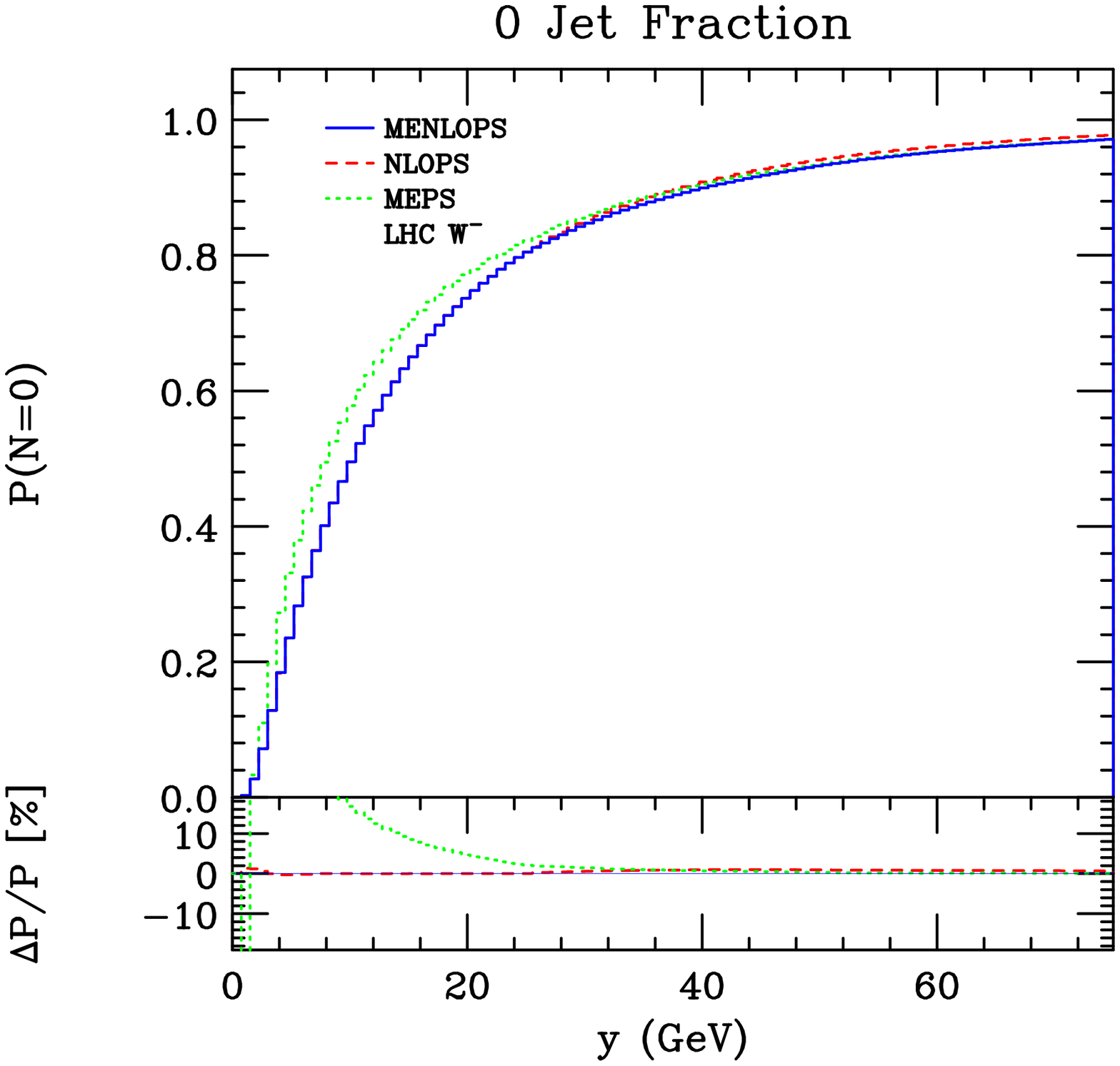}\hfill{}\includegraphics[scale=0.37,angle=90]{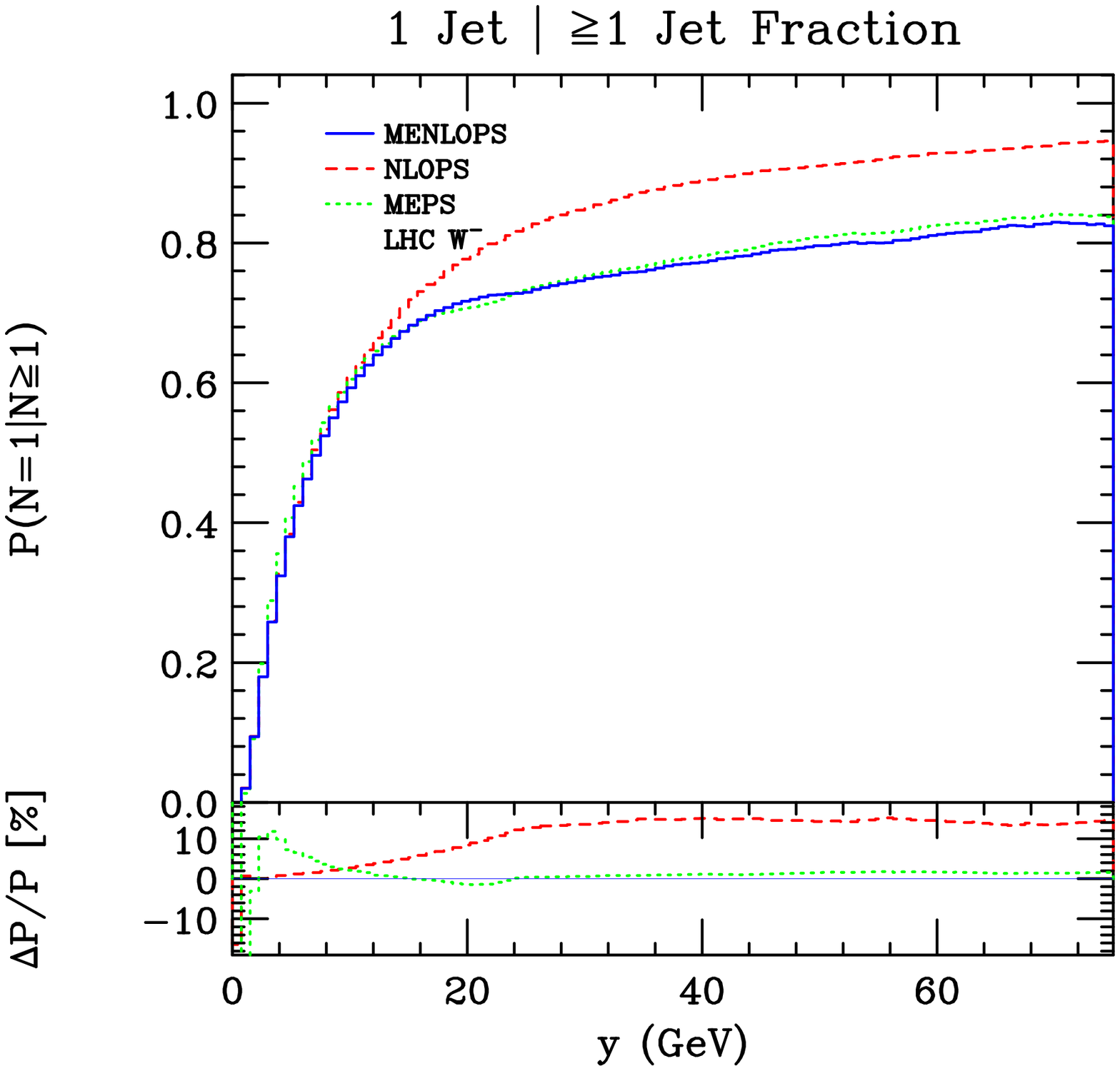} 
\par\end{centering}

\caption{In the left plot, the dashed (red), dotted (green) and solid
(blue) lines show the 0-jet
fractions in the \textsc{Nlops}, \textsc{Meps} and \textsc{Menlops}
$\mathrm{pp}\rightarrow\mathrm{W}^{-}\rightarrow\overline{\nu}_{\mathrm{e}}e^{-}$
samples respectively, as a function of the jet resolution scale $y$,
defined according to the Durham $k_{\perp}$ jet measure.
On the right plot, the number of 1-jet events over
the total number of events with at least one jet is reported.}

\label{fig:W_jet_fractions_1} 
\end{figure}

\begin{figure}[H]
\begin{centering}
\includegraphics[scale=0.4,angle=90]{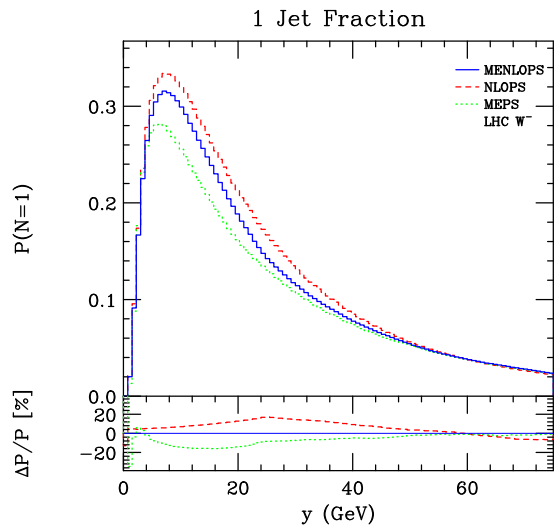} 
\par\end{centering}

\caption{The fraction of 1-jet events of the \textsc{Nlops},
\textsc{Meps} and \textsc{Menlops} full samples,
as a function of the jet clustering scale $y$. The convention for the line
types (and colours) are the same as in the previous plots.}

\label{fig:W_jet_fractions_2} 
\end{figure}

In Figures~\ref{fig:W_jet_fractions_1} and \ref{fig:W_jet_fractions_2}
are the fractions of events in each of the samples, for various values
of the jet clustering scale (see Sect.\,\ref{sub:MENLOPS-implementation}).
The solid (blue) histogram shows the results of this analysis procedure
to a \textsc{Menlops} sample constructed as described in Sect.\,\ref{sub:MENLOPS-implementation}
using the default $\mathrm{W}$ production merging scale 25 GeV. The
corresponding results for the pure \textsc{Nlops} and \textsc{Meps}
samples are shown in the dashed (red) and dotted (green) lines respectively. 

These plots serve to emphasize the physics behind the 0- and 1-jet
cross sections as written in Sect.\,\ref{sec:Hardest-emission-xsecs}.
Were it not for the formal technical difference between the jet
measure and the $k_{T}$ evolution variable, Eqs.\,(\ref{eq:sec3_powheg_0_jet_b})
and (\ref{eq:sec3_meps_0_jet}) could be
rewritten as
\begin{align}
  \mathrm{d}\sigma_{\mathrm{PW}}\left(0\right) &
  =\overline{B}\left(\Phi_{B}\right)\,\Delta_{R}\left(y\right)\,\mathrm{d}\Phi_{B}
  & \mathrm{d}\sigma_{\mathrm{ME}}\left(0\right) &
  =\overline{B}_{\mathrm{ME}}\left(\Phi_{B}\right)\,\Delta_{\widehat{R}}\left(y\right)\,\mathrm{d}\Phi_{B},\label{eq:ttbar_zero_jet_xsecs}
\end{align}
where the Sudakov form factors here correspond to the probability that
no emission occurs in the hard region $y>y_{0}$. The appearance of the
0-jet fraction histogram is then no surprise, having the
characteristic Sudakov form factor shape.

One also can see the characteristic Sudakov form factor shape
in the conditional 1-jet rate from the subsample of events with at
least one jet. We remind the reader that this basically represents the
probability for a 1-jet event to remain resolved as a 1-jet event, in
evolving down from larger values of $y$. We see that in the
pure \textsc{Nlops} case this relative rate is significantly higher.
This is a clear signal that additional jets are missing from the
\textsc{Nlops} simulation. These structures are
as expected according to the arguments in
Sect.\,\ref{sec:Combining-Powheg-and-Meps} and the analysis
surrounding
Eqs.\,\ref{eq:sec3_powheg_0_jet_a}-\ref{eq:sec3_powheg_1_jet}.

The full 1-jet fraction in Fig.~\ref{fig:W_jet_fractions_2} is a
combination of the complement to 1 of the 0-jet fraction and of the
conditional one jet fraction (i.e. it is one minus the left plot times
the right plot in Fig.~\ref{fig:W_jet_fractions_1}).

\begin{figure}[H]
\begin{centering}
\includegraphics[width=0.4\textwidth,angle=90]{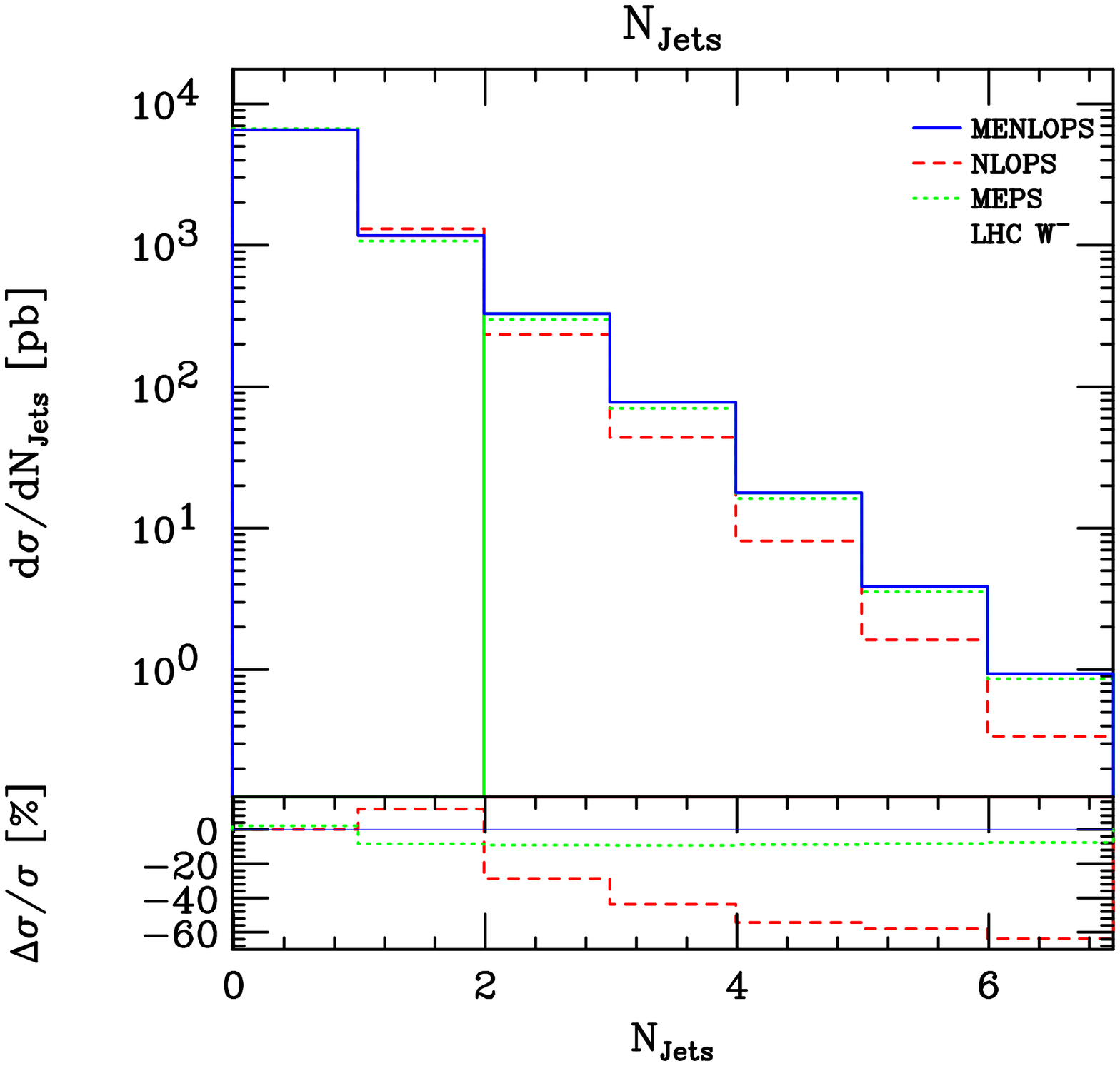}\hfill{}\includegraphics[width=0.4\textwidth,angle=90]{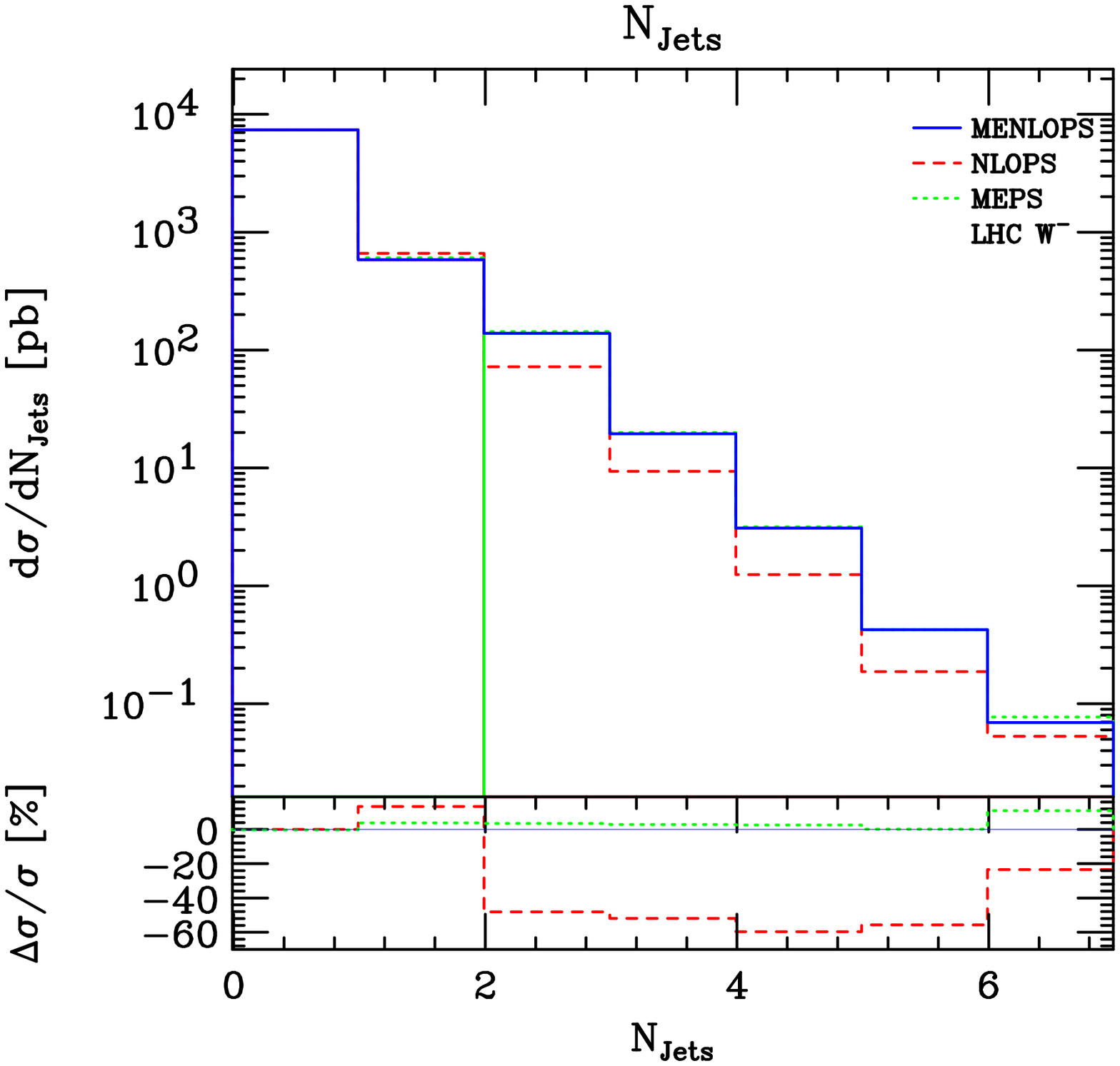} 
\par\end{centering}

\vspace{5mm}

\caption{The jet multiplicity distributions for $\mathrm{W}^{-}\rightarrow\mathrm{e}^{-}\bar{\nu}_{\mathrm{e}}$
using two different choices of the \textsc{Menlops}
merging scale: 25 GeV (left) and 40 GeV (right).}

\label{fig:W_jet_multiplicities} 
\end{figure}

Looking at the 0-jet fractions in Fig.\,\ref{fig:W_jet_fractions_1}
one sees that there is a tendency for the \textsc{Meps} sample
to contain fractionally more soft events than the other two. This
may be understood
as being due to differing approaches to the soft resummation in the
\textsc{Powheg} simulation, with respect to the P{\footnotesize YTHIA}
virtuality ordered shower. The description of the soft
region obtained from the transverse momentum ordered shower is theoretically
much closer to that in \textsc{Powheg}, producing results in much
better agreement in that region. We hasten to add that the choice
of scales used in the evaluation of the \textsc{PDF}s in the transverse
momentum ordered shower is also theoretically more sound \cite{Dokshitzer:1978hw,Nason:2006hfa}.
However, from the point of view of the \textsc{Meps} merging aspect,
on the whole we have found better results with the virtuality ordered
shower, for which the \textsc{Madgraph} \textsc{MLM} implementation
is more mature. Also we favor maximizing the corrective effects which
arise from the exact matrix elements. With this in mind the virtuality
ordered description is preferable since it allows a lower value of
the \textsc{Meps} merging scale.

Turning to the 1-jet fractions the picture is somewhat different.
Here the fraction of 1-jet events, in the soft region, is around 20\%
lower in the \textsc{Meps} and \textsc{Menlops} samples, for larger
values of the clustering scale. This is indicative of the fact that
the fraction of events with more than one jet, with respect to the
fraction with at least one jet, is higher in the \textsc{Meps} sample,
through the inclusion of $\mathcal{O}\left(\alpha_{\mathrm{S}}^{2}\right)$
tree level matrix elements%
\footnote{Note that the inclusion of the higher order tree level matrix elements
can equally lead to a \emph{reduction }in\emph{ }the 1-jet fraction
with respect to that in the shower approximation.\emph{ }%
}. 

At this point we feel it may be useful to put these figures in context
with regard to the \textsc{Menlops} algorithm and the theoretical
arguments surrounding it in Sect.\,\ref{sec:Combining-Powheg-and-Meps}.
Specifically, recall that in \textsc{Powheg}, for small values of
the clustering scale, the distribution of radiation in the 0-jet sample
is dominated by large logarithms
(Eqs.\,\ref{eq:sec3_powheg_0_jet_a},\ref{eq:ttbar_zero_jet_xsecs})
and is therefore, formally, no worse in accuracy than any parton shower.
Conversely, at higher values of the clustering scale the large logarithms
are suppressed and the full \textsc{NLO} accuracy of the \textsc{Powheg}
simulation therefore gives a much better prediction. From the 0-jet
fraction plot one can see that the \textsc{Menlops} sample is combined,
from the point of view of the cross section, at a high value of the
clustering scale, with 80\% of events containing no extra jet activity.
In any case, the 0-jet cross section, which plays a key role in determining
the content of the \textsc{Menlops} sample, is \emph{always} described
by the \textsc{Nlops} prediction at least as well as the \textsc{Meps}
one; the \textsc{Menlops} sample contains the same fraction of 0-jet
events as the \textsc{Nlops} one by design. 

Displayed in Figure~\ref{fig:W_jet_multiplicities} are the jet multiplicity
distributions obtained by merging the \textsc{Nlops} and \textsc{Meps}
samples with \textsc{Menlops} merging scales of 25 GeV and 40 GeV,
using the same scale to define the jets in each case. Here
one sees that the cross section for lower jet multiplicities is larger
in the sample with the 40 GeV merging scale than in the 25 GeV one,
while the opposite is true for the higher multiplicities. 
The nature of the results shown here can be easily explained.
The gap between the pure \textsc{Meps} and \textsc{Menlops}
results, as seen in the fractional difference plots, is small,
indicating that the \emph{K-}factor for the total cross section is
in close agreement with the \emph{K}-factor associated with the production
of at least one jet. This can also be understood by considering that
the size of the gap is proportional to, amongst other things, the
difference in the 0-jet fractions, which can be seen to vanish at
40 GeV in Figure~\ref{fig:W_jet_fractions_1}.

\subsubsection{Inclusive observables}

\begin{figure}[H]
\begin{centering}
\includegraphics[scale=0.37,angle=90]{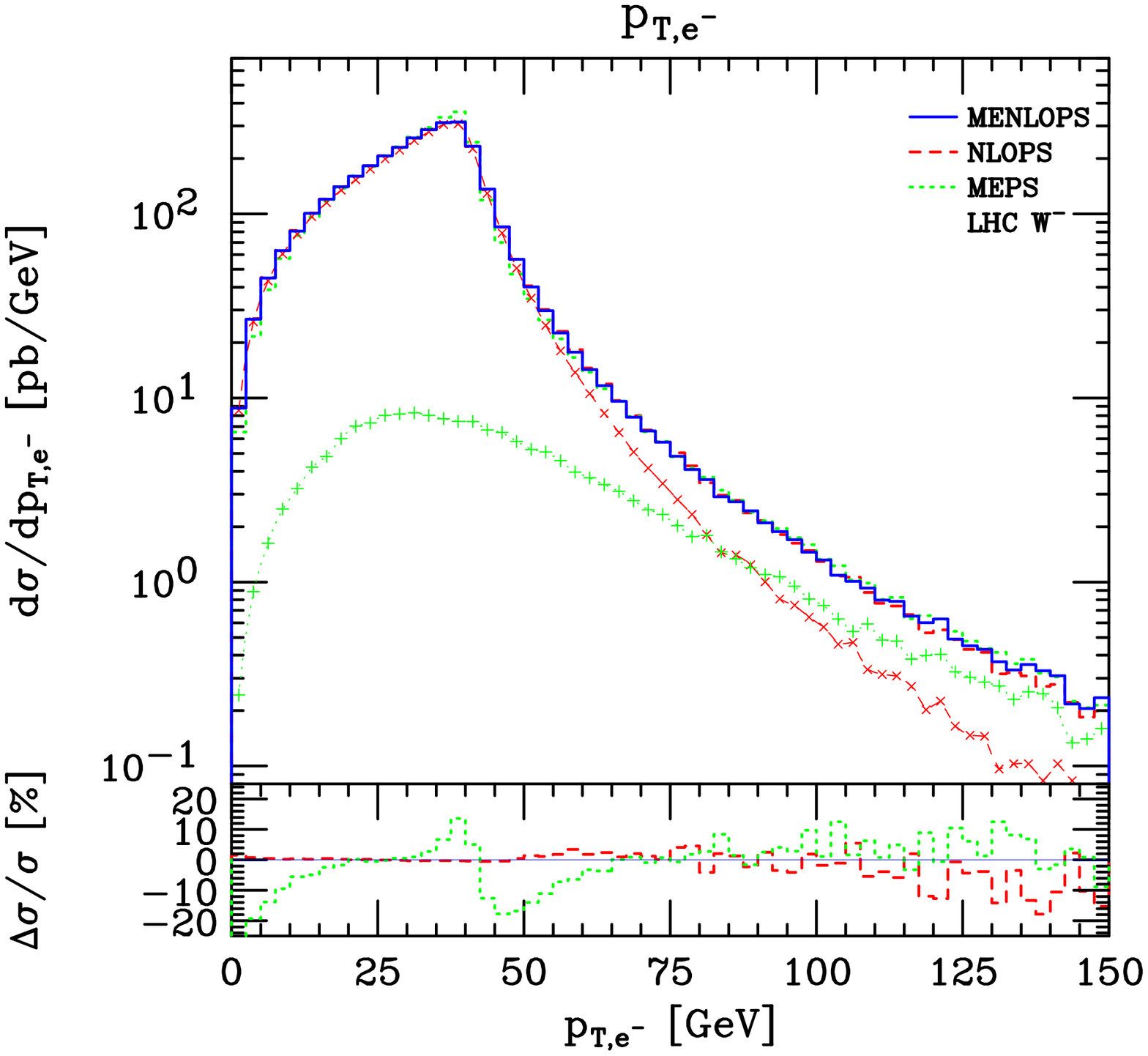}\hfill{}\includegraphics[scale=0.37,angle=90]{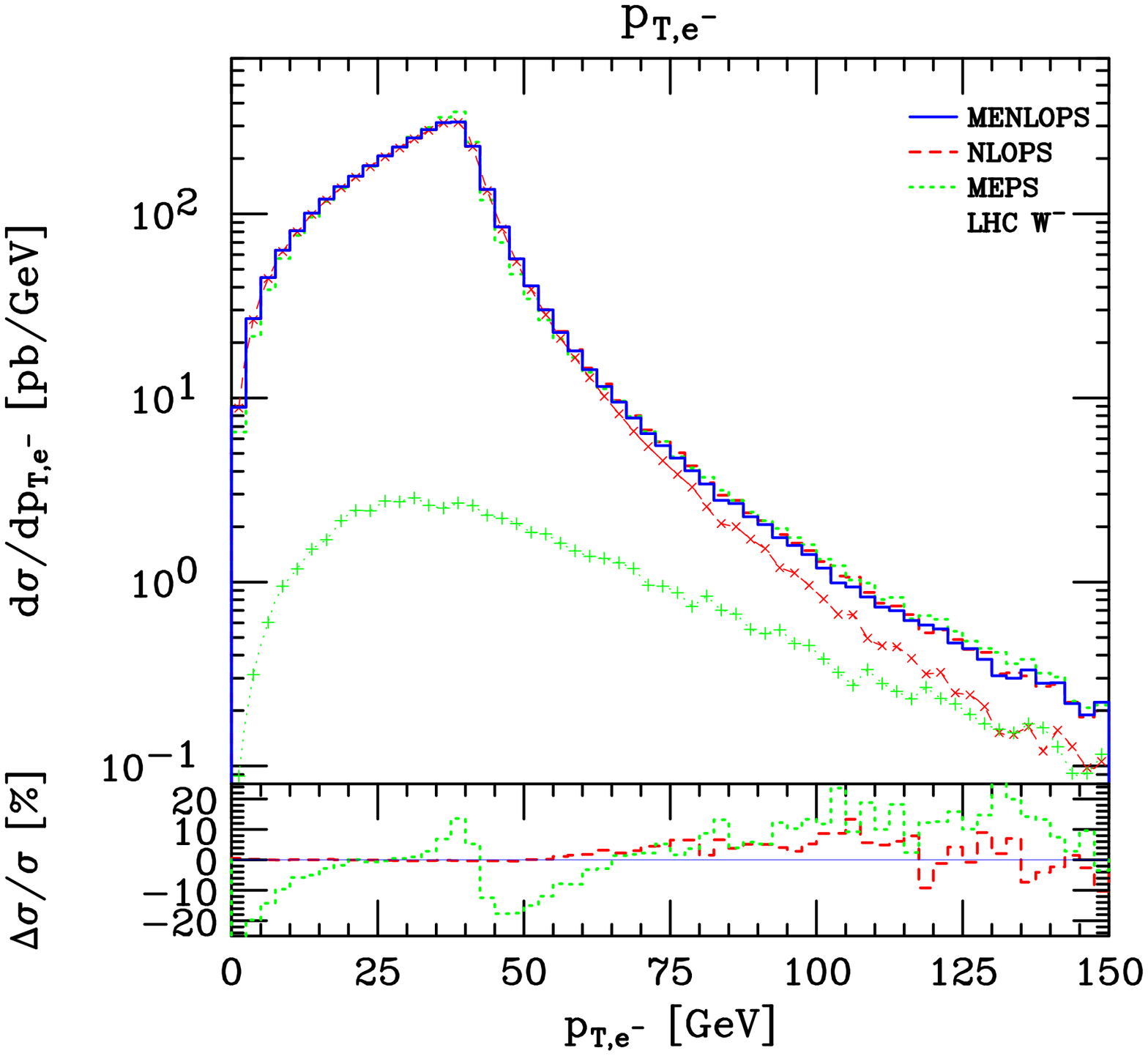} 
\par\end{centering}

\vspace{5mm}

\begin{centering}
\includegraphics[scale=0.37,angle=90]{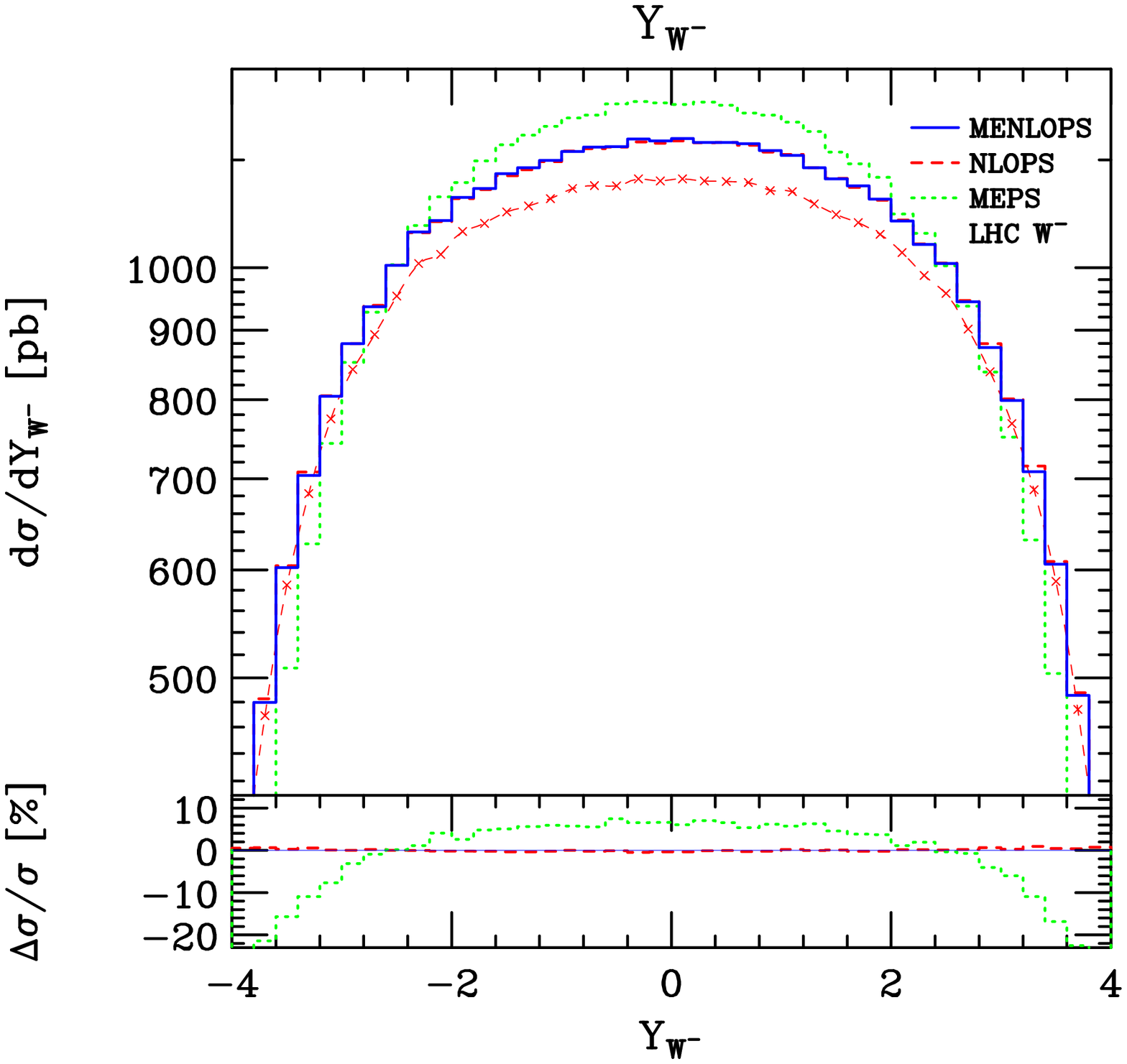}
\hfill{}\includegraphics[scale=0.37,angle=90]{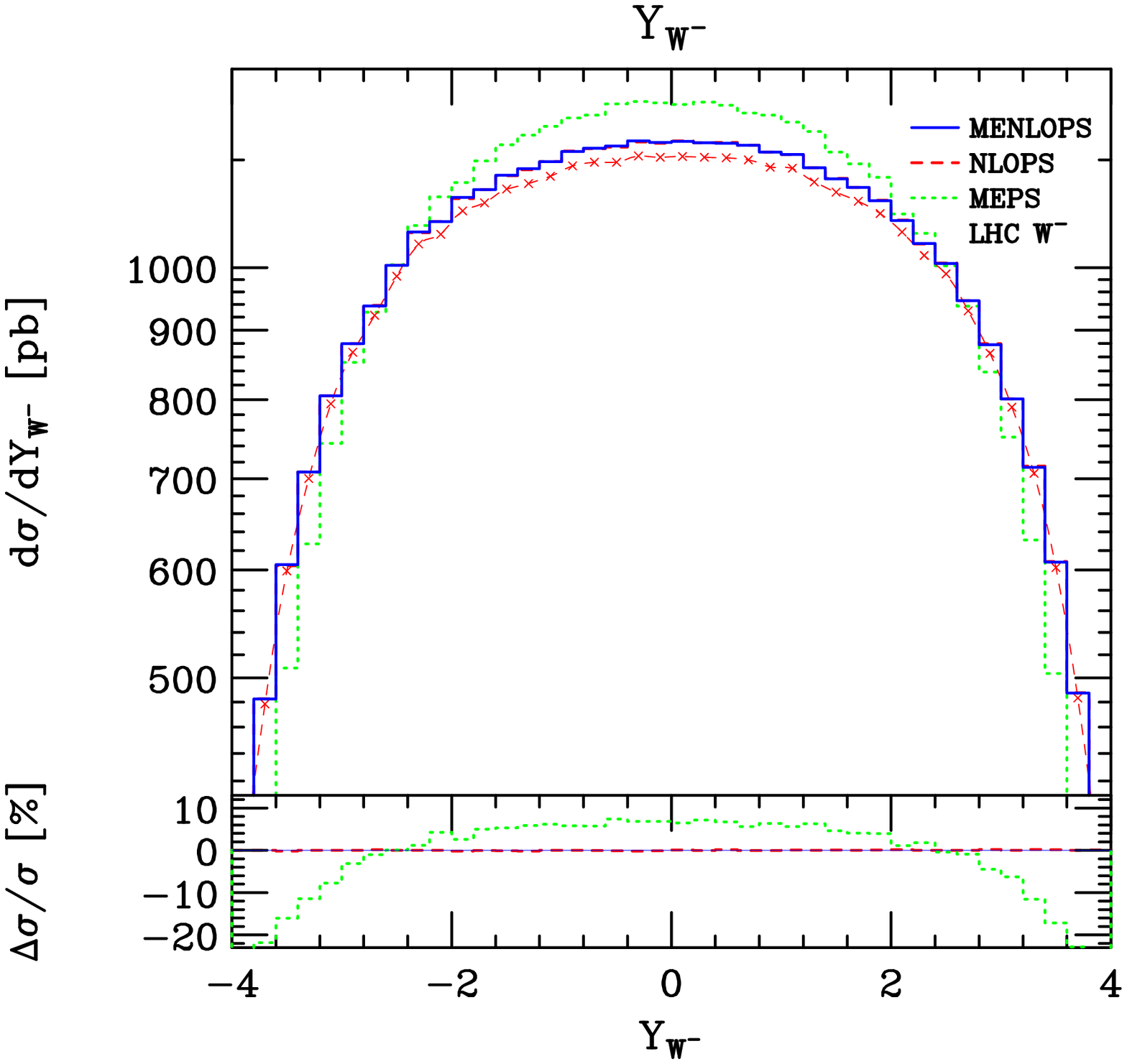} 
\par\end{centering}

\caption{In the upper half of this figure we show the transverse momentum of
the electron in $\mathrm{W}^{-}\rightarrow\mathrm{e}^{-}\bar{\nu}_{\mathrm{e}}$
using a 25 GeV (left) and 40 GeV (right) jet resolution
scale in performing the \textsc{Menlops} merging. The lower pair of
plots shows, analogously, the rapidity distribution of the $\mathrm{W}^{-}$.
Despite the relatively large difference in the merging scales the
combined \textsc{Menlops} prediction is stable with respect to the
changing scale, showing deviations from the \textsc{NLO} result at
the level of only 1 or 2\% in both cases.}

\label{fig:W_inclusive_observables} 
\end{figure}

In Figure~\ref{fig:W_inclusive_observables} we show the transverse
momentum spectrum of the $\mathrm{e}^{-}$ from the $\mathrm{W}^{-}$
decay, as well as the rapidity of the $\mathrm{W}^{-}$, for two different
choices of the \textsc{Menlops} merging scale, our default value of
25 GeV and also 40 GeV. In respect of these quantities there are two
main points two consider: the stability and composition of
the \textsc{Menlops} sample with respect to changing the merging scale,
and also potential differences due to NLO effects.

In all cases these inclusive \textsc{Menlops} predictions are shown to
be insensitive to the change in the merging scale.  We draw attention
to the fact that the high $p_{T}$ tail of the electron transverse
momentum spectrum, for the 25 GeV scale choice, is entirely due to
events from the \textsc{Meps} sample, \emph{i.e. }events with at least
two jets,, while for the 40 GeV choice it is given by an even mixture
of \textsc{Meps} and \textsc{Nlops} events. The stability of the result
follows from the fact that the two types of simulation are in good agreement
regarding the shapes of this distribution.

The distribution for the rapidity of the $\mathrm{W}^{-}$ is also
interesting. Here again we see that the distribution is stable with
respect to changing the \textsc{Menlops} merging scale from 25 to
40 GeV, and in both cases only exhibits $\mathcal{O}\left(1\%\right)$
level fluctuations with respect to the \textsc{Nlops} prediction.
This behavior is rather unsurprising, since the \textsc{Meps} contribution
constitutes no more than $5\%$ of the total number of events in the
\textsc{Menlops} sample for the case of 25 GeV merging scale and hence
even less for the 40 GeV case.

Note that the shape of the \textsc{Meps} and \textsc{Nlops}/\textsc{Menlops}
rapidity spectra show rather large differences, up to 20\%. Such differences
can occur through the use of different \textsc{PDF}s. However, in
our case, the \textsc{MRST 2002 NLO} parton density functions were
used in all aspects of the generation process. Inconsistencies in
the treatment of the CKM matrix and proton flavour content could also
cause some discrepancies. However we have checked that the inputs to
the \textsc{Meps} and \textsc{Nlops} simulations are compatible in
this respect; $b$-quarks are included in both cases and the $\mathrm{V}_{\mathrm{ud}}$
CKM matrix element is set to $0.975$ (differences due to the last
two factors should in any case be very small; in fact, we found
that using a diagonal CKM matrix has a completely negligible
effect on distributions).

Intuitively one might expect that the addition of multiple hard jets
to a leading order parton shower simulation, as in the \textsc{Meps}
case, would require the produced system, and thus also the
$\mathrm{W}$, to be more central. Exactly this
behavior can be seen in, for example, Figure~3 of Ref.\,\cite{Krauss:2005nu},
where it is clear that the pseudorapidity of the $\mathrm{Z}$ boson
in the Drell-Yan process is less central when only the leading order
matrix element is used in generating the \textsc{Meps} sample. This
is in line with the differences we see in the rapidity distributions
in$ $ Figure~\ref{fig:W_inclusive_observables}. For this observable,
the results in Ref.\,\cite{Krauss:2005nu} only have leading order
accuracy, meaning that the effect witnessed there is beyond the remit
of that study and is duly neglected there. In the context of our theoretical
analysis in Section~\ref{sec:Hardest-emission-xsecs}, this effect
is contained in the ratio of $\overline{B}_{\mathrm{ME}}(\Phi_B)/B(\Phi_B)$.
Hence we conclude that the differences may be attributed to the inclusion
of NLO terms in the \textsc{Nlops}/\textsc{Menlops} samples that are not
present in the \textsc{Meps} one. This conclusion is
supported by the result displayed in Fig.~13 of Ref.\,\cite{Alioli:2008gx},
where it is shown that, if the same parton densities
are used, there is no difference in the shape of the LO and NLO
rapidity distribution of the vector boson. This means that the effect
of real corrections, that would make the distribution more central, are
exactly compensated by NLO virtual effects.

\begin{figure}[H]
\begin{centering}
\includegraphics[width=0.4\textwidth,angle=90]{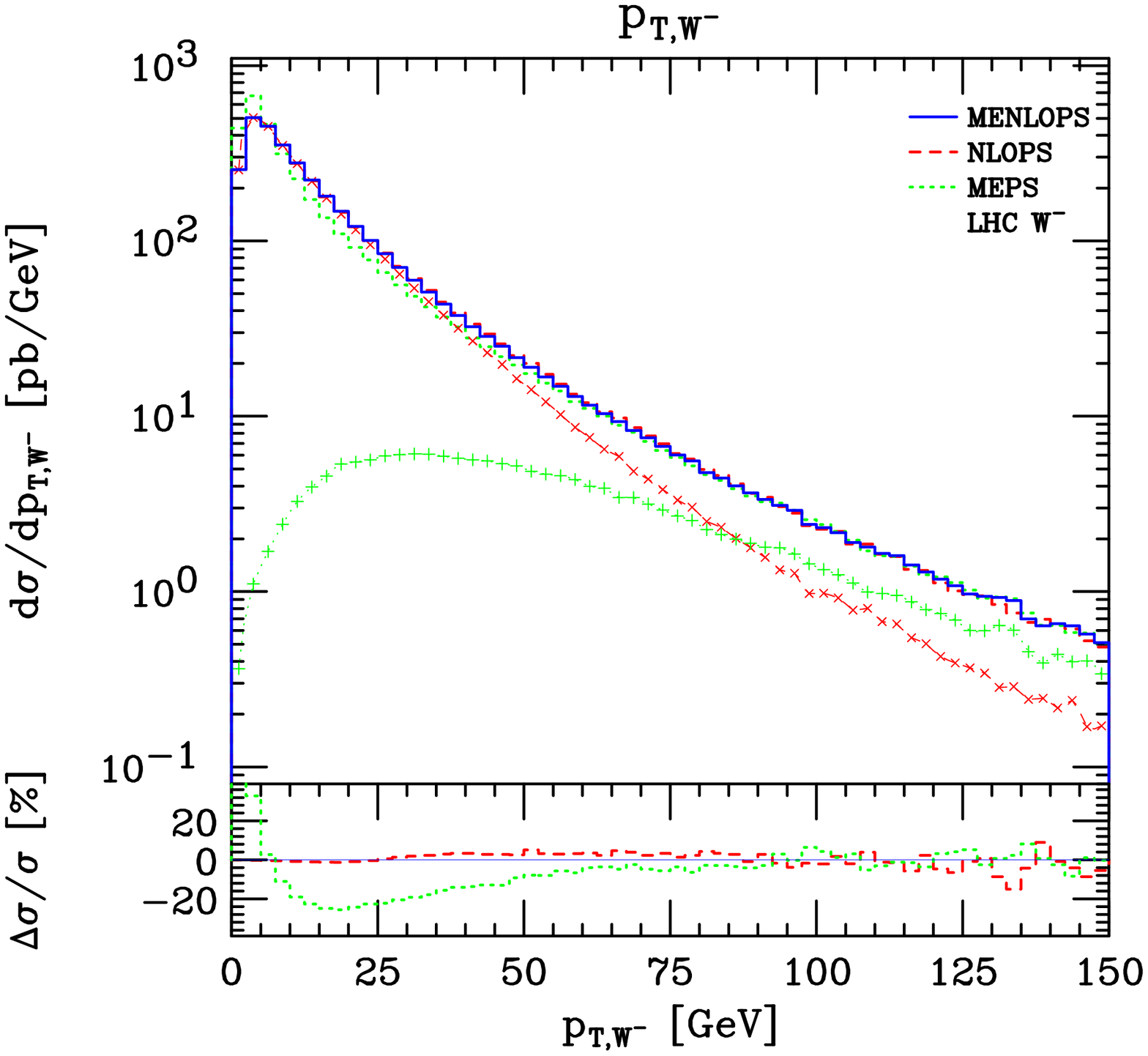}\hfill{}\includegraphics[width=0.4\textwidth,angle=90]{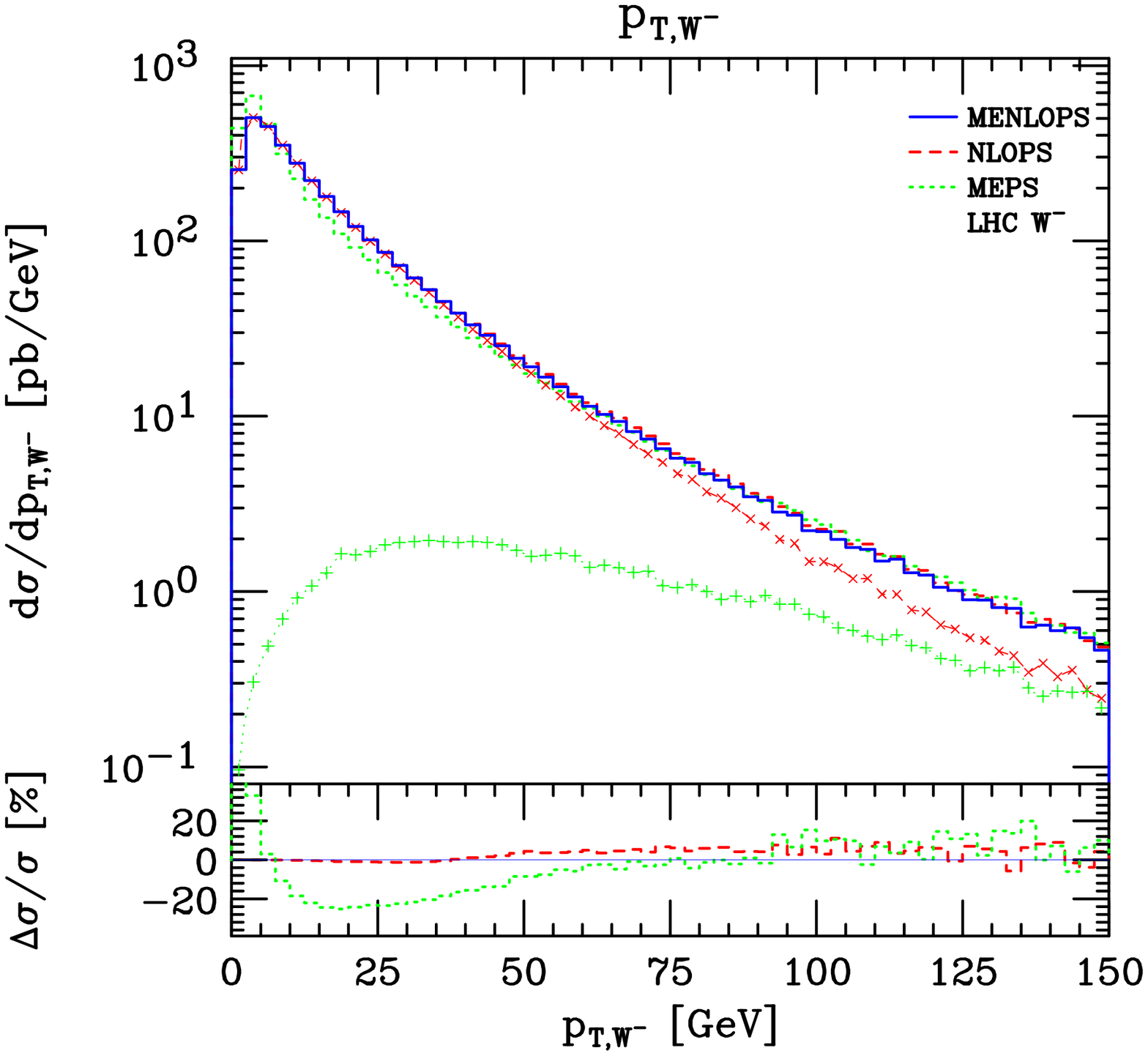} 
\par\end{centering}

\vspace{5mm}

\caption{The transverse momentum spectrum of the $\mathrm{W}^{-}$
using a 25 GeV (left) and 40 GeV (right) jet resolution scale as the
\textsc{Menlops} merging scale. As in Figure~\ref{fig:W_inclusive_observables},
the greater resolution scale used in producing the \textsc{Menlops}
sample (solid) on the right hand side results in the \textsc{Meps}
component (dotted) being greatly diminished. Nevertheless, the merged
distribution very much assumes the form of the pure \textsc{Nlops}
prediction (dashed) to within $\mathcal{O}\left(1\%\right)$, with
deviations only beginning to become noticeable in the high $p_{T}$
tail, where contributions from events containing more than one jet
become more important. }

\label{fig:W_pT} 
\end{figure}

We now turn our attention to the $\mathrm{W}$ boson $p_{T}$ spectra
seen in Figure~\ref{fig:W_pT}. We observe that
the distribution is essentially stable with respect to the change
in \textsc{Menlops} merging scale, from 60 to 100 GeV, with the \textsc{Menlops}
prediction being indistinguishable from the \textsc{Nlops} one. This
again reflects the good agreement in the shape of the \textsc{Meps}
and \textsc{Nlops} predictions since the latter dominate the \textsc{Menlops}
sample in the region above 75 GeV in the default sample.

We ascribe the increasing \textsc{Meps} content of the \textsc{Menlops}
sample, at high $p_{T}$, as being due to the fact that such events
naturally involve more energy transfer, and therefore they
should be associated with more jet activity. More
technically, consider that
if a high transverse momentum $\mathrm{W}$ boson is observed, momentum
conservation requires that there be an equally significant amount
of momentum in the form of QCD radiation to balance it. From our earlier
expressions for the 1-jet cross section Eq.\,\ref{eq:sec3_powheg_1_jet}
one can deduce that the probability
for a 1-jet event, with the jet produced at some high scale $p_T$,
to remain resolved as a 1-jet event at the lower scale $y_{0}$, is
given by the effective Sudakov form factor for that configuration.
It follows that the higher is the initial value of $p_T$, the less
likely one is to still observe just one jet at $y_{0}$. This argument
is of course rather general and the structure of the \textsc{Menlops}
predictions for all of the $p_{T}$ spectra in these results
can be understood in these
terms. 

\subsubsection{Jet activity}

\begin{figure}[H]
\begin{centering}
\includegraphics[width=0.4\textwidth,angle=90]{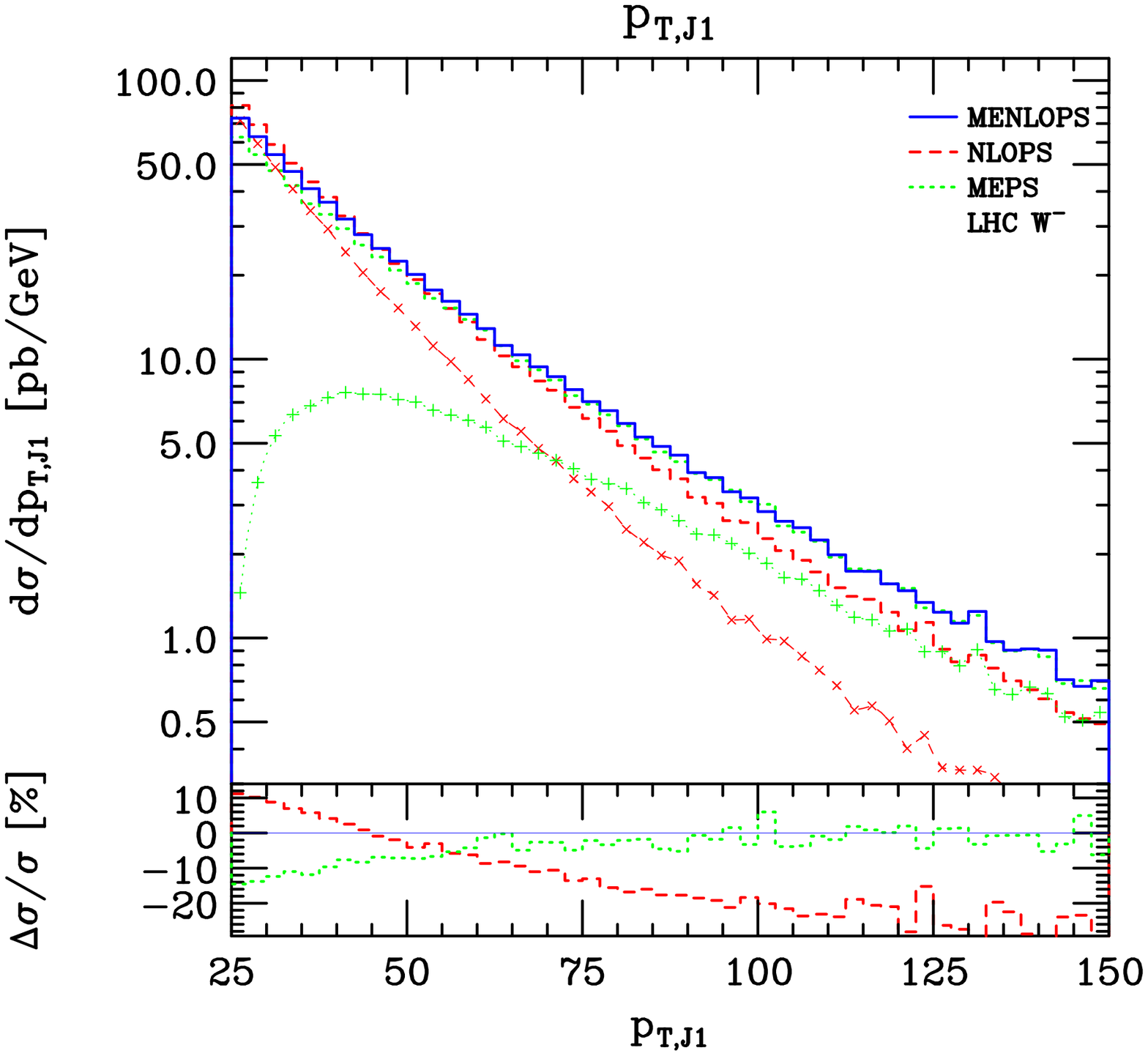}\hfill{}\includegraphics[width=0.4\textwidth,angle=90]{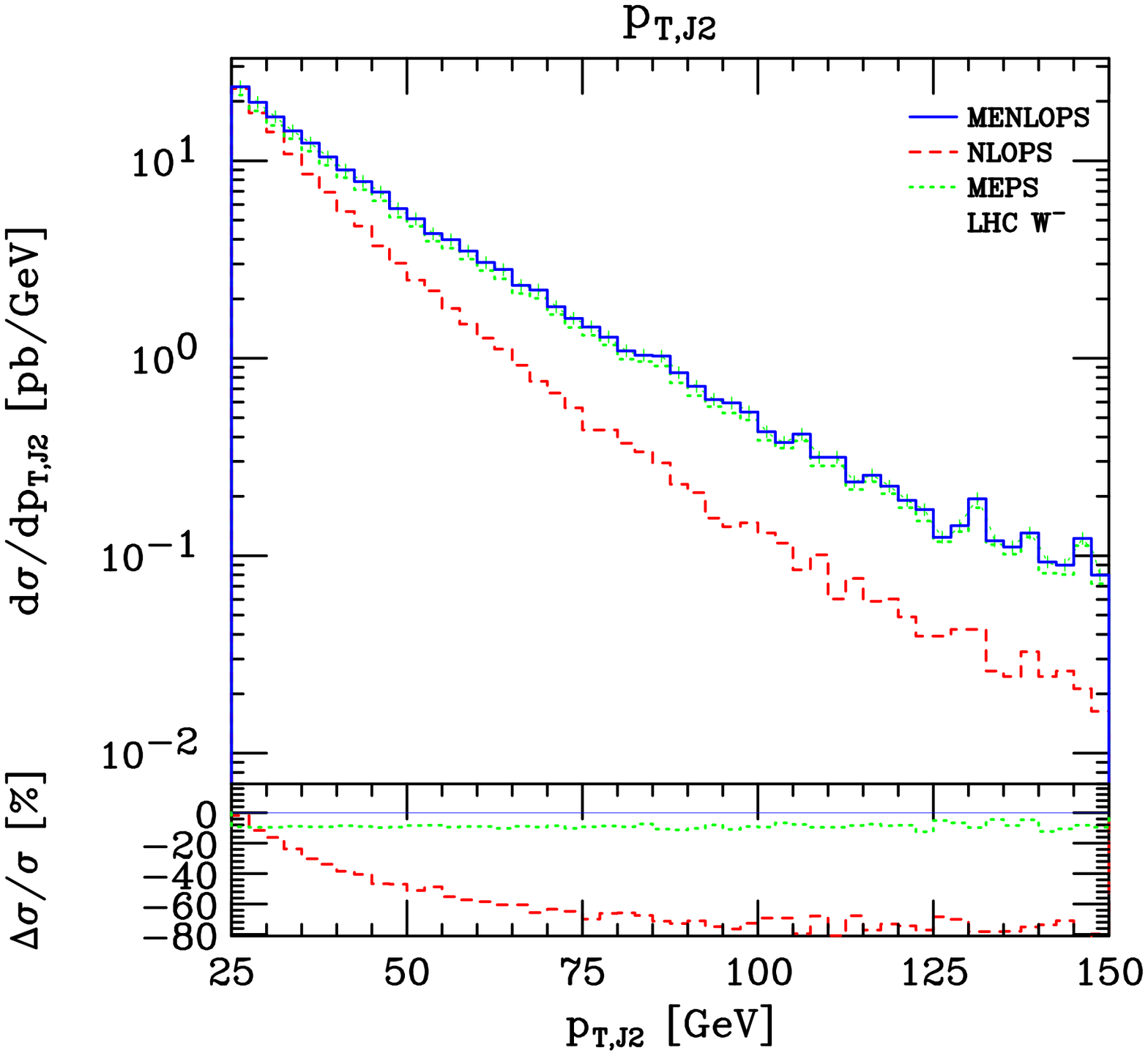} 
\par\end{centering}

\vspace{5mm}

\begin{centering}
\includegraphics[width=0.4\textwidth,angle=90]{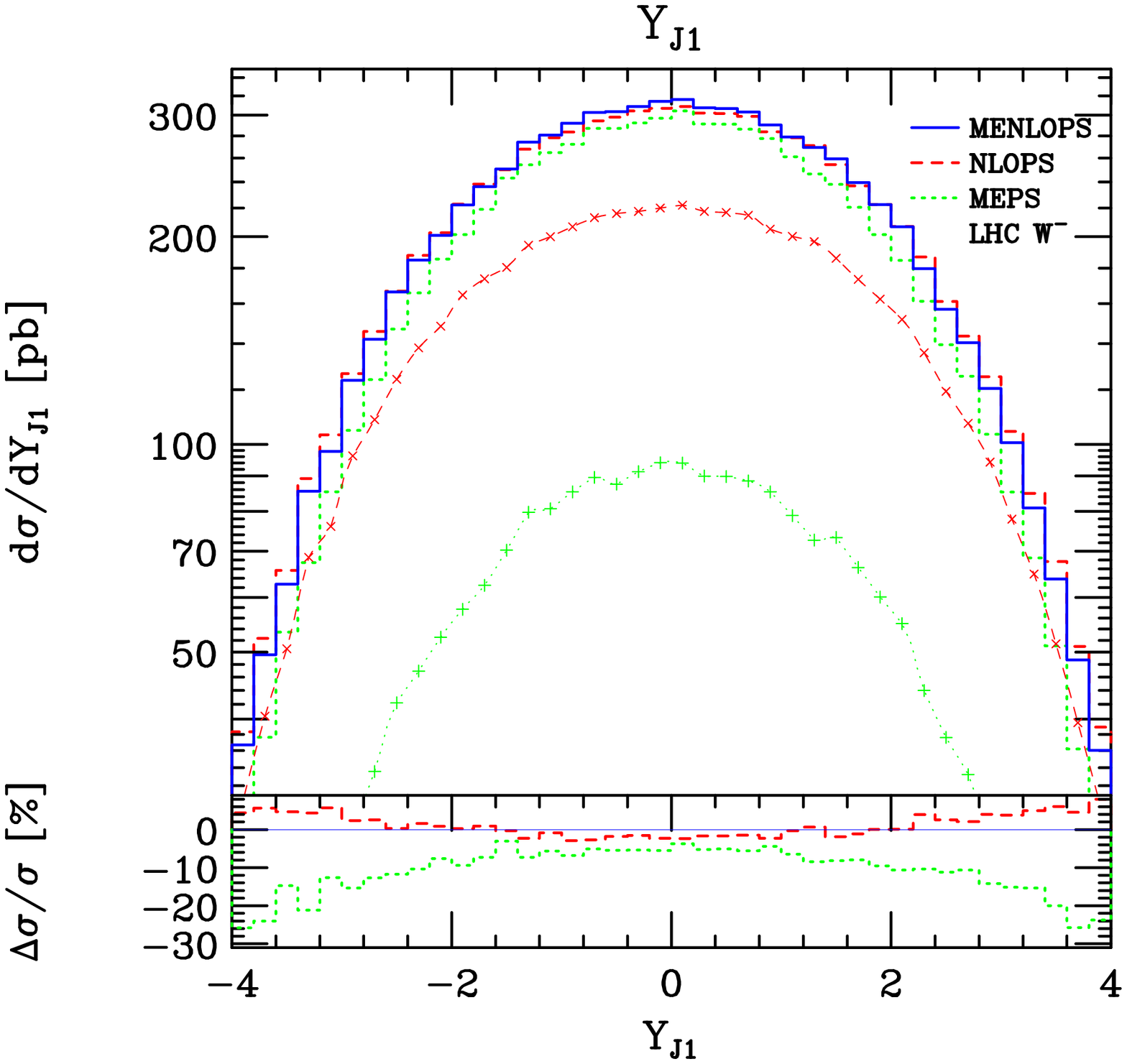}
\hfill{}\includegraphics[width=0.4\textwidth,angle=90]{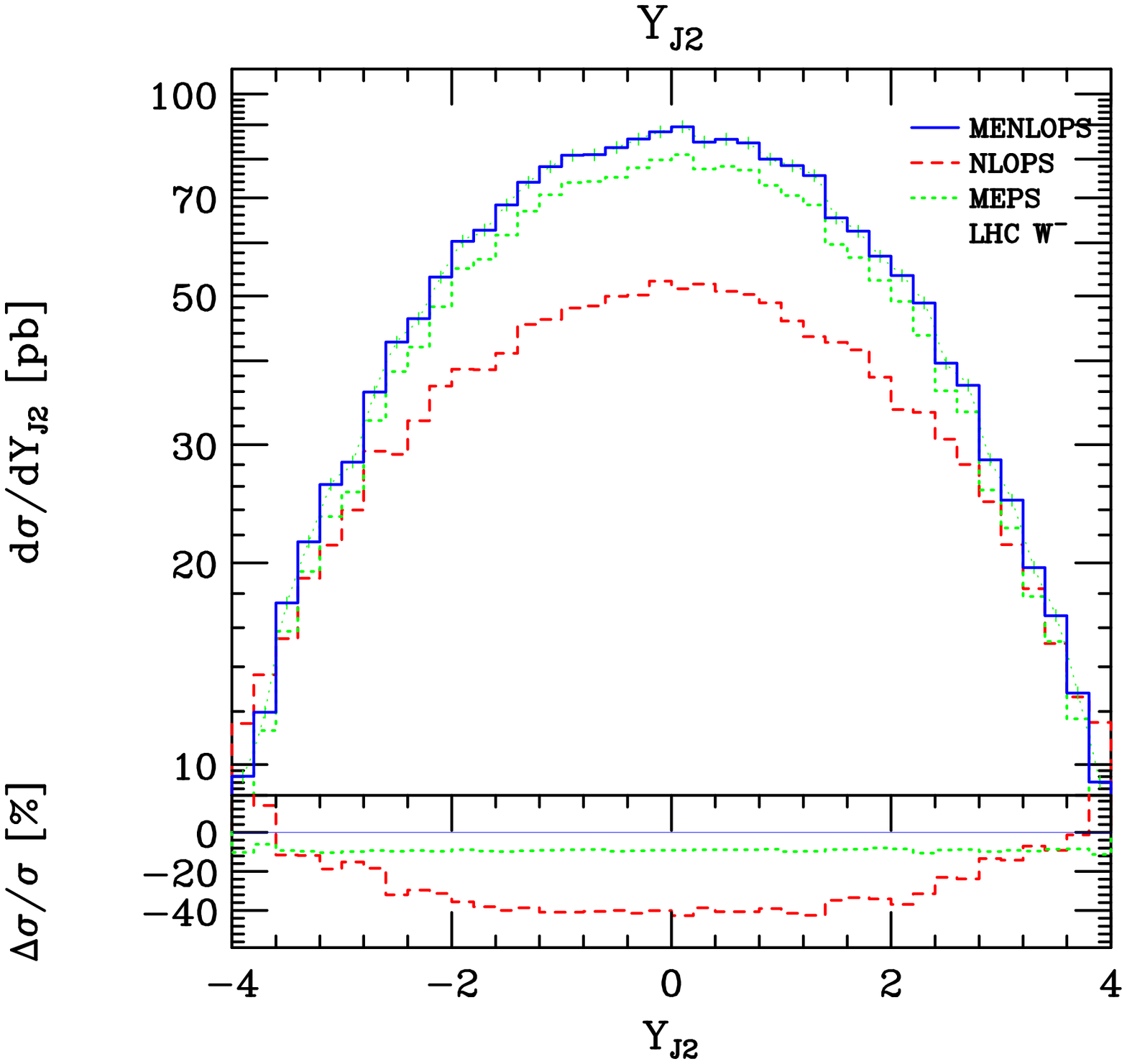} 
\par\end{centering}

\caption{ In the upper
half of this figure we show the transverse momentum distribution of
the hardest (left) and second hardest (right) jets, with the corresponding
rapidity distributions shown underneath.
The \textsc{Menlops} predictions (solid) shown here and their
\textsc{Nlops} (dashed) and \textsc{Meps} (dotted) 
components were obtained from a \textsc{Meps}-\textsc{Nlops}
combination with a merging scale of $25\,\mathrm{GeV}$. }

\label{fig:W_1st_2nd_jet_pts_and_ys} 
\end{figure}

In Figure~\ref{fig:W_1st_2nd_jet_pts_and_ys} we show the transverse
momentum and rapidity distributions of the first and second highest
$p_{T}$ jets in $\mathrm{p}\mathrm{p}\rightarrow\mathrm{W}\left(\rightarrow\mathrm{e}^{-}\overline{\nu}_{\mathrm{e}}\right)+\mathrm{jets}$.
The distributions for the leading jet mirror the corresponding
ones for the $\mathrm{W}^{-}$ boson which it recoils against. The
composition of the \textsc{Menlops} $p_{T}$ spectrum result can be
understood in much the same way as was just discussed for the case
of the $\mathrm{W}^{-}$ boson $p_{T}$, with one key difference being the
degree of exclusivity of the observable.
Whereas the $\mathrm{W}^{-}$ transverse momentum includes contributions
from all jet multiplicities, and is therefore predominantly based
on 0-jet \textsc{Nlops} events, the leading jet $p_{T}$ spectrum,
obviously, includes no contributions from 0-jet events. Hence, a greater
fraction of events with at least two jets (\textsc{Meps} events) enter
this prediction. This explains why, in the high $p_{T}$ region, the
\textsc{Menlops} prediction for the $\mathrm{W}^{-}$ transverse momentum
spectrum is equal to that of the \textsc{Nlops} sample, while for
the leading jet it is instead equal to the \textsc{Meps} one. 

The fact that the \textsc{Meps} and \textsc{Nlops} results are different
by 25\% in this tail region is an entirely separate issue. We iterate
that, formally, both \textsc{Meps} and \textsc{Nlops} simulations
are only capable of describing predictions for the leading jet with
leading order, leading-log, accuracy. With this in mind the difference
seen is basically of higher order in $\alpha_{\mathrm{S}}$.
Having said that, we note that the \textsc{Meps} result tends to overestimate
the \textsc{Nlops} one.

Finally, we remark that it may seem puzzling that the \textsc{Meps}
and \textsc{Nlops} results agree very well for the $\mathrm{W}^{-}$
transverse momentum spectrum and yet not so well for that of the leading
jet. This is explainable by considering that the \textsc{Nlops} simulation
will prefer to produce additional radiation in the shower approximation,
in the direction of the leading jet or of the incoming beams,
whereas the \textsc{Meps} simulation
is more capable of producing additional radiation closer in angle
to the $\mathrm{W}^{-}$ boson, thus requiring the leading jet to
recoil more.

The predictions for the second jet are completely determined, by construction,
by the \textsc{Meps} sample. The corrective effects of the \textsc{Meps}
contributions in the jet rapidity distributions have a more intuitive
understanding; since the second hardest jet in the \textsc{Powheg}
simulation originates from the parton shower, the subset of two jet
events generated in this way will tend to have proportionally more
events in which the second jet is more collinear with the beam axis,
than in the \textsc{Meps} case.
\begin{figure}[H]
\begin{centering}
\includegraphics[width=0.4\textwidth,angle=90]{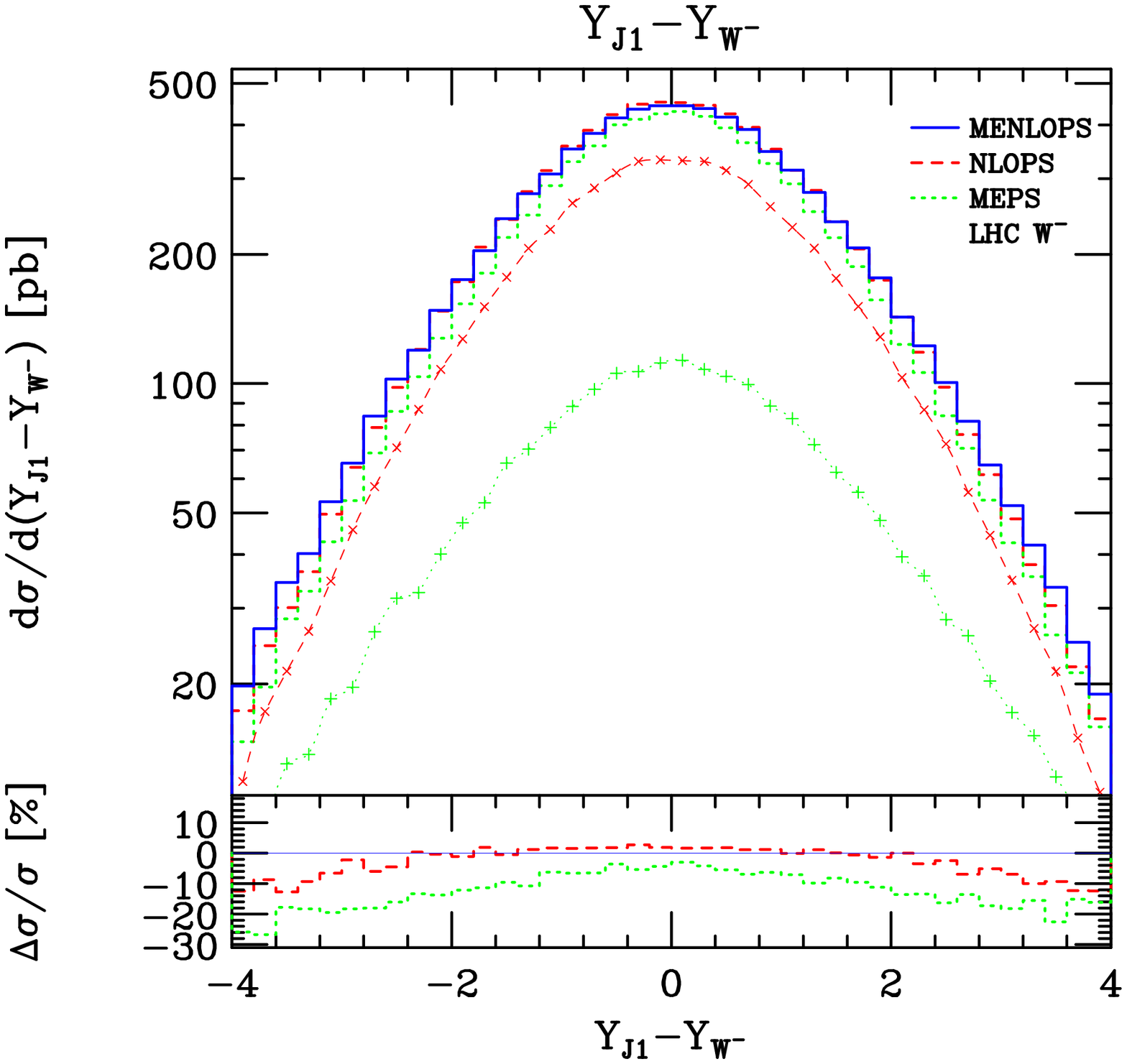}\hfill{}\includegraphics[width=0.4\textwidth,angle=90]{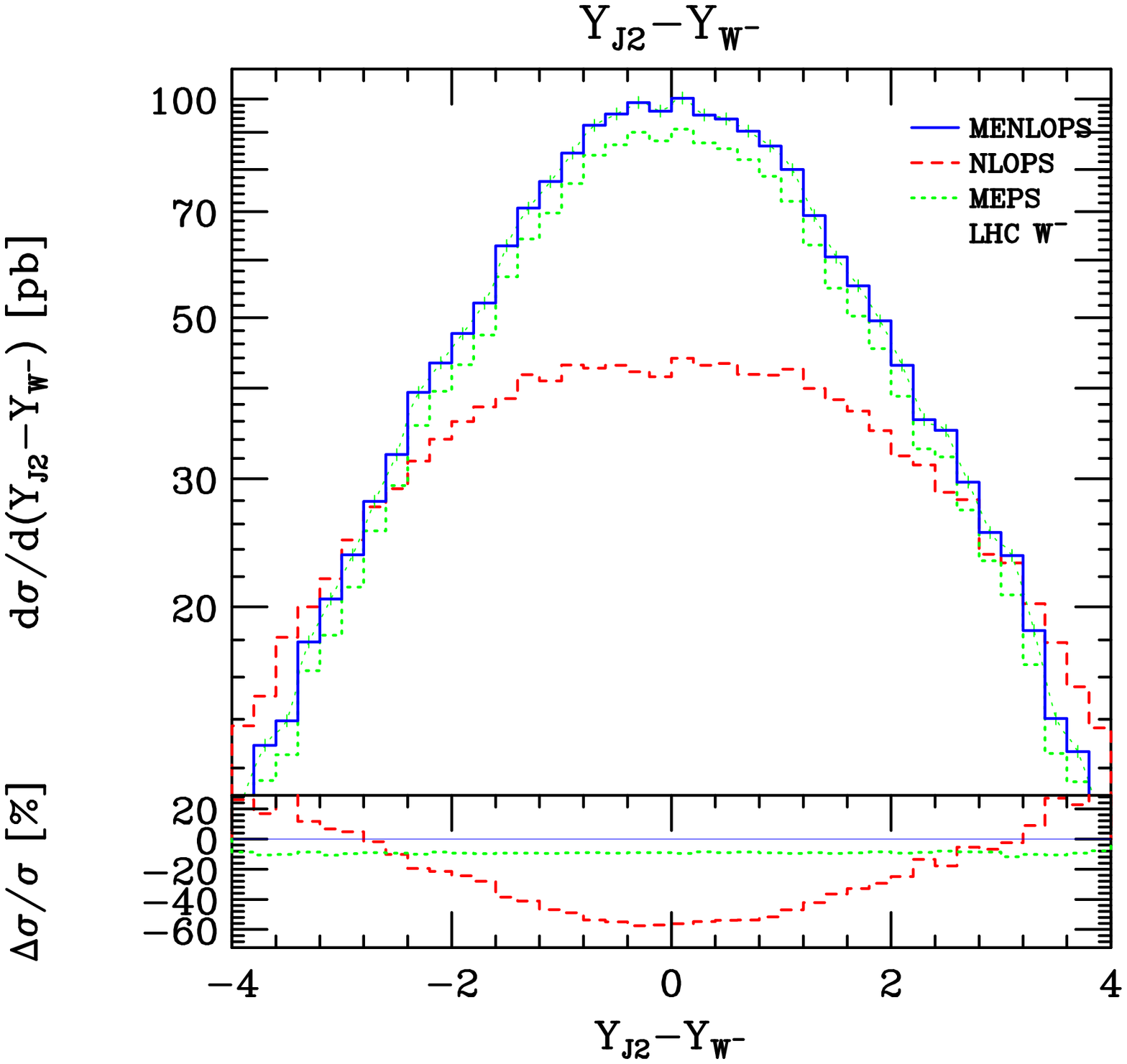} 
\par\end{centering}

\caption{In this figure we show the rapidity difference between hardest jet
and $\mathrm{W}^{-}$ boson (left) and the second hardest jet and
the $\mathrm{W}^{-}$ boson (right). The first plot requires the presence
of at least one jet in the event, precluding contributions from 0-jet
\textsc{Nlops} events. Hence, the relative \textsc{Meps} component
of the \textsc{Menlops} sample (solid green) is increased with respect
to the case of the $\mathrm{W}^{-}$ rapidity distribution (Fig.\,\ref{fig:W_inclusive_observables}).
In the other plot the \textsc{Menlops} sample comprises of only \textsc{Meps}
events and a considerable correction to the \textsc{Nlops} result
can be seen. This correction is due to the fact that the parton shower
approximation is used to generate jets beyond the leading jet in the
\textsc{Nlops}
simulation, whereas the \textsc{Meps} result is better, giving a
leading order prediction for this distribution. }

\label{fig:W_jet_rapidity_corrs} 
\end{figure}

\begin{figure}[H]
\begin{centering}
\includegraphics[width=0.4\textwidth,angle=90]{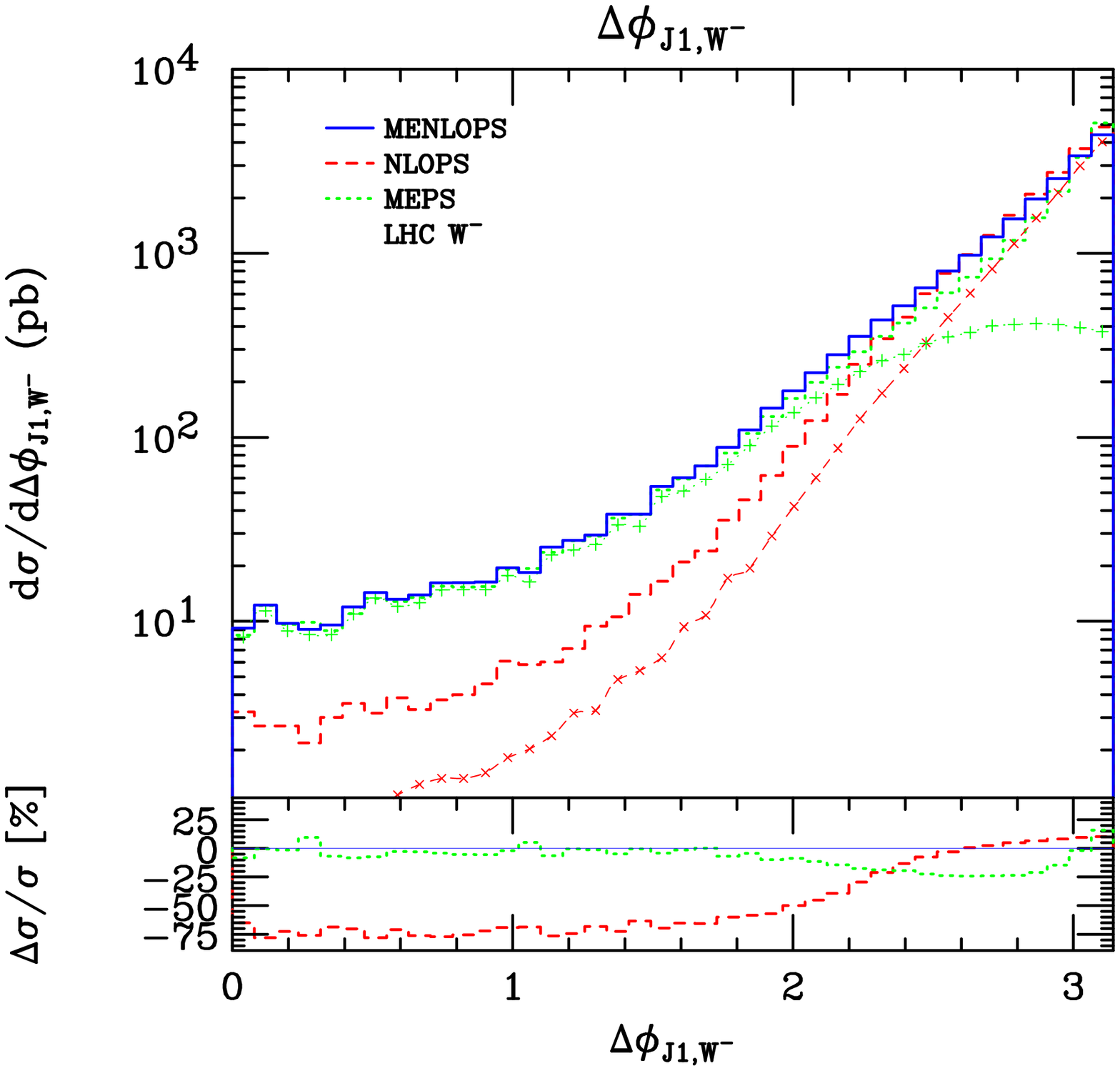}\hfill{}\includegraphics[width=0.4\textwidth,angle=90]{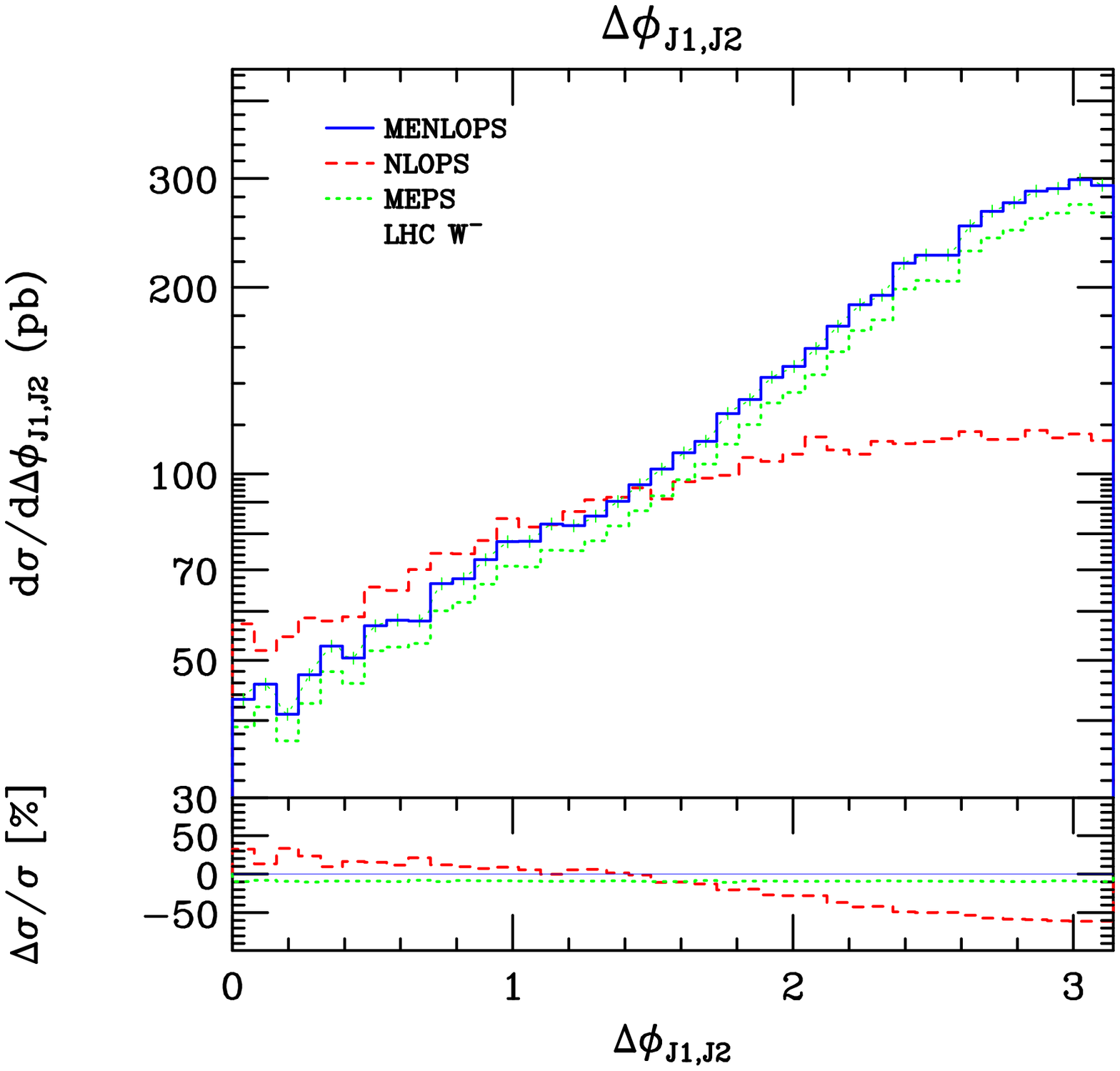} 
\par\end{centering}

\caption{In this figure we show two distributions further illustrating how
the description of additional jet activity compares in the \textsc{Nlops},
\textsc{Meps} and \textsc{Menlops} event samples. On the left we show
the difference in azimuth between the leading jet and the $\mathrm{W}^{-}$
boson, while on the right we show the difference in azimuth between
the two leading jets. These distributions show large differences by
virtue of the fact that the description of the second jet in the \textsc{Nlops}
simulation is given by the parton shower approximation. The parton
shower approximation strictly only contains information on the collinear
limits of matrix elements and, furthermore, it does not propagate
spin correlation information along the shower.}

\label{fig:W_delta_phis_1} 
\end{figure}

Substantial improvements in the description of the second jet can
be seen again very clearly in Figures~\ref{fig:W_jet_rapidity_corrs}
and \ref{fig:W_delta_phis_1}. Figure~\ref{fig:W_jet_rapidity_corrs}
shows a correction of 60\% in the \textsc{Meps}/\textsc{Menlops}
predictions with respect to the \textsc{Nlops} result in the rapidity
of the second jet with respect to the $\mathrm{W}^{-}$ boson.
The second jet in the \textsc{Meps} and \textsc{Menlops} samples
is significantly more central than in the \textsc{Nlops} case, which
is probably due to the fact that the \textsc{Meps} approach is more likely
to produce central jets than the shower algorithm.

Similarly large corrections can be seen in the azimuthal correlations
shown in Figure~\ref{fig:W_delta_phis_1}. On account of the fact that
the second hardest jet in the \textsc{Nlops} approach originates from
the shower approximation, any additional radiation from the incoming
legs is essentially distributed uniformly in azimuth, while final
state radiation is strongly correlated with the direction of the
leading jet. Given this fact one expects a deficit of events in the
\textsc{Nlops} sample for which the difference in azimuth between the
$\mathrm{W}^{-}$ and the leading jet,
$\Delta\phi_{\mathrm{J1},\mathrm{W}^{-}}$, is small. This is indeed
seen to be the case in Fig.\,\ref{fig:W_delta_phis_1}, which reveals
that the deficit is a rather significant one. A similar trend can be
seen later, for the case of $\mathrm{t}\bar{\mathrm{t}}$ pair production,
concerning the $\Delta\phi_{\mathrm{J1},\mathrm{t}\bar{\mathrm{t}}}$ correlation
(Fig.\,\ref{fig:tt_j2_rapidity_corr_and_delta_phi}).

Figure~\ref{fig:W_delta_phis_1} also shows the azimuthal correlation
between the two leading jets.
The \textsc{Meps} and \textsc{Menlops} predictions exhibit a much
higher degree of correlation in the back-to-back region. 
 In the \textsc{Nlops} simulation the only correlations which
may be present there are those due to kinematics and
momentum recoil effects, as opposed to genuine dynamics, since the
shower Monte Carlo produces secondary radiation that either follows the direction
of the leading jet (and thus has small azimuth), or is emitted by the incoming
partons, and is thus uniform in azimuth.

\begin{figure}[H]
\begin{centering}
\includegraphics[width=0.4\textwidth,angle=90]{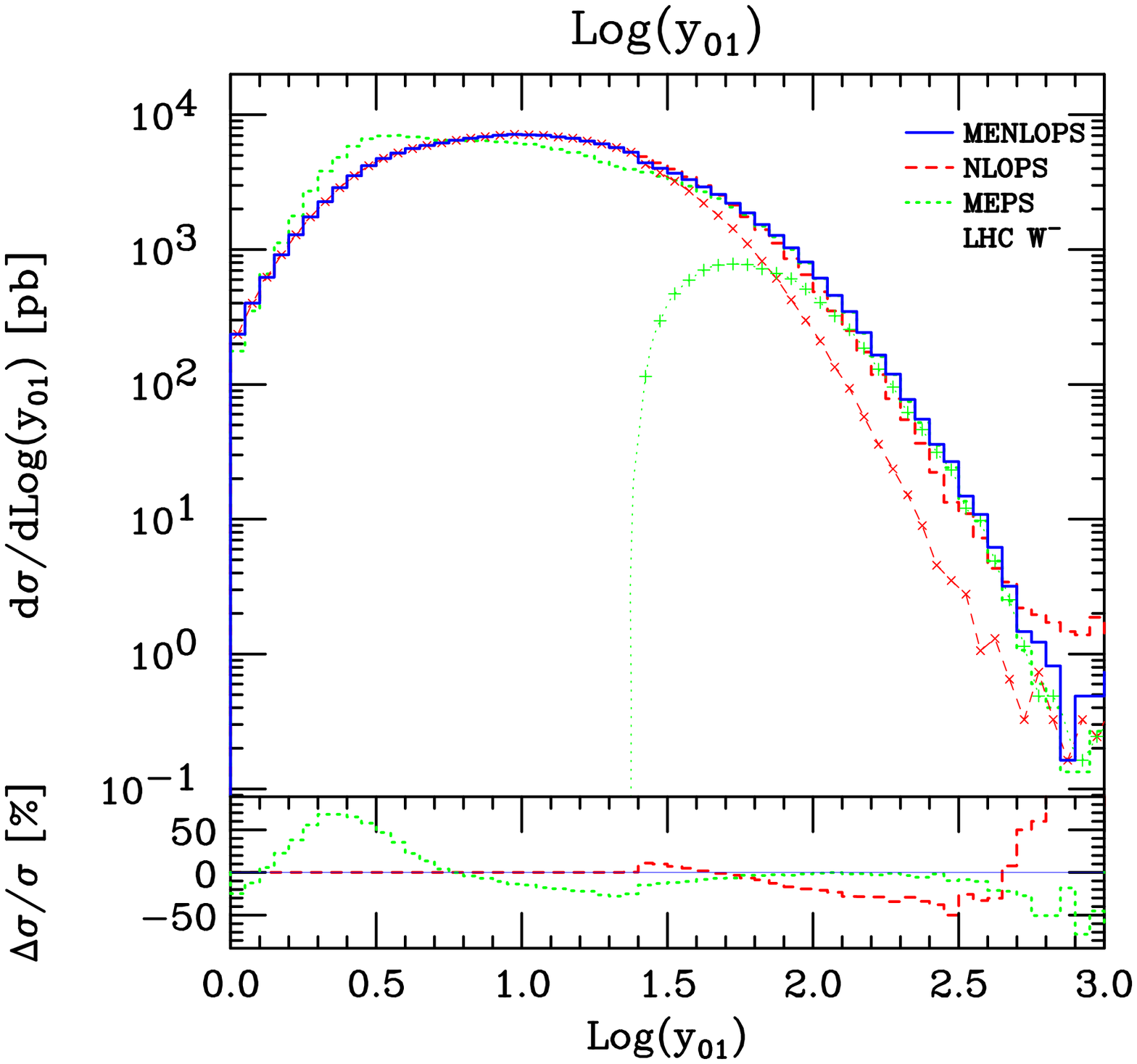}\hfill{}\includegraphics[width=0.4\textwidth,angle=90]{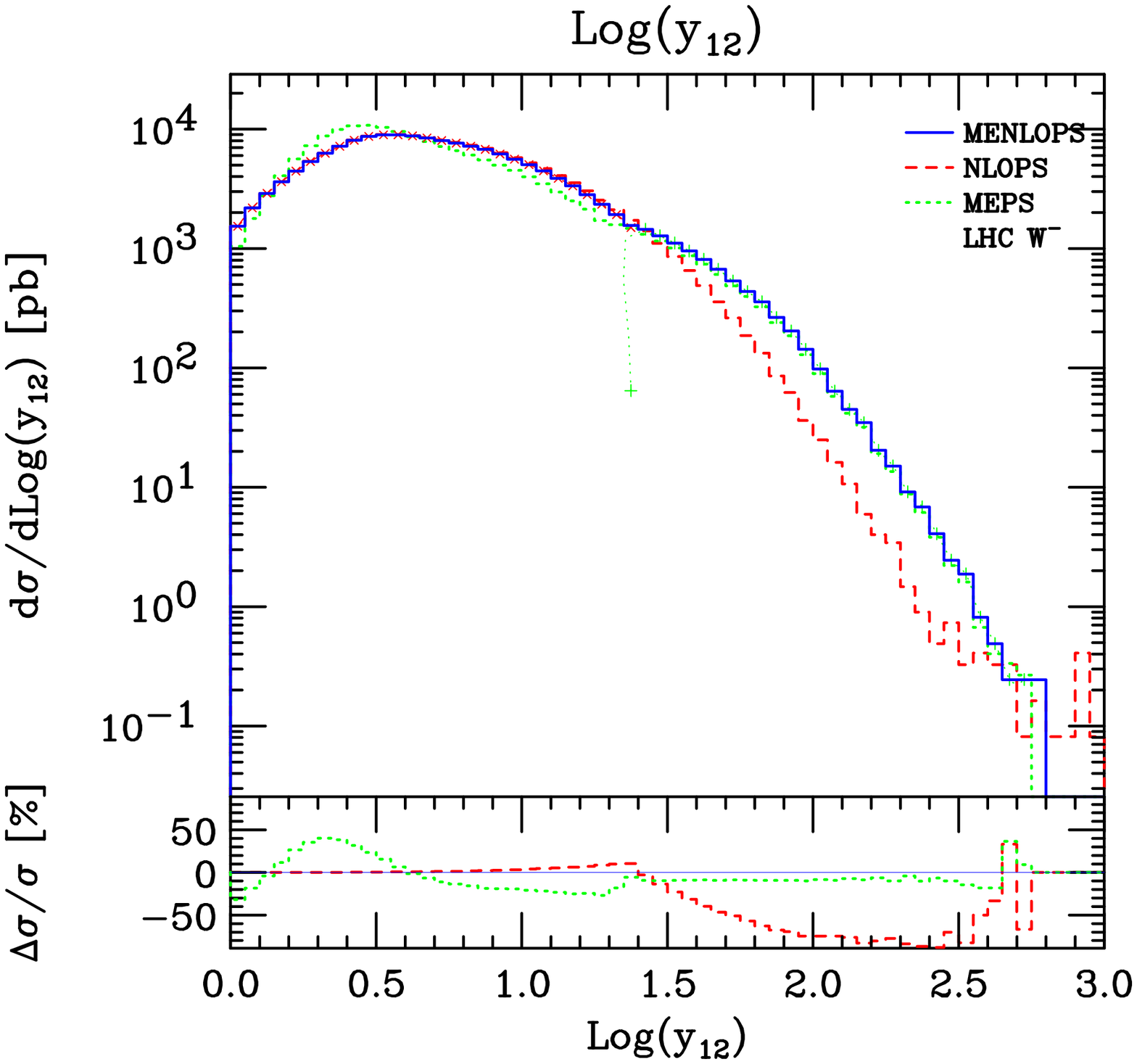} 
\par\end{centering}

\vspace{5mm}

\begin{centering}
\includegraphics[width=0.4\textwidth,angle=90]{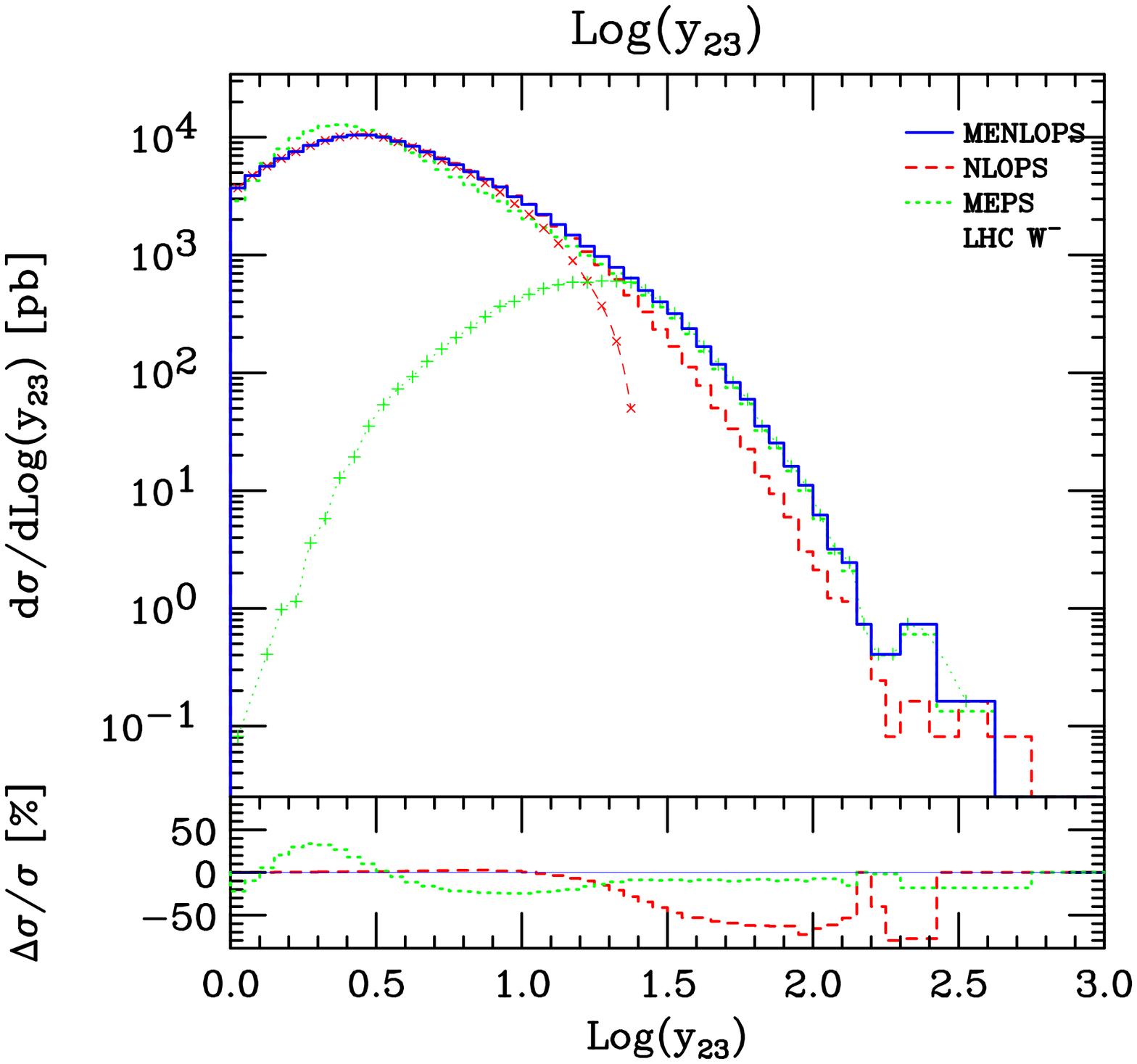}
\hfill{}\includegraphics[width=0.4\textwidth,angle=90]{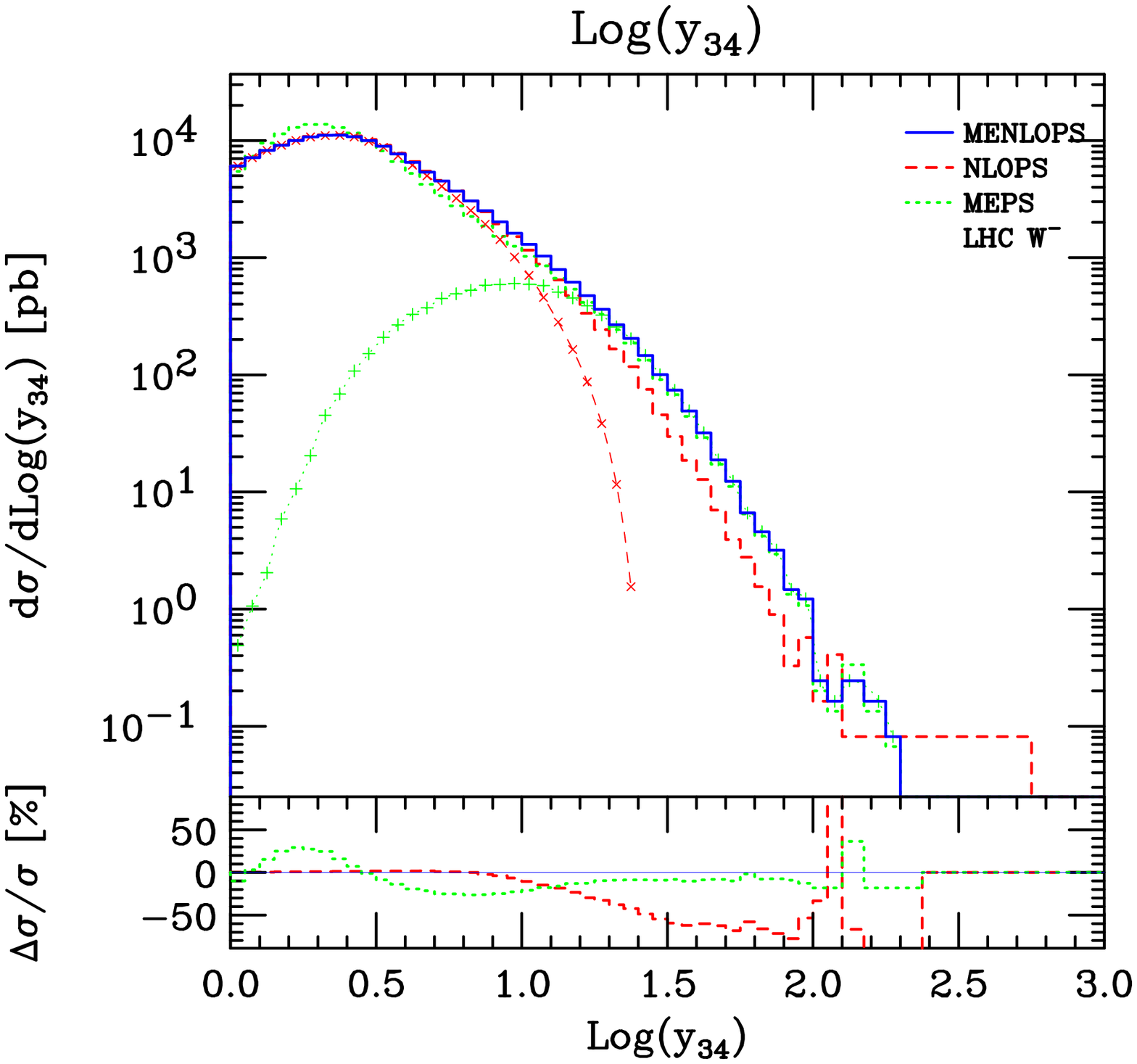} 
\par\end{centering}

\caption{Differential jet rates for $\mathrm{p}\mathrm{p}\rightarrow\mathrm{W}^{-}\left(\rightarrow\mathrm{e}^{-}\overline{\nu}_{\mathrm{e}}\right)+\mathrm{jets}$.
These plots show the logarithm of the value of the jet clustering
scale $y_{nm}$ at which an $n$-jet event is resolved as an $m=n+1$-jet
event.}

\label{fig:W_djrs} 
\end{figure}

Lastly we consider the differential jet rates displayed in
Figure~\ref{fig:W_djrs}.  Recall that these distributions directly
probe the behavior of the \textsc{Meps} and \textsc{Menlops} samples
around the phase space partitions in these two approaches. We recall
that the merging
scale used to make the \textsc{Meps} combination was taken to be 20
GeV, while in making the default \textsc{Menlops} sample we use a
value of 25 GeV.

In the \textsc{Meps} case the merging between the parton shower and
the matrix elements involves a phase space partition for every different
multiplicity. In the \textsc{Menlops} case all events with 0 or 1
jet are described by the one \textsc{Nlops} simulation, with the \textsc{Meps}
sample alone describing the rest. It follows that the \textsc{Menlops}
approach should not induce the appearance of discontinuities in the differential
jet rates,
with the exception of the $y_{12}$ jet rate, where there is a complete
transition at 25 GeV from the \textsc{Meps} description to the \textsc{Nlops}
one.

In all cases one can see that the predictions from the pure \textsc{Meps}
sample (green dots) are smooth with no evidence of any merging scale
dependence. The two distributions exhibit some differences in the
soft region, the \textsc{Meps} sample favoring more soft emission.
This behavior has already been noted in the discussion
of the 0-jet fraction in  Fig.\,\ref{fig:W_jet_fractions_1},
where it was attributed to differences in the \textsc{Meps}
Sudakov form factors with respect to \textsc{Powheg}.

The \textsc{Menlops} predictions (blue) are also smooth across the
25 GeV boundary in the $y_{12}$ distribution, in spite of the abrupt
transition from the \textsc{Nlops} to the \textsc{Meps} samples. Some
relatively minor distortion in the first derivative can be seen at
this point in the $y_{12}$ plot. However, it is similar in magnitude
to that seen in the pure \textsc{Meps} results.

In general one should not expect that either the \textsc{Menlops}
or the \textsc{Meps} differential jet rates be completely smooth,
since the distributions either side of the boundary are only formally
equivalent at the leading-log level, differences by terms of $\mathcal{O}\left(\alpha_{\mathrm{S}}\right)$
should be expected in both cases. However, since in both procedures
one aims to merge at the lowest scale allowed the large logarithms
dominate the distributions at the merging partition(s). 

In principle there is no reason why one may not construct the \textsc{Menlops}
sample using a \emph{floating} value of the merging scale. In fact it is 
testament to the flexibility and transparency of the whole approach that this floating
scale can be implemented with great ease, in dividing the \textsc{Meps} and
\textsc{Nlops} samples. We have also performed such a merging taking the scale
to be Gaussian distributed about 25 GeV with a standard deviation of 5 GeV. The
results of doing this are the same as above, although the $y_{12}$ jet rate
naturally appears smoother this way. However, we prefer to be prudent and present
our results in such a way as to be open about the presence of the unphysical scale.

\subsection{Top quark pair production\label{sub:Top-quark-pair}}

In this subsection we present the results of applying our method to
the top quark pair production process. The simulation of the top quarks
in both \textsc{Meps} and \textsc{Nlops} samples does not include
their decay. In addition, the final-state top quarks are not input
to the jet clustering process; thus when no additional radiation occurs
the jet finding algorithm will return 0 jets.

In producing the \textsc{Meps} sample with \textsc{Madgraph} we have
set the $k_{\perp}$ jet measure cut on the tree level event generation
to be 20 GeV, and we have chosen the corresponding \textsc{Meps}
merging scale to be 30 GeV. These values are recommended for the production
of inclusive $\mathrm{t}\bar{\mathrm{t}}$ pair production in Refs.\,\cite{Alwall:2008qv,Madgraph:2007},
when using the virtuality ordered P{\footnotesize YTHIA }parton shower.
We note that the recommended \textsc{Meps} merging scale in the case
of the transverse momentum ordered shower is 100 GeV, much larger than our
chosen value.

The default \textsc{Menlops} sample used to produce the results in
this subsection was constructed by combining the \textsc{Nlops} and
\textsc{Meps} samples with a merging scale of 60 GeV. This scale is
30 GeV above the merging scale in our \textsc{Meps} sample but still
lower than that recommended in the case of merging with the transverse
momentum ordered shower \cite{Alwall:2008qv}. The \textsc{Menlops}
cross section is equal to that of the \textsc{Nlops} event generation,
817 pb, and the fraction of \textsc{Meps} events in the total sample
is 12.5\%, marginally above $\alpha_{\mathrm{S}}$. In looking at
the jet multiplicity distributions and inclusive observables it is
useful to consider the effects of varying the merging scale, hence,
some results are also obtained using a greater merging scale of 100
GeV, for which the fraction of \textsc{Meps} events in the \textsc{Menlops}
sample drops to 4\%. 

As explained in Section~\ref{sec:Hardest-emission-xsecs}, and as
will be demonstrated in the following, the fraction $\alpha_{\mathrm{S}}$
of \textsc{Meps} events in the sample has a completely negligible
impact on the \textsc{NLO} accuracy of inclusive observables. However,
to lower the merging scale further will lead to an increased number
of \textsc{Meps} events and formally compromise \textsc{NLO} accuracy.
Hence the \textsc{Menlops} description here can be understood as offering
the best of the \textsc{Nlops} and \textsc{Meps} descriptions except,
arguably, in the $k_{\perp}$-jet measure window $30<y_{ij}<60\,\mathrm{GeV}$,
for jet multiplicities higher than two, which is populated by events
taken from the \textsc{Nlops} sample.

\subsubsection{Jet multiplicities}

\begin{figure}[H]
\begin{centering}
\includegraphics[scale=0.37,angle=90]{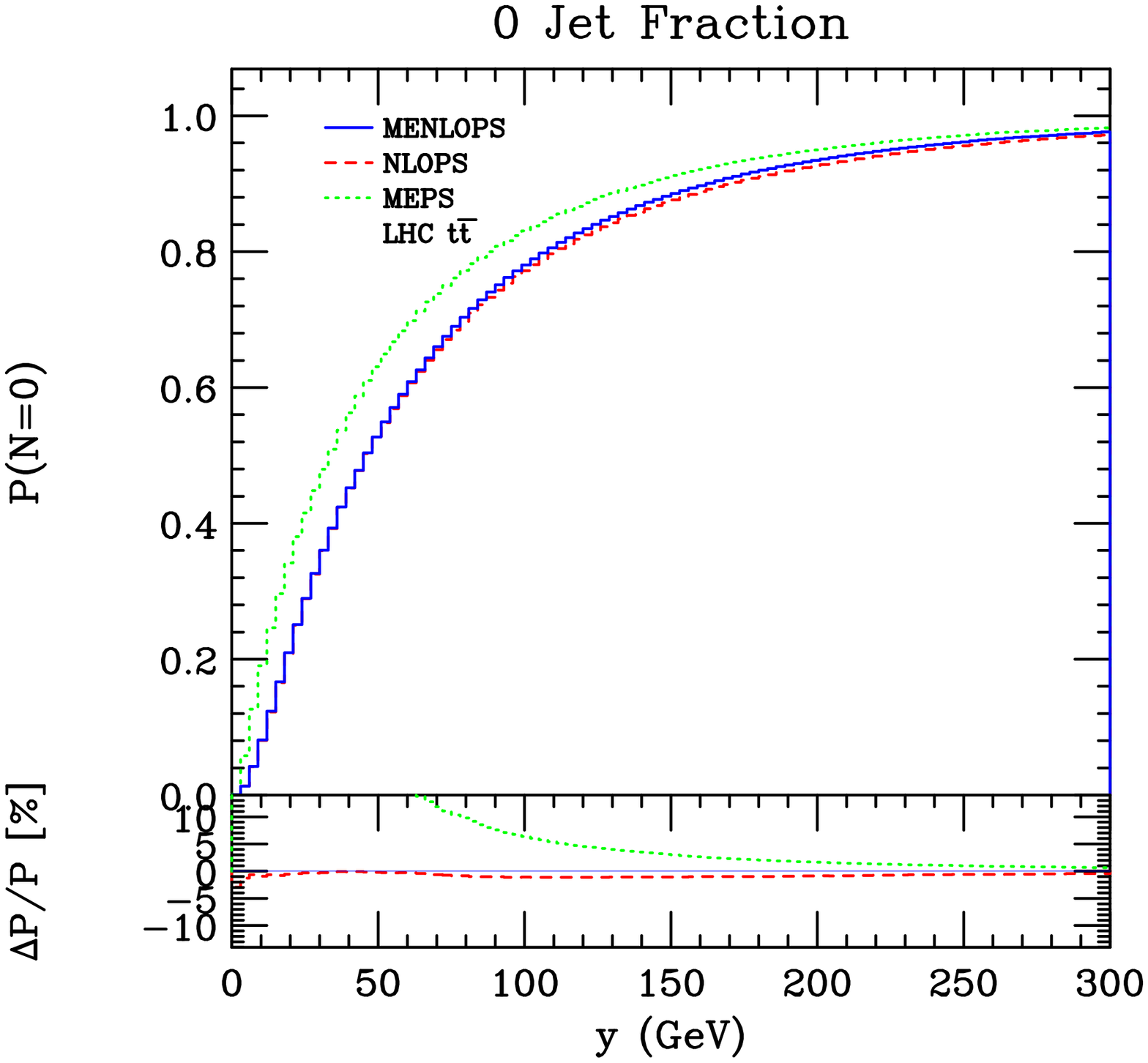}\hfill{}\includegraphics[scale=0.37,angle=90]{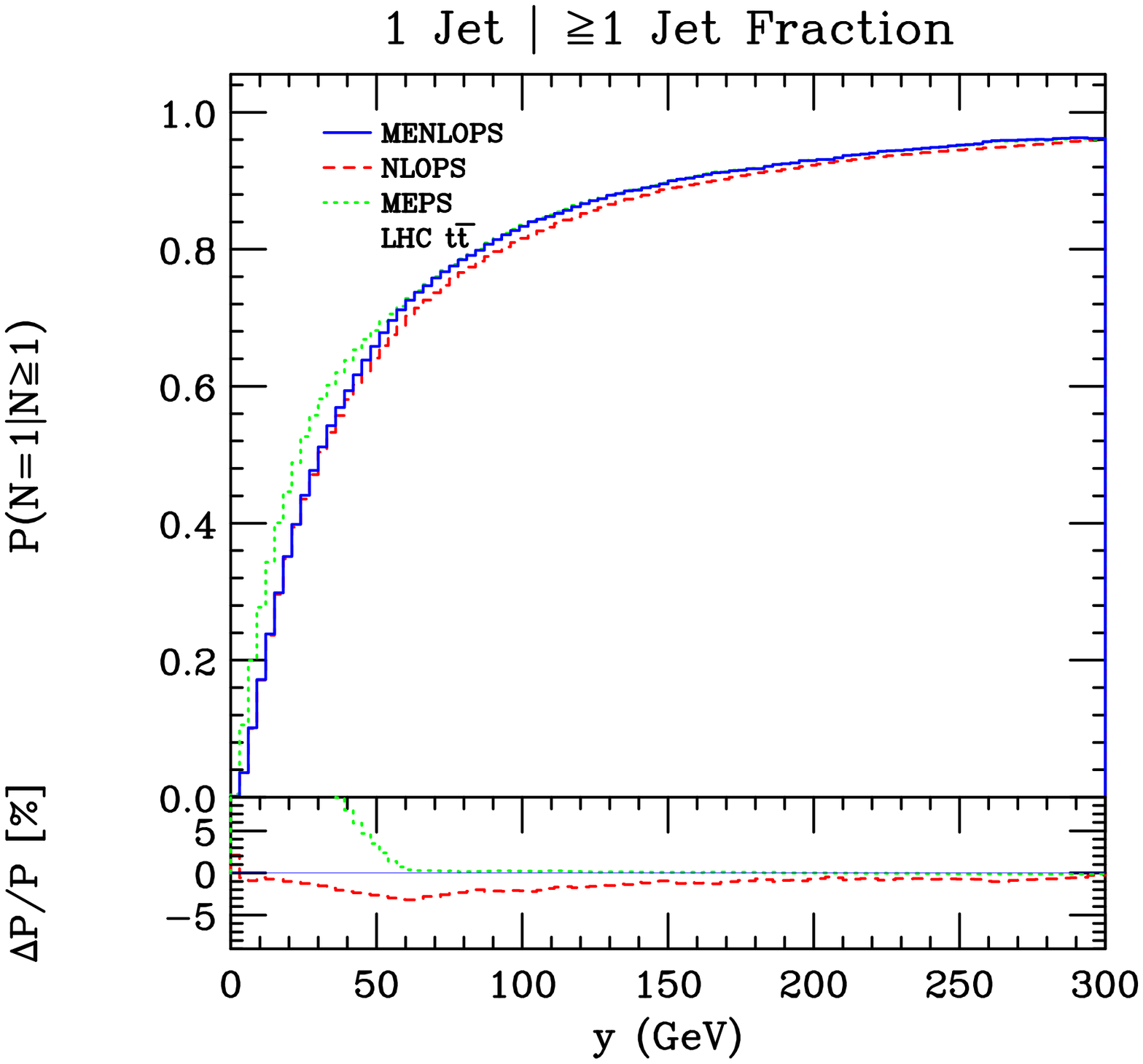} 
\par\end{centering}

\caption{In the left plot we show the 0-jet fractions in the event
samples, as a function of the jet resolution scale $y$, defined according
to the Durham $k_{\perp}$ jet measure. The dashed (red) and dotted (green) lines
correspond to the pure \textsc{Nlops} and \textsc{Meps} predictions
respectively. On the right plot we show
the fraction of 1-jet events in the subset consisting of events with
at least one jet.}

\label{fig:tt_jet_fractions_1} 
\end{figure}

\begin{figure}[H]
\begin{centering}
\includegraphics[scale=0.37,angle=90]{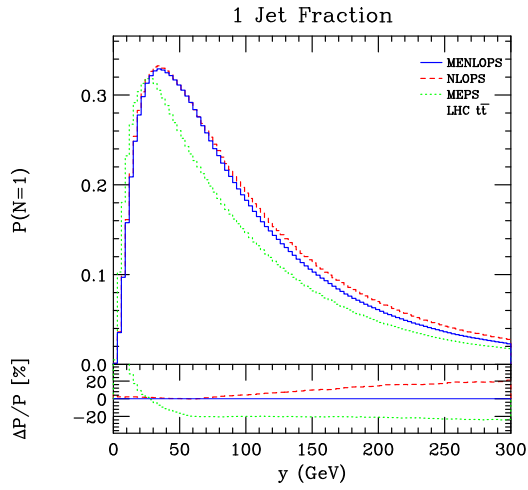} 
\par\end{centering}

\caption{The 1-jet fractions in the event
samples, as a function of the jet resolution scale $y$, defined according
to the Durham $k_{\perp}$ jet measure. The dashed (red) and dotted (green) lines
correspond to the pure \textsc{Nlops} and \textsc{Meps} predictions
respectively. }

\label{fig:tt_jet_fractions_2} 
\end{figure}

Shown in Figures~\ref{fig:tt_jet_fractions_1} and \ref{fig:tt_jet_fractions_2}
are the 0- and 1-jet fractions in each of the samples, as a function
of our jet resolution parameter $y$ (Sect.\,\ref{sub:MENLOPS-implementation}).
These histograms were made by applying the exclusive jet finding algorithm
to each of the samples at different values of the clustering cut parameter.
The solid (blue) histogram shows the results of this analysis procedure when
applied to a \textsc{Menlops} sample constructed as described in Sect.\,\ref{sub:MENLOPS-implementation}
with a merging scale of 60 GeV, the corresponding results for the
pure \textsc{Nlops} and \textsc{Meps} samples are shown in the dashed (red)
and dotted (green) lines respectively. 
The 0-jet fraction has the typical shape of a Sudakov form factor.
This is in keeping with the fact that the latter has an interpretation
as the probability for not emitting any radiation above a given scale.  
We also show for each sample the conditional probability for obtaining
a 1-jet event from a sample of events where each contains at least
one jet. The ratio of the latter quantity in the \textsc{Meps} and
\textsc{Nlops} samples is used in constructing the \textsc{Menlops}
sample Eq.\,\ref{eq:sec3_menlops_master_formula}. Using similar
reasoning to that above, this quantity can be thought of as representing
the Sudakov form factor probability for a 1-jet event to evolve into
a $\geqslant 2$-jet event at the given scale. As with the 0- and 1-jet distributions
this Sudakov form factor should be understood in the sense of having
been averaged over the underlying Born variables. This is clear from
the form of the distribution in Figure~\ref{fig:tt_jet_fractions_1}.
The 1-jet fraction of Figure~\ref{fig:tt_jet_fractions_2} results from the
combined effects of the Sudakov form factor for 0-jet emission, that differs
from 1 by the probability to emit 1 or more jets, and the probability
to find a 1-jet event out of the sample of events with at least one jet.

Having elaborated on the general dynamics behind the jet fractions,
we now move to discuss the finer details of the distributions, comparing the
predictions from each merging scheme.
It is clear from the plots that the largest differences between the
\textsc{Nlops} and \textsc{Meps} samples occur in the region of the
Sudakov peak, where the rates are governed by the all orders resummation
of large logarithms. In the case of the 0- and 1-jet fractions, below
the \textsc{Meps} merging scale (30 GeV), this is therefore the difference
between the P{\footnotesize YTHIA} virtuality ordered parton shower
and the \textsc{Powheg} hard emission generator. Since the resummation
in the \textsc{Powheg} case is nearly \textsc{NLL} accurate,
this quantity is better determined by it than by P{\footnotesize YTHIA}.
In fact, as we will discuss in more detail later in the context of
the $\mathrm{t}\bar{\mathrm{t}}$ $p_{T}$ spectrum, the \textsc{Powheg}
prediction offers a substantial improvement over that of the virtuality
ordered shower from the point of view of the treatment of the scales
used in the evaluation of the \textsc{PDF}s.
Note that below the \textsc{Menlops} merging scale (60 GeV) the 0-jet
fraction in the \textsc{Menlops} sample is identical, by construction,
to the \textsc{Nlops} result, while in the case of the 1-jet fraction
they are different by a constant factor 
(see Eq.\,\ref{eq:sec3_menlops_master_formula}).

Going above the 60 GeV merging scale, as we resolve the events over increasingly
large $y$ values, one sees that the \textsc{Menlops} 0- and 1-jet fractions begin
to include proportionally more \textsc{Meps} events, since, for example,
2-jet events at the \textsc{Menlops} merging scale are resolved as
0- and 1-jet events at these higher scales. Also, above the 60 GeV
merging scale, the conditional probability for obtaining a 1-jet event
given a set of events each with at least 1-jet, is by default equal
to that in the \textsc{Meps} sample. Recall that this is very closely
related to the Sudakov form factor probability for a 1-jet event to
evolve to a scale below the jet resolution scale, without emitting
any radiation. In the \textsc{Nlops} case this is given by P{\footnotesize YTHIA}
alone, while in the \textsc{Meps} case corrections from exact, higher
multiplicity, tree level matrix elements are included. 
Whereas, for small values of $y$, these probabilities are controlled by
the Sudakov form factors in P{\footnotesize YTHIA}  and in the \textsc{Meps},
and thus the \textsc{Nlops} and the \textsc{Meps} have the
same accuracy, for relatively large $y$ the \textsc{Meps} value should be
preferred, since it relies upon the exact matrix element result. This is
why the \textsc{Meps} value of this fraction is adopted in our
\textsc{Menlops} method.

\begin{figure}[H]
\begin{centering}
\includegraphics[width=0.4\textwidth,angle=90]{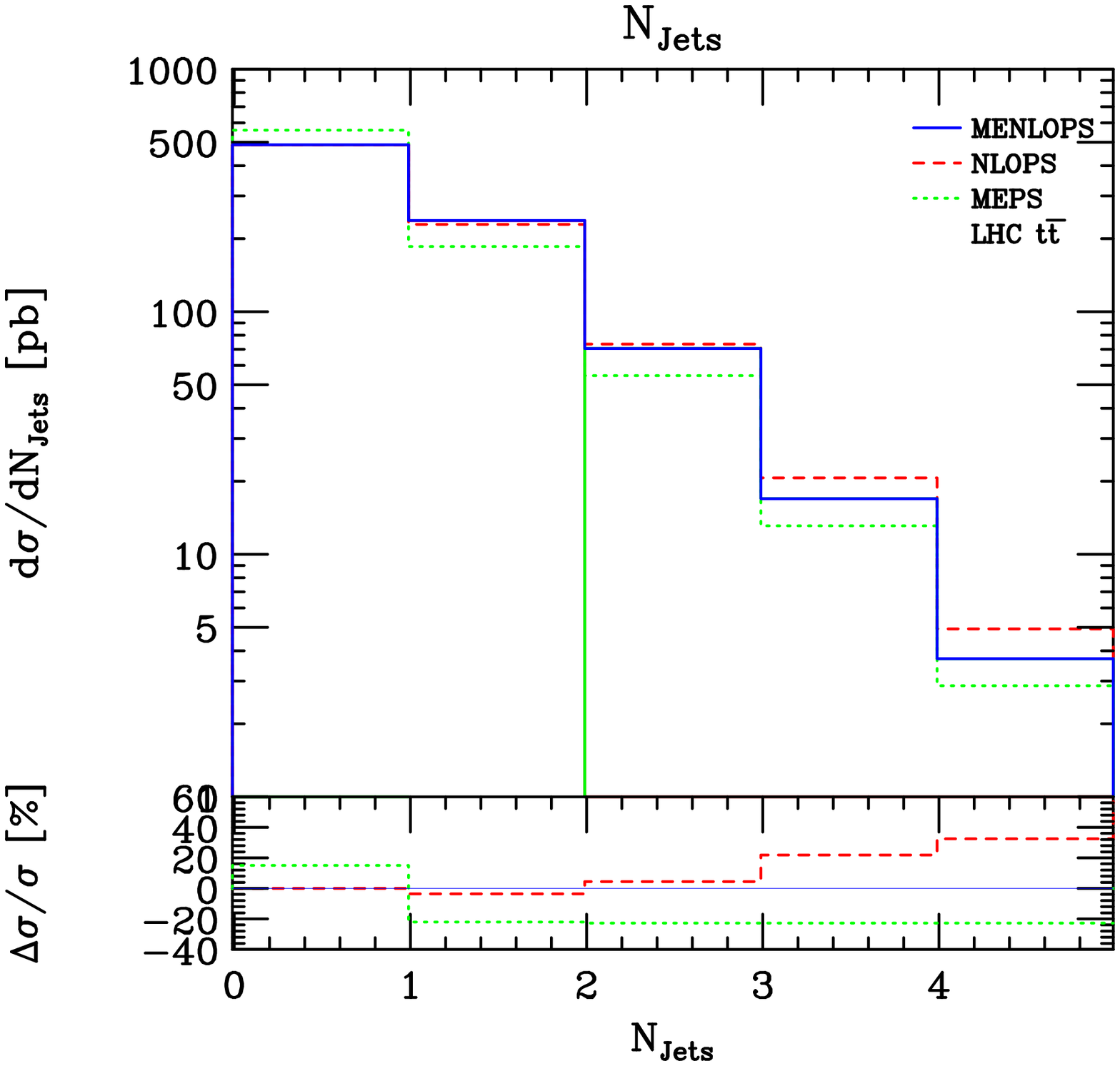}\hfill{}\includegraphics[width=0.4\textwidth,angle=90]{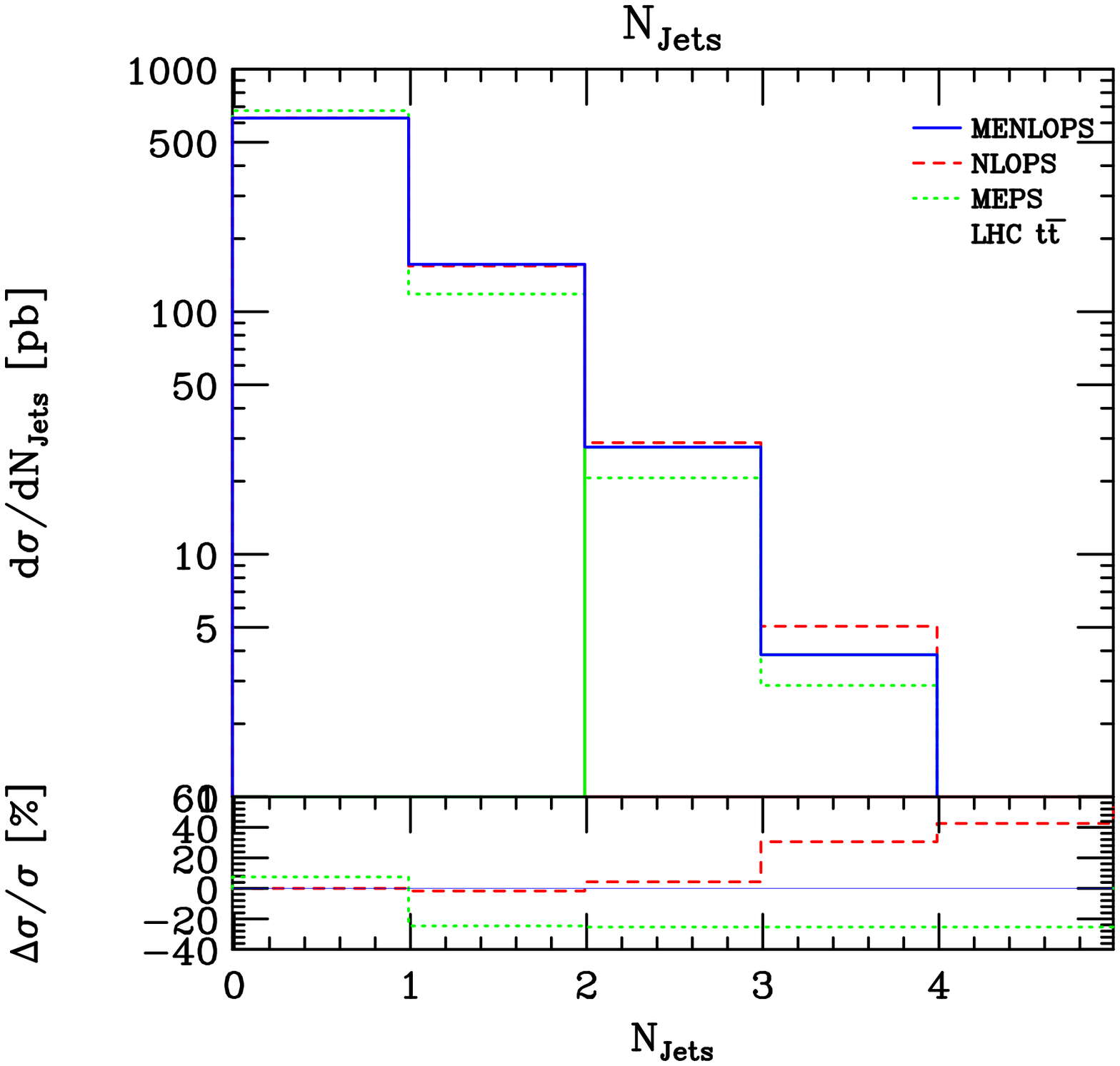} 
\par\end{centering}

\vspace{5mm}

\caption{The jet multiplicity distributions for $\mathrm{t}\bar{\mathrm{t}}$
pair production events using two different choices of the \textsc{Menlops}
merging scale: 60 GeV (left) and 100 GeV (right). In both cases we
have used this merging scale as the jet resolution parameter in order
to calculate the multiplicities. The 0- and 1-jet events in the \textsc{Menlops}
sample are taken solely from the \textsc{Nlops} sample, while those
with higher multiplicities come from the \textsc{Meps} one.}

\label{fig:tt_jet_multiplicities} 
\end{figure}

Figure~\ref{fig:tt_jet_multiplicities}
is meant to illustrate the working of the merging procedure.
It shows the jet multiplicity
distributions when the \textsc{Nlops} and and \textsc{Meps} samples
are combined with \textsc{Menlops} merging scales of 60 GeV and 100
GeV, with the jet resolution scale taken equal to the \textsc{Menlops}
merging scale. As expected, one sees that the cross section for each
value of the multiplicity is lower in the latter case.

Since the jet resolution scale here is equal to the \textsc{Menlops}
merging scale, the \textsc{Menlops} histogram entry (solid) corresponding
to events with no additional jets, as well as that of its \textsc{Nlops}
component, is exactly equal, by construction, to the pure
\textsc{Nlops} result (dashes). Equally, for jet multiplicities
greater than one, the overall \textsc{Menlops} rates are exactly equal
to those of the \textsc{Meps} distribution (dots) multiplied
by the \textsc{NLO} \emph{K}-factor associated with the production
of at least one jet, $\sigma_{\mathrm{PW}}\left(\ge1\right)/\sigma_{\mathrm{ME}}\left(\ge1\right)$,
as described in Section~\ref{sec:Combining-Powheg-and-Meps} (Eq.\,\ref{eq:sec3_menlops_master_formula}). 

These observations are most obvious from the ratio plots showing the
fractional difference of the pure \textsc{Meps} and \textsc{Nlops}
samples with respect to the \textsc{Menlops} one. We remind the reader
that the pure \textsc{Meps} prediction has been rescaled by a different
\emph{K}-factor, $\sigma_{\mathrm{PW}}\left(\ge0\right)/\sigma_{\mathrm{ME}}\left(\ge0\right)$,
the ratio of the total cross sections. It follows that the dotted (green)
line in the lower panels is constant with respect to the blue \textsc{Menlops}
reference line, but not equal to it, the gap being given by the total
\emph{K}-factor divided by the \emph{K-}factor for the production
of at least one jet, minus one.

\subsubsection{Inclusive observables}

\begin{figure}[H]
\begin{centering}
\includegraphics[width=0.4\textwidth,angle=90]{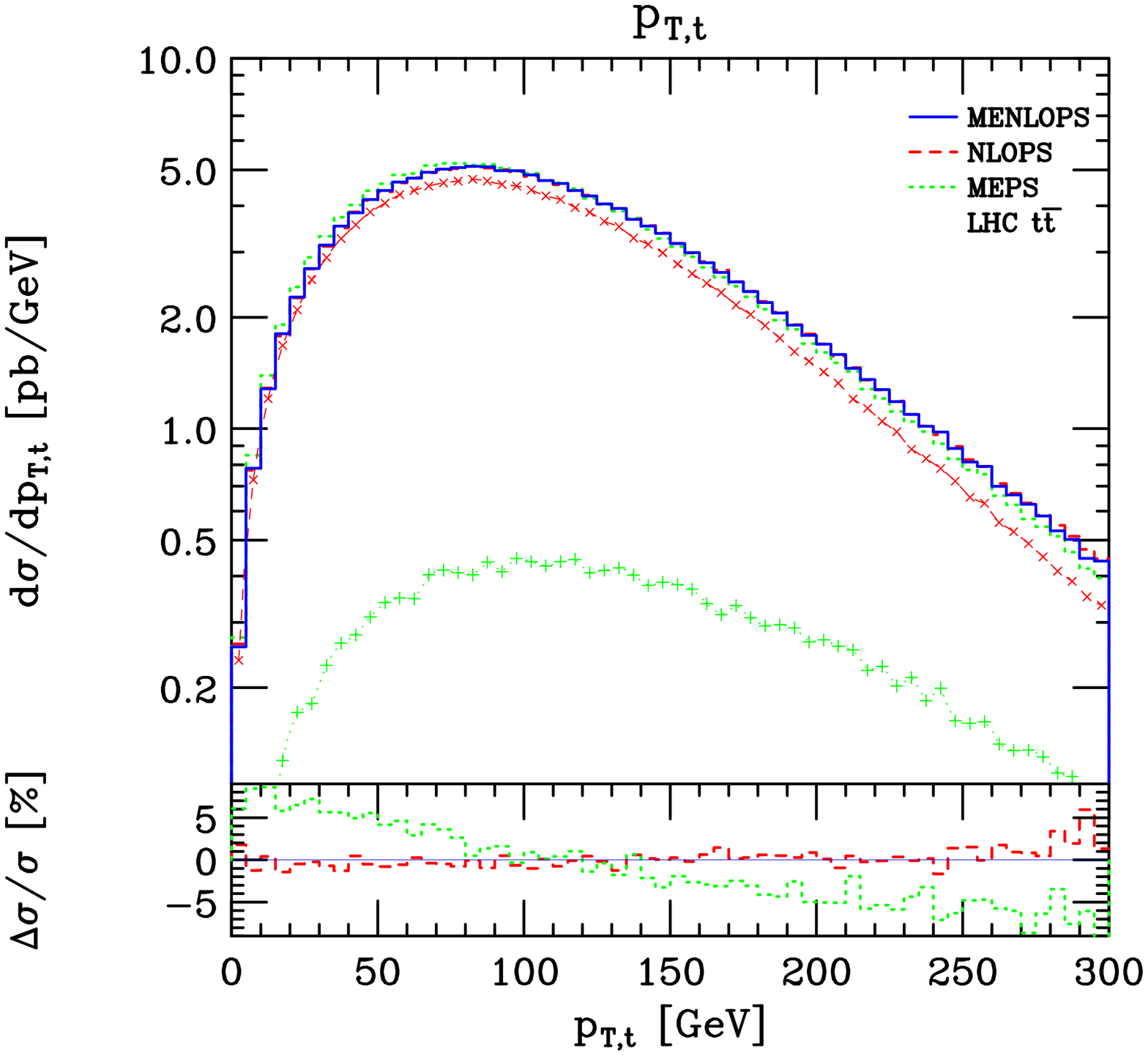}\hfill{}\includegraphics[width=0.4\textwidth,angle=90]{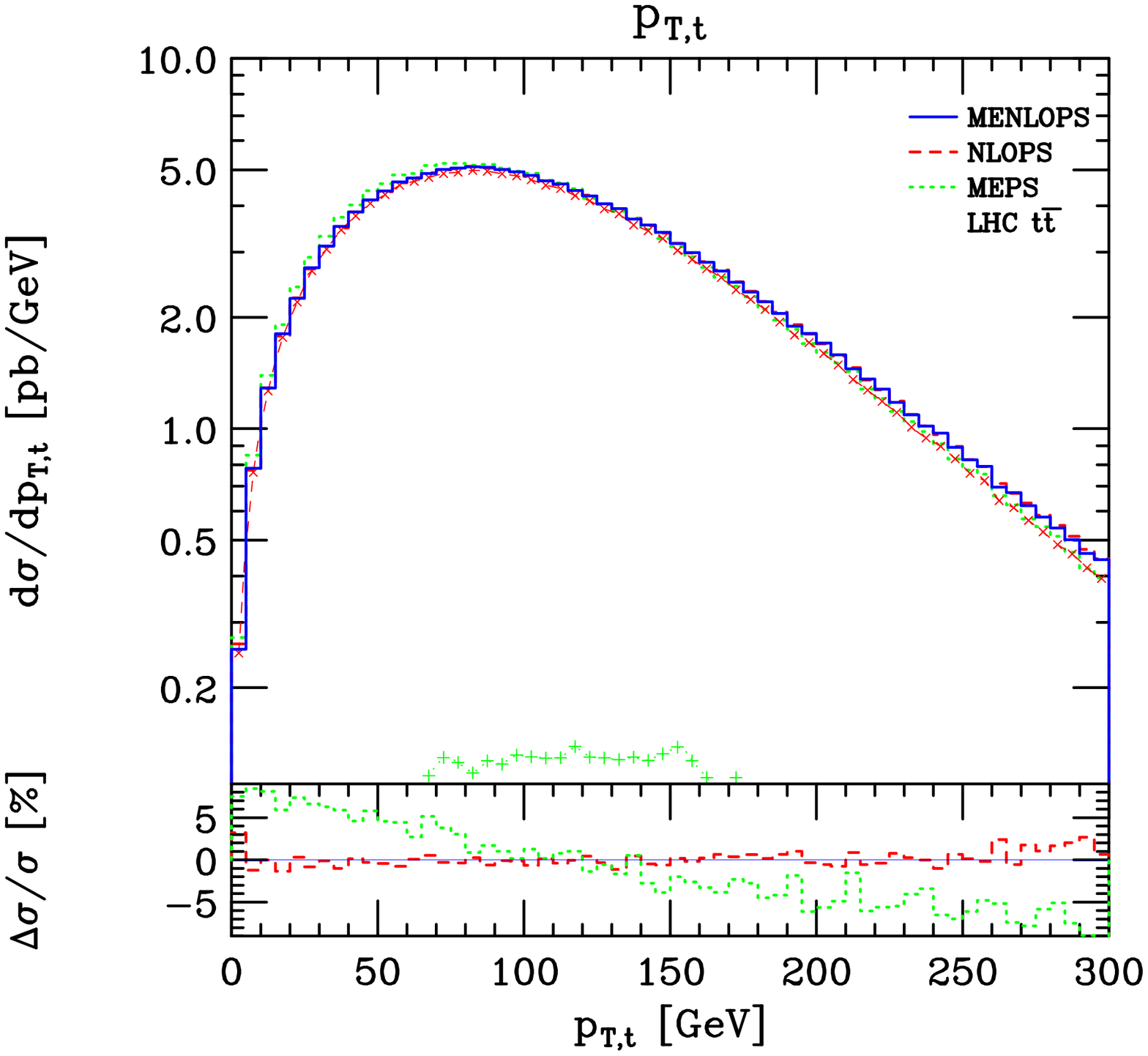} 
\par\end{centering}

\begin{centering}
\includegraphics[width=0.4\textwidth,angle=90]{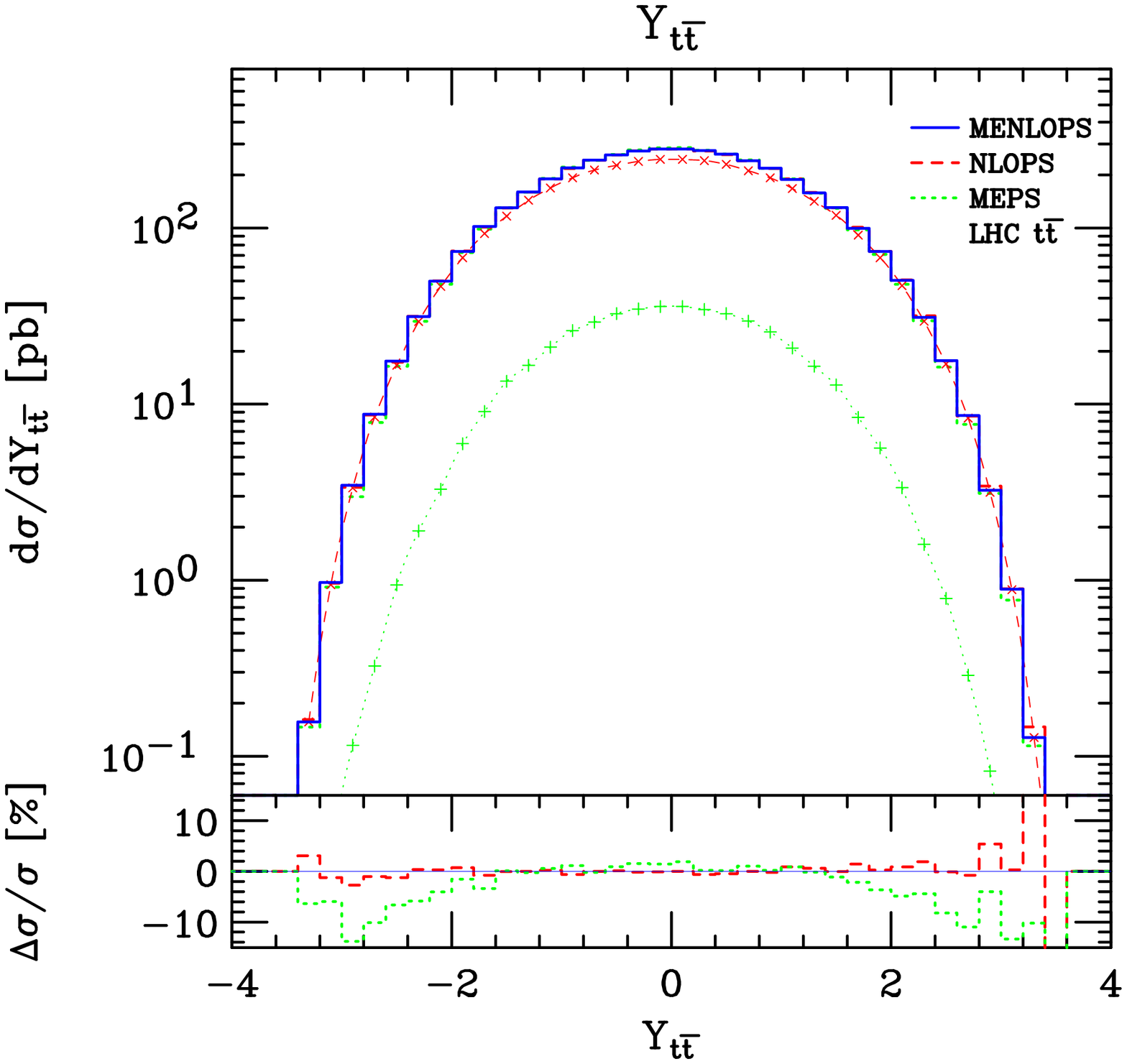}
\hfill{}\includegraphics[width=0.4\textwidth,angle=90]{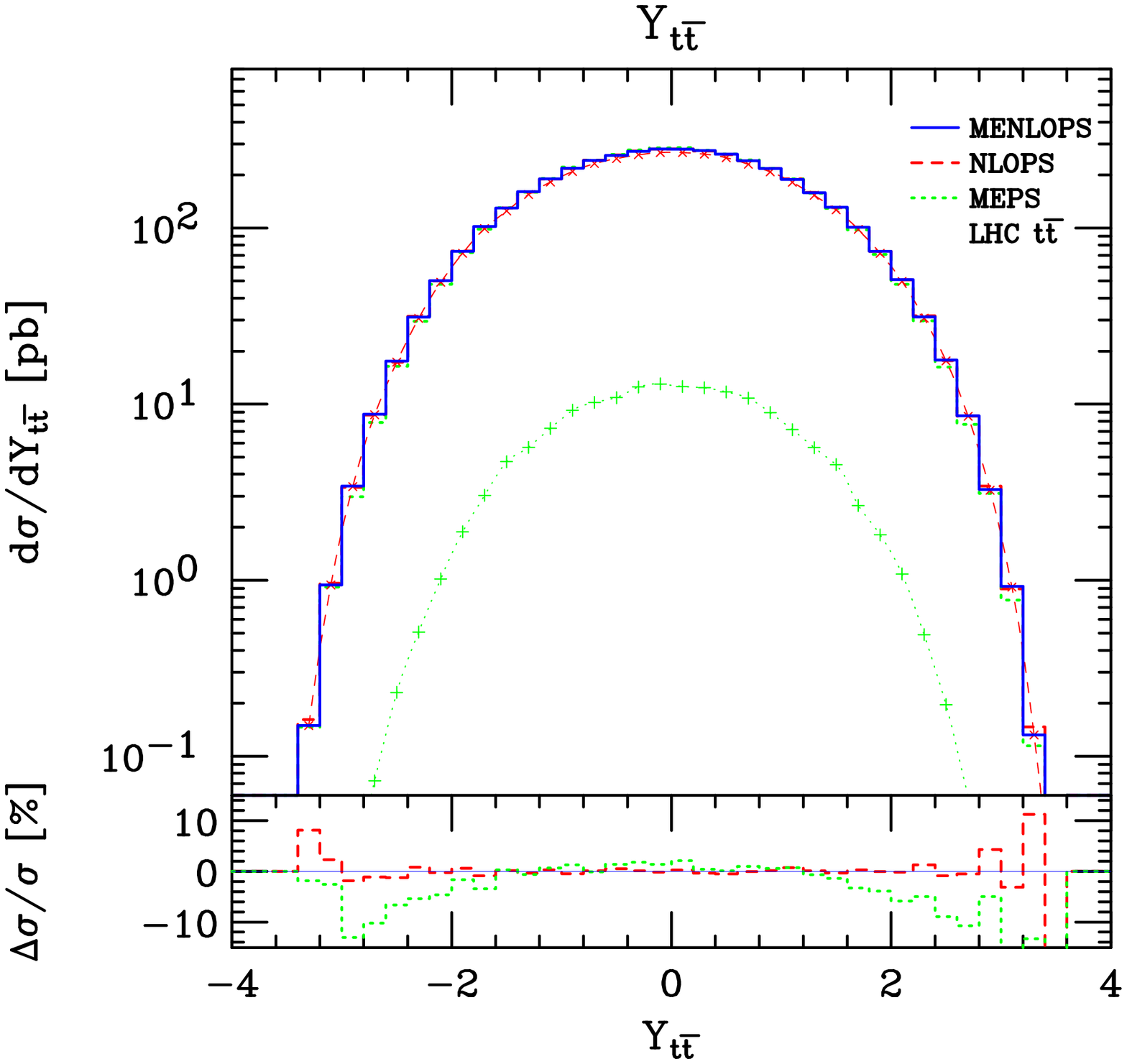} 
\par\end{centering}

\caption{In the upper half of this figure we show the top quark transverse
momentum in $\mathrm{t}\bar{\mathrm{t}}$ pair production using a
60 GeV (left) and 100 GeV (right) jet resolution scale in the \textsc{Menlops}
merging procedure. The lower pair of plots shows, analogously, the
rapidity distribution for the combined $\mathrm{t}\bar{\mathrm{t}}$
system. Despite the relatively large difference in the merging
scales the combined \textsc{Menlops} prediction is stable with respect
to the changing scale, showing deviations from the \textsc{NLO} result
at the level of only 1 or 2\% in both cases.}

\label{fig:tt_inclusive_observables} 
\end{figure}

In Figure~\ref{fig:tt_inclusive_observables} we show the transverse
momentum distribution of the top quark in $\mathrm{t}\overline{\mathrm{t}}$
pair production as well as the rapidity of the top quark pair, using
two different scales to carry out the merging of the \textsc{Meps} and
\textsc{Nlops} samples (60 GeV and 100 GeV). 

The \textsc{Menlops} results are found to be blind to the variation
in this unphysical merging scale, exhibiting no discernible deviation
from the pure \textsc{Nlops} result. This is particularly evident
from considering the tail of the top quark $p_{T}$ distributions.
These observations are reassuring and completely understandable given
the total content of the \textsc{Menlops} sample; for the 60 GeV merging
scale the \textsc{Meps} subsample comprises only 12.5\% of the total,
while in the 100 GeV case it is only 4\%, therefore deviations from
the pure \textsc{Nlops} result are restricted to be of order 10\%
times $\alpha_{\mathrm{S}}$. 

Appreciable differences of $\mathcal{O}\left(\alpha_{\mathrm{S}}\right)$
can be seen in the shapes of the \textsc{Meps} prediction of $Y_{\mathrm{t}\overline{\mathrm{t}}}$
with respect to those of the \textsc{Nlops}/\textsc{Menlops} samples.
We attribute this discrepancy to the absence of exact \textsc{NLO}
corrections in the \textsc{Meps} case. A similar trend was observed
in the case of the W boson rapidity spectrum, for which the deviation
in the shape was more prominent. Here, as in that case, we propose 
that these differences arise from a systematic bias of the \textsc{Meps}
approach, which produces the leading order final-state system more centrally
than one expects on the grounds of pure LO and NLO computations 
(Sect.\,\ref{sub:W-boson-production}). 

\begin{figure}[H]
\begin{centering}
\includegraphics[width=0.4\textwidth,angle=90]{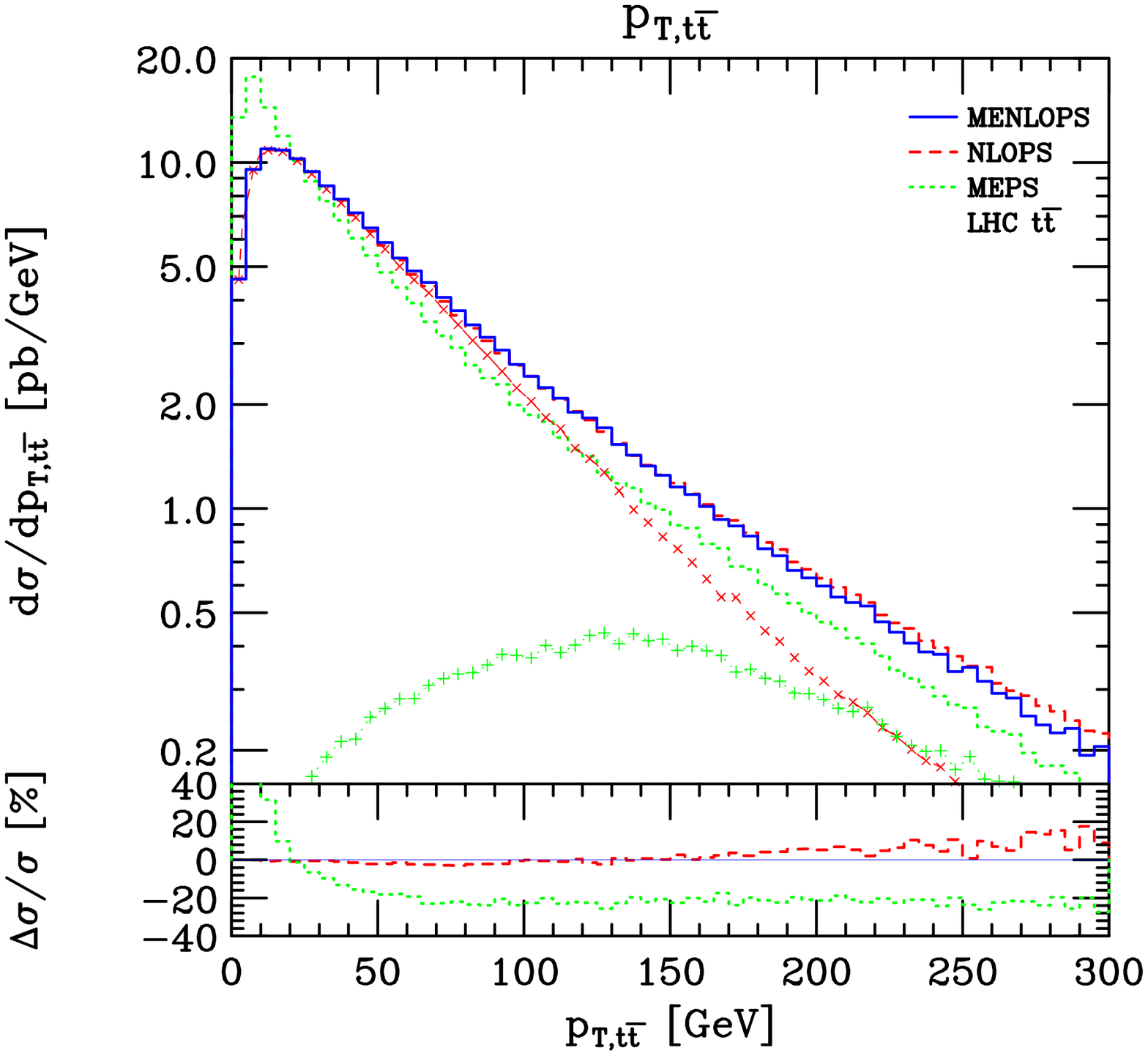}\hfill{}\includegraphics[width=0.4\textwidth,angle=90]{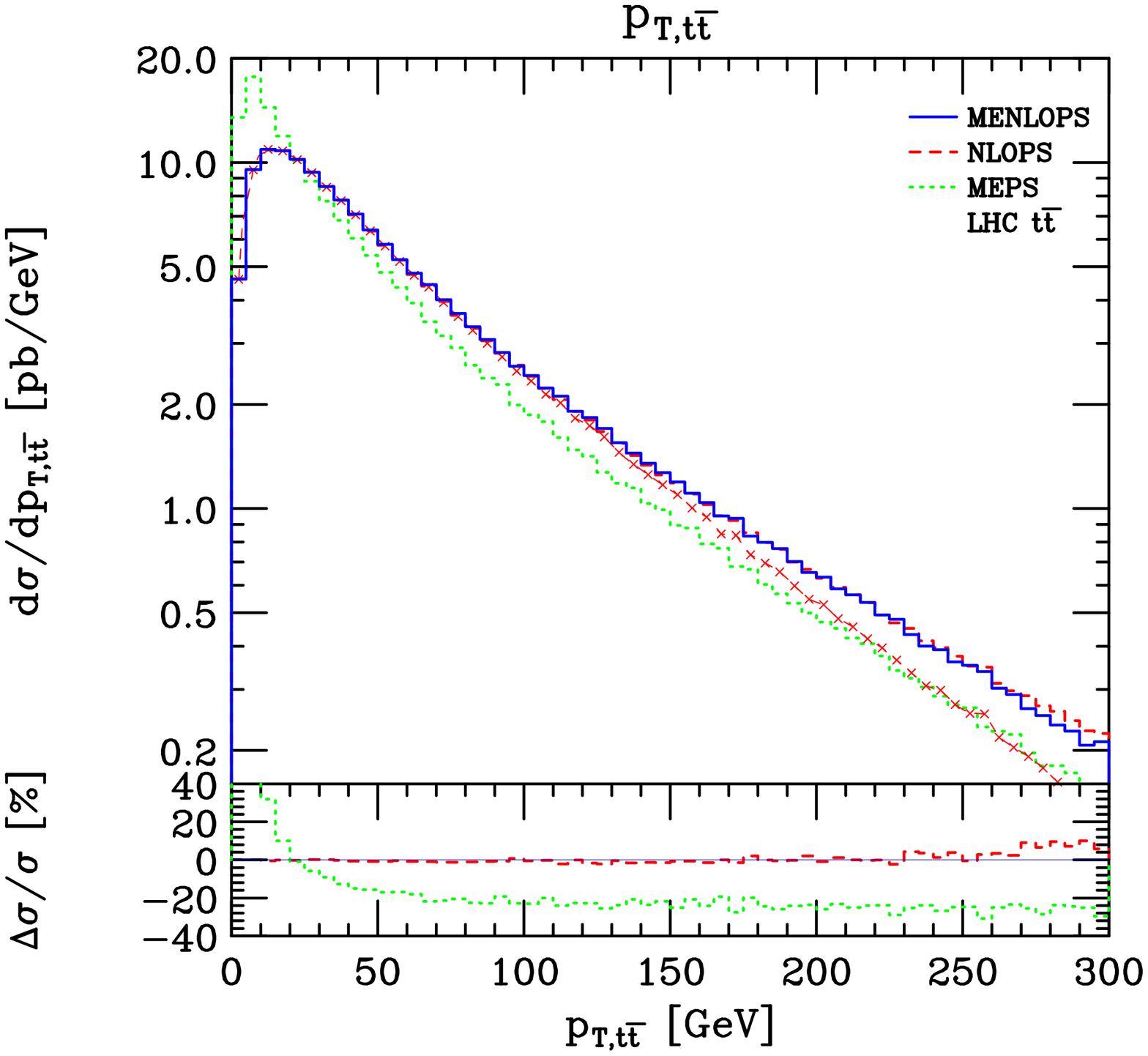} 
\par\end{centering}

\vspace{5mm}

\caption{Above we show the transverse momentum distribution of the $\mathrm{t}\bar{\mathrm{t}}$
pair system using a 60 GeV (left) and 100 GeV (right) jet resolution
scale as the \textsc{Menlops} merging scale. As in Figure~\ref{fig:tt_inclusive_observables},
the greater resolution scale used in producing the \textsc{Menlops}
sample (solid) on the right hand side, results in the \textsc{Meps}
component (dotted with $+$ symbols)
being greatly diminished. Nevertheless, the merged
distribution very much assumes the form of the pure \textsc{Nlops}
prediction (dashed) to within $\mathcal{O}\left(1\%\right)$, with
deviations only beginning to become noticeable in the high $p_{T}$
tail, where contributions from events containing more than one jet
become more important.}

\label{fig:tt_pT_of_tt} 
\end{figure}

Figure~\ref{fig:tt_pT_of_tt} shows the transverse momentum distribution
of the $\mathrm{t}\bar{\mathrm{t}}$ pair system. Here again the \textsc{Menlops}
distribution is basically insensitive to the \textsc{Meps}-\textsc{Nlops}
merging scale, always being within a few percent of the \textsc{Nlops}
prediction. It is clear, from the plot of the fractional differences,
that the spectrum of the \textsc{Meps} sample is around 20\% lower
than that of the \textsc{Nlops} simulation. Note, however, that this
difference is almost constant from around 50 GeV upwards, that is,
the \textsc{Meps} and \textsc{Nlops} description of the \emph{shape}
\emph{in} \emph{that} \emph{region} is in much better agreement than
the 20\% offset suggests. As noted above, the correspondence in the
shapes of the distributions is important for the stability of the
\textsc{Menlops} prediction there. The principal cause of the offset
is in fact due to the 60\% \textsc{Meps} excess in the vicinity of
the Sudakov peak which, on account of our rescaling the \textsc{Meps}
results so as to have the same weight as \textsc{Nlops} events, makes
the distribution of the former appear lower at larger
$p_{T}$. This excess has actually already manifested
itself in the form of an increased fraction of 0-jet events in the \textsc{Meps}
sample (Fig.\,\ref{fig:tt_jet_fractions_1}).
We have also reproduced this plot with the native P{\footnotesize YTHIA}
$\mathrm{t}\overline{\mathrm{t}}$ simulation alone, without \textsc{Meps}
merging. Using the virtuality ordered shower, as for the \textsc{Meps}
sample, we find the same excess around the Sudakov peak. If, on the
other hand, we use the transverse momentum ordered shower, we do not
observe the excess, in fact we see nice agreement with \textsc{Powheg}
in the peak region. This is perhaps not surprising given that the
generation of radiation in the \textsc{Powheg} sample is wholly done
according to evolution in $p_{T}$. Based on these observations we
attribute the excess in the peak region to the different choices of
evolution variables. In particular, we note that the scale used to
evaluate the \textsc{PDF} factors in the P{\footnotesize YTHIA} veto
algorithm are the evolution variables themselves. However, just as
$p_{T}$ is the correct scale to use as the argument of the running
coupling in the shower (the default in P{\footnotesize YTHIA} \cite{Sjostrand:2006za}),
it is also the correct argument to use in evaluating the \textsc{PDF}s
\cite{Nason:2006hfa}. We therefore expect that the mismatch in scales
leads to the sizeable differences in the soft region, at the parton
level.

Having now taken this point into consideration, looking more closely
at the plot of the fractional difference in Fig.\,\ref{fig:tt_pT_of_tt},
one can see that the shapes spectrum in the \textsc{Meps} and \textsc{Menlops}
are \emph{still}, very slightly, softer than the pure \textsc{Nlops}
prediction at high $p_{T}$. To understand this, one should consider
that the region of phase space in which the $\mathrm{t}\overline{\mathrm{t}}$
pair has a large transverse momentum is, naturally, to be associated
with more energetic events and hence events with higher jet multiplicities.
It follows from the \textsc{Menlops} algorithm, that in the limit
that the $\mathrm{t}\overline{\mathrm{t}}$ pair transverse momentum
tends to large values, the \textsc{Meps} component of the \textsc{Menlops}
sample will dominate. However we stress that, from the point of view
of this observable, in this region of phase space, the higher order
effects in the \textsc{Meps} distribution in no way represent an improvement
on the \textsc{Nlops} prediction since the corresponding virtual corrections
to $\mathrm{t}\overline{\mathrm{t}}+\mathrm{jet}$ production are
missing. Formally both approaches have the same degree of accuracy
in describing the high $p_{T}$ tail, with differences in the shapes
being of higher order in $\alpha_{\mathrm{S}}$.

\subsubsection{Jet activity}

\begin{figure}[H]
\begin{centering}
\includegraphics[width=0.4\textwidth,angle=90]{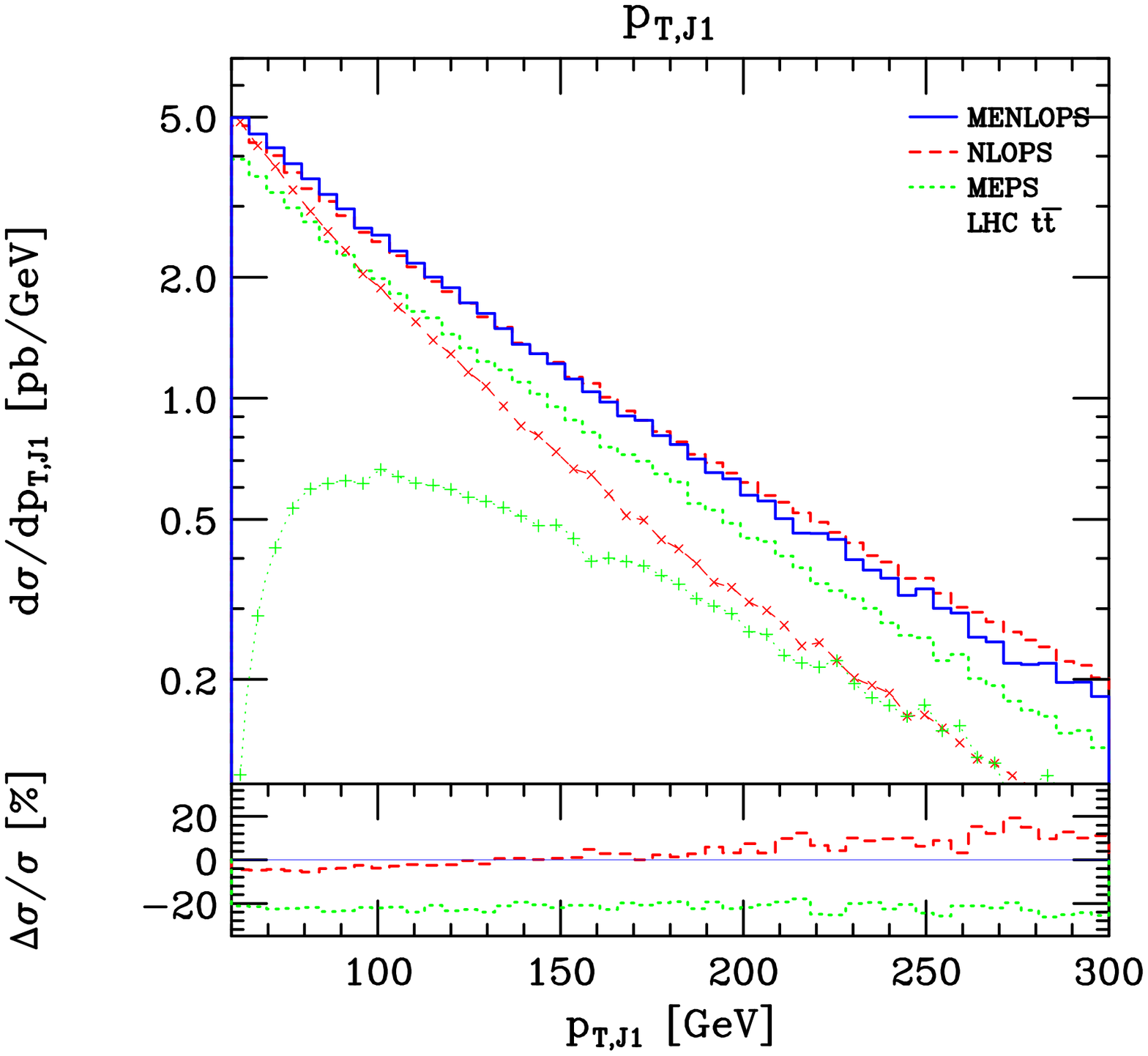}\hfill{}\includegraphics[width=0.4\textwidth,angle=90]{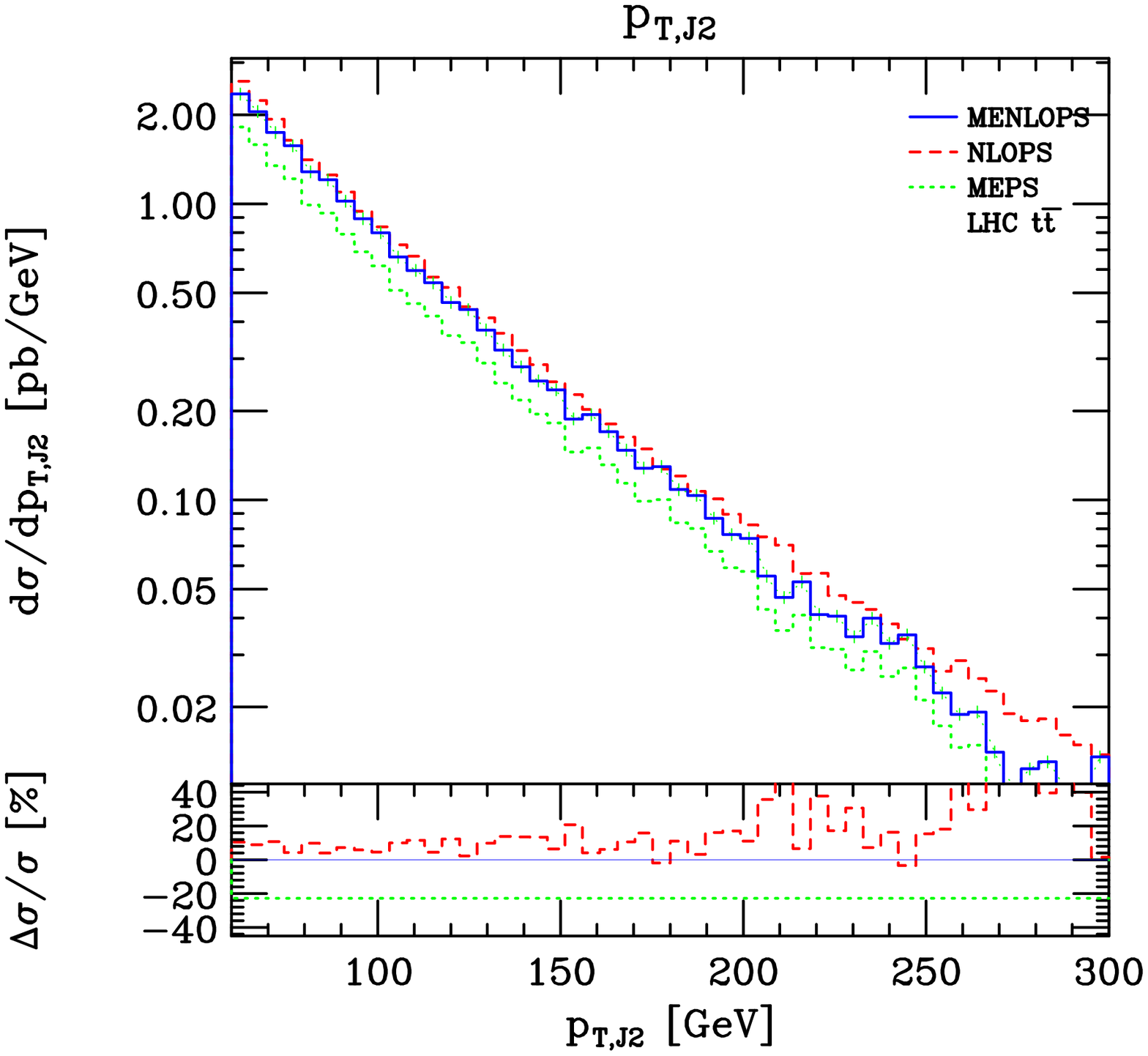} 
\par\end{centering}

\vspace{5mm}

\begin{centering}
\includegraphics[width=0.4\textwidth,angle=90]{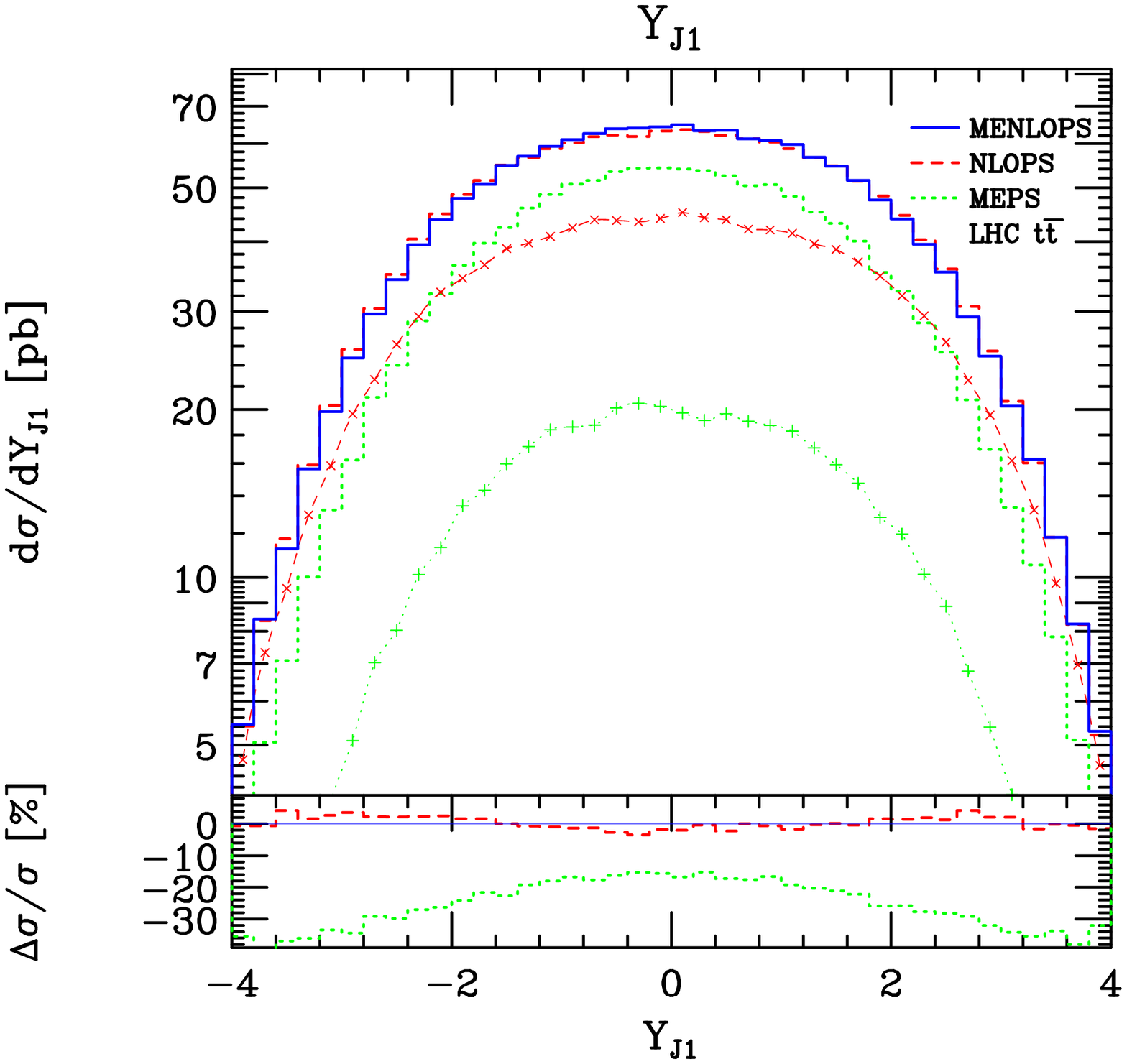}
\hfill{}\includegraphics[width=0.4\textwidth,angle=90]{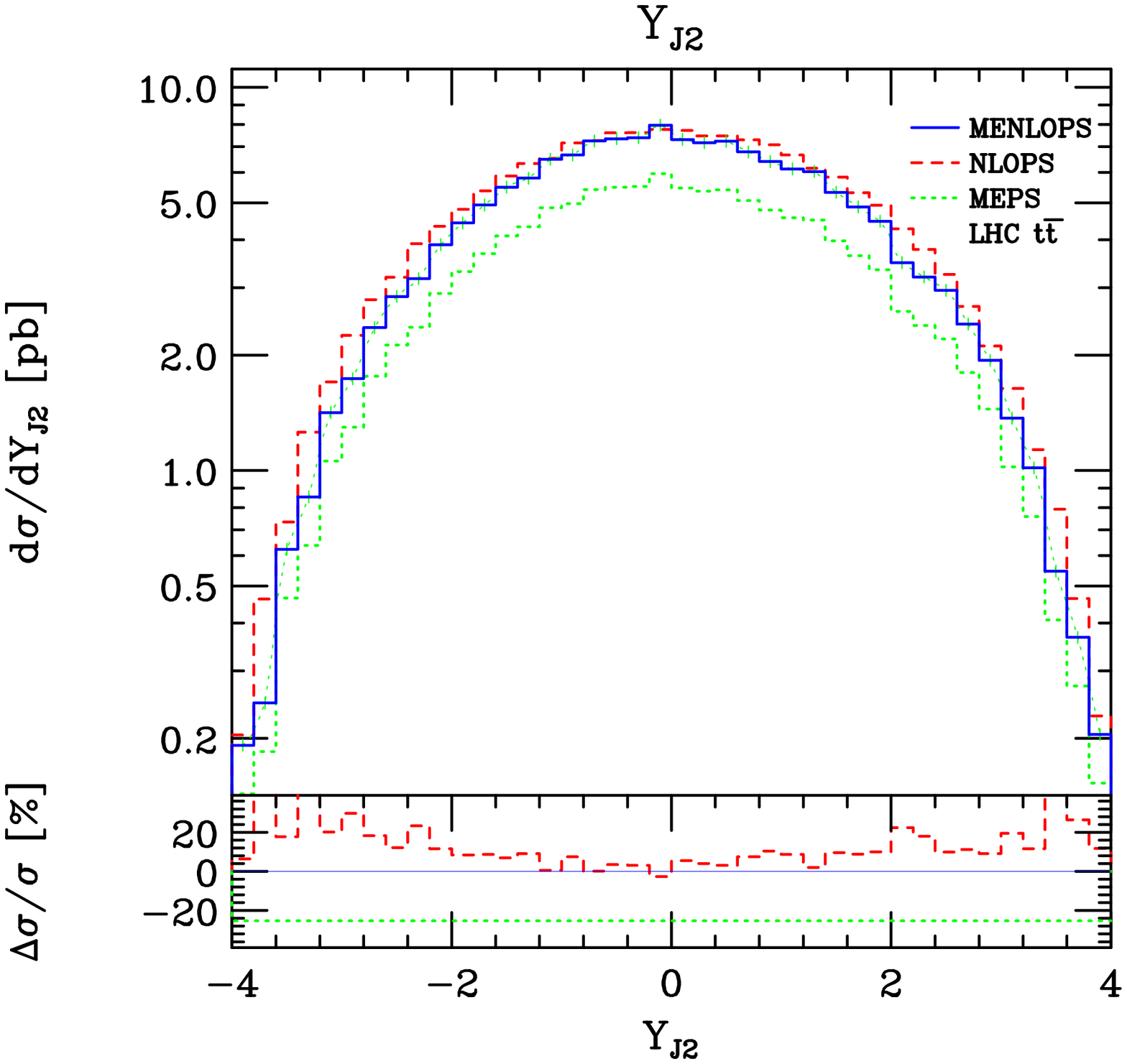} 
\par\end{centering}

\caption{ In the upper half of
this figure we show the transverse momentum distribution of the hardest
(left) and second hardest (right) jets, with the corresponding rapidity
distributions shown underneath. 
The \textsc{Menlops} predictions (solid) shown here and their \textsc{Nlops}
(dashed) and \textsc{Meps} (dotted) components were obtained from a \textsc{Meps}-\textsc{Nlops}
combination with a merging scale of $60$~GeV. }

\label{fig:tt_1st_2nd_jet_pts_and_ys} 
\end{figure}

In Figure~\ref{fig:tt_1st_2nd_jet_pts_and_ys} we show the transverse
momentum spectra of the hardest (left) and second hardest (right)
jets, together with their corresponding rapidity distributions. 

The distributions for the leading jet are predominantly given by that
of the \textsc{Nlops} simulation (dashed), with a structure reflecting
the analogous $\mathrm{t}\bar{\mathrm{t}}$ pair distributions; moreover,
the explanations for the structure are largely the same as in that
case. The main difference between the leading jet transverse momentum
and rapidity spectra, compared to those of the $\mathrm{t}\bar{\mathrm{t}}$
system, lies in the increased \textsc{Meps} contribution to the \textsc{Menlops}
predictions in the case of the leading jet. This is simply due to
the fact that this is a less inclusive quantity. Whereas the
$\mathrm{t}\bar{\mathrm{t}}$
pair distributions receive contributions from events with any number of jets
(including no jets at all), those of the leading jet can only be
constructed from events with at least one jet. Since
the \textsc{Nlops} contribution to the inclusive \textsc{Menlops}
sample is made of 0- and 1-jet events only, it is then natural that
one sees a substantial decrease in the fractional \textsc{Nlops} component
contributing to observables which exclude 0-jet events. Of, course,
this does not represent any kind of problem since both \textsc{Meps}
and \textsc{Nlops} simulations are only capable of describing such
distributions with leading order, leading-log, accuracy.

By design, the \textsc{Menlops} algorithm completely excludes
\textsc{Nlops} events with two or more jets from the final sample,
replacing them with \textsc{Meps} events, since, for observables concerning
the second jet, \textsc{Nlops} predictions are not even accurate at
leading order. The predictions for the second jet are therefore completely
determined by the \textsc{Meps} sample. This is easily seen to be
the case by looking at the plots of the fractional differences,
where one can see that the \textsc{Menlops} result is above the \textsc{Meps}
one by a constant factor (Eq.\,\ref{eq:sec3_menlops_master_formula}).
As in the case of the leading jet, there is a tendency for the $p_{T}$
spectra of the second hardest jet to be softer in the
\textsc{Meps}/\textsc{Menlops} sample. This softening effect may run somewhat
counter to the common lore surrounding shower Monte Carlo. However,
it is well established that the parton shower approximation can lead
to spectra harder than the true
one.\footnote{In fact, the ability of the shower
to overestimate the rate of hard emissions is fundamental to \emph{Matrix
Element Correction} procedures, the forerunners of \textsc{Meps} merging
schemes, used to correct the pattern of radiation in the shower \cite{Sjostrand:2006za,Seymour:1994df,Seymour:1994we}.} 
We also notice
(see Figure~\ref{fig:tt_jet_multiplicities}) that for large jet multiplicities
the \textsc{Nlops} result is higher than the \textsc{Meps} result, even if the
latter is multiplied by an appropriate $K$-factor. We conclude that in
$\mathrm{t}\bar{\mathrm{t}}$ production the \textsc{Meps} method leads to
slightly softer jets and slightly reduced activity in the event.

In the jet rapidity distributions we see a few puzzling features that may even seem
to contradict some previous conclusions. Intuitively one expects, from simplified
kinematical reasoning, that harder radiation and higher multiplicities should be
associated with more activity in the central rapidity region. Since, in the present
case, the \textsc{Nlops} prediction exhibits the first two of these traits, albeit
slightly, naively one may assume that the corresponding jet rapidity distributions
should be more central than in the \textsc{Meps} sample. However, in
Figure~\ref{fig:tt_1st_2nd_jet_pts_and_ys} (by looking at the insert
with the relative difference) we see the opposite behavior. We wish to quickly point
out that the relative sizes of the differences in the $p_{T}$ spectra should be borne
in mind when considering these points. In the case of $\mathrm{W}$ production the second jet
$p_{T}$ spectrum was found to be approximately \emph{five} times harder in the
\textsc{Meps} case with respect to the \textsc{Nlops} one, and so the factor of two
excess seen in the corresponding rapidity spectrum follows from simple kinematics
considerations alone. On the contrary, here the differences are much smaller: the jet
$p_{T}$ spectra agree to within 30\% in terms of their shape, as do their rapidity
distributions. The same basic kinematic arguments are therefore not, by themselves,
applicable in explaining the relationship between the jet $p_{T}$ and rapidity
distributions and the trends therein.

\begin{figure}[H]
\begin{centering}
\includegraphics[width=0.4\textwidth,angle=90]{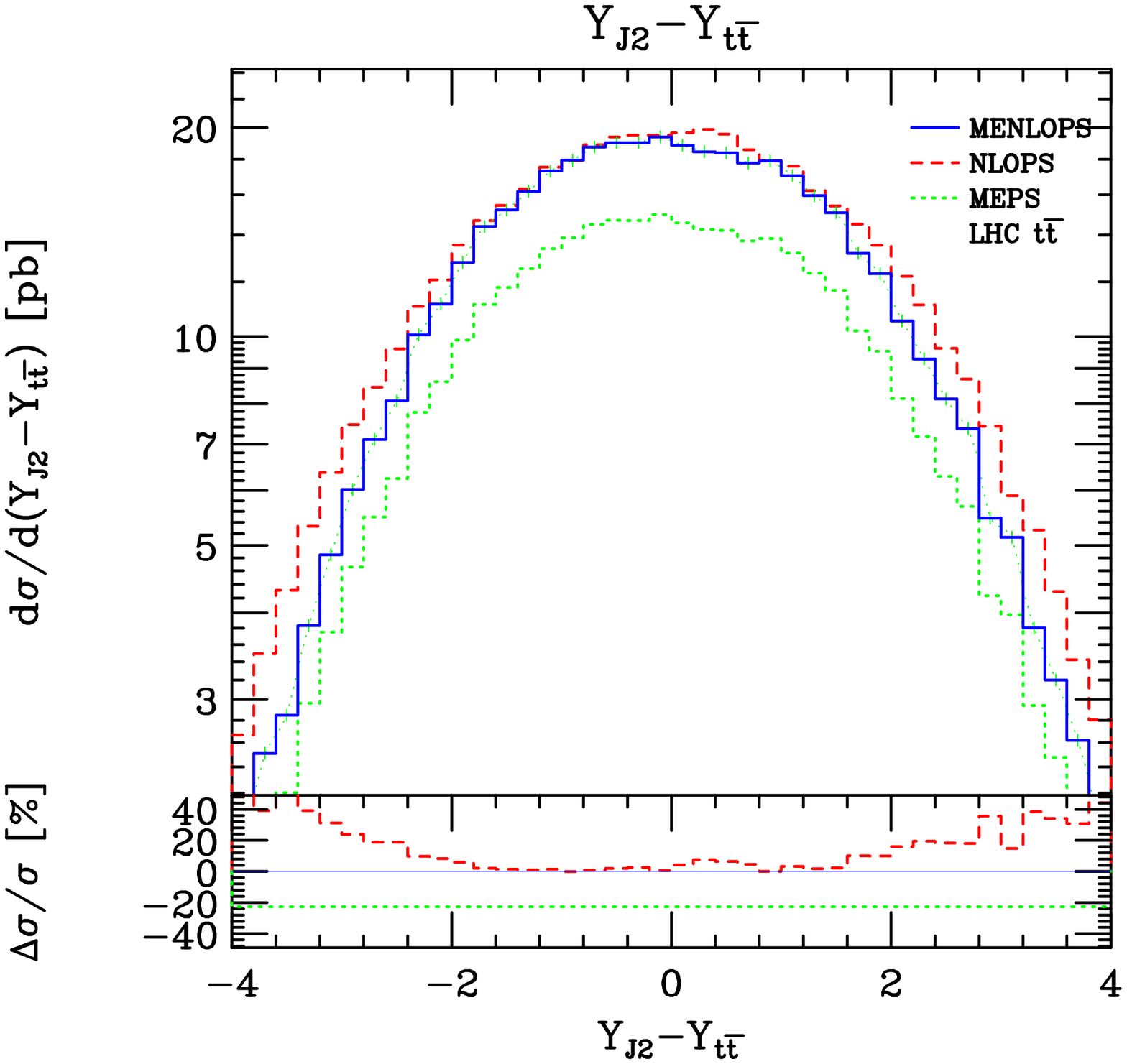}\hfill{}\includegraphics[width=0.4\textwidth,angle=90]{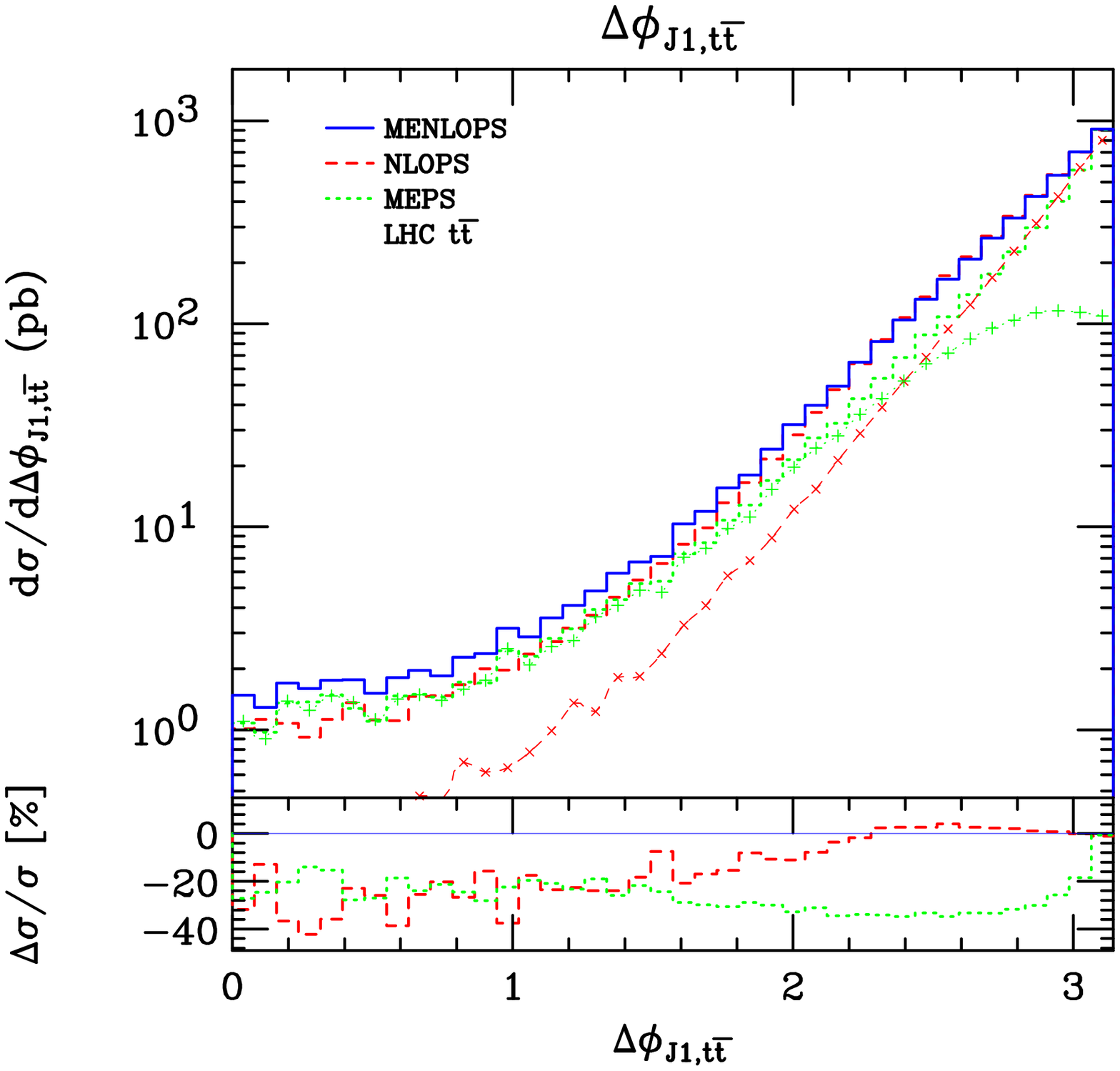} 
\par\end{centering}

\caption{On the left we show the rapidity of the second hardest jet with respect
to the $\mathrm{t}\bar{\mathrm{t}}$ pair, and on the right is the difference
in azimuth between the leading jet and the $\mathrm{t}\bar{\mathrm{t}}$
system. Both of these distributions directly probe the jet structure
in the events beyond that of the leading jet.}

\label{fig:tt_j2_rapidity_corr_and_delta_phi} 
\end{figure}

Whereas, in the case of the leading jet, the \textsc{Menlops} prediction is
predominantly shaped by the \textsc{Nlops} distribution, for the second jet it
takes its form exclusively from the \textsc{Meps} one. In both cases the
\textsc{Nlops} distributions are proportionally larger at larger rapidities with
respect to the \textsc{Meps} ones. We thus infer that some dynamical mechanism
widens the rapidity spectrum in the \textsc{Nlops} sample. One plausible mechanism
must have to do with the virtual corrections implemented in the \textsc{Nlops}
simulation. In fact, in \textsc{Powheg} a $\overline{B}/B$ factor is present in
the generation of \emph{all} events, which depends upon the rapidity of the
$\mathrm{t}\bar{\mathrm{t}}$ system; this may slightly suppress the central region
with respect to the \textsc{Meps} case. This hypothesis is substantiated by the
known fact that NLO inclusive quantities in heavy flavour production display a
remarkable proportionality to the corresponding LO ones, provided the same PDF
sets are used\footnote{We have explicitly confirmed this using independent code,
in which we found the rapidity distribution of the top-quark pair at \textsc{LO}
and \textsc{NLO} agreed to within $\pm$3\% in the region 
$|Y_{\mathrm{t}\mathrm{\bar{t}}}|\,<\,3$, once the \textsc{LO} result was rescaled.}.
Since \textsc{NLO} results include real emissions, and those should make the
rapidity distributions more central, we must conclude that virtual corrections
counteract this effect, and widen the rapidity spectrum. This line of argument is
essentially the same as that taken earlier, in explaining the similar broadening
effects seen in the case of the $\mathrm{W}$ boson and $\mathrm{t}\bar{\mathrm{t}}$
rapidity distributions. Put differently, in simpler terms: since the rapidity
distribution of the $\mathrm{t}\bar{\mathrm{t}}$ system itself is broader in the
\textsc{Nlops} case, due to the inclusion of virtual corrections, given that the
rapidity of the jets are strongly correlated with it (as can be seen in \emph{e.g.}
Fig.\,\ref{fig:tt_j2_rapidity_corr_and_delta_phi}), it is natural to expect that they
too have broader distributions.

The rapidity correlation between the second jet and the $\mathrm{t}\bar{\mathrm{t}}$
pair, shown in Fig.\,\ref{fig:tt_j2_rapidity_corr_and_delta_phi},
echoes the tendency for the second jet to be more central in Fig.\,\ref{fig:tt_1st_2nd_jet_pts_and_ys}.
Also this result may be related to the fact that large rapidities
of the $\mathrm{t}\bar{\mathrm{t}}$ pair may be enhanced by virtual corrections, and that
large differences between the second jet and the $\mathrm{t}\bar{\mathrm{t}}$ system rapidity
may require a relatively large $\mathrm{t}\bar{\mathrm{t}}$ rapidity.

Figure~\ref{fig:tt_j2_rapidity_corr_and_delta_phi} also shows the
difference in azimuth between the leading jet and the $\mathrm{t}\bar{\mathrm{t}}$
system.
Also this figure seems to suggest, that the radiation that accompanies the
first jet is more collinear in the \textsc{Nlops} sample than in the
\textsc{Meps} sample. In order to understand the plot, one should also keep
in mind that the \textsc{Meps}
has a smaller fraction of single jet events than the \textsc{Nlops}, due to the
shape of the $\mathrm{t}\bar{\mathrm{t}}$ $p_T$ spectrum displayed in
Fig.~\ref{fig:tt_pT_of_tt}.
Also, from Fig.~\ref{fig:tt_jet_fractions_1}, we see that, for $y<60\,$GeV,
the \textsc{Meps} has a depleted single jet event multiplicity and an enhanced
0-jet fraction.

 As with the distributions of the rapidity of the second jet
and its rapidity correlation with respect to the $\mathrm{t}\bar{\mathrm{t}}$
system, this observable allows us to further probe the direction
in which additional radiation is emitted. Recall that, in the \textsc{Nlops}
approach, the distribution of any jets in the event beyond the leading
one originate from a parton shower description.
Thus we expect the  \textsc{Nlops} sample to underpopulate the small
$\Delta \Phi_{\mathrm{J1, t\bar{t}}}$ region, an effect that was also seen
in the corresponding $\mathrm{W}$ production distribution.
In the \textsc{Meps} case additional initial state radiation
will tend to be more correlated with the direction of the incoming
partons, while final state radiation will instead be correlated in
angle with that of the leading jet (the progenitors), hence one expects,
and indeed finds, proportionally more \textsc{Nlops} events for which
the jet and $\mathrm{t}\bar{\mathrm{t}}$ pair are back-to-back in
azimuth than in the \textsc{Meps}/\textsc{Menlops} samples.

\begin{figure}[H]
\begin{centering}
\includegraphics[width=0.4\textwidth,angle=90]{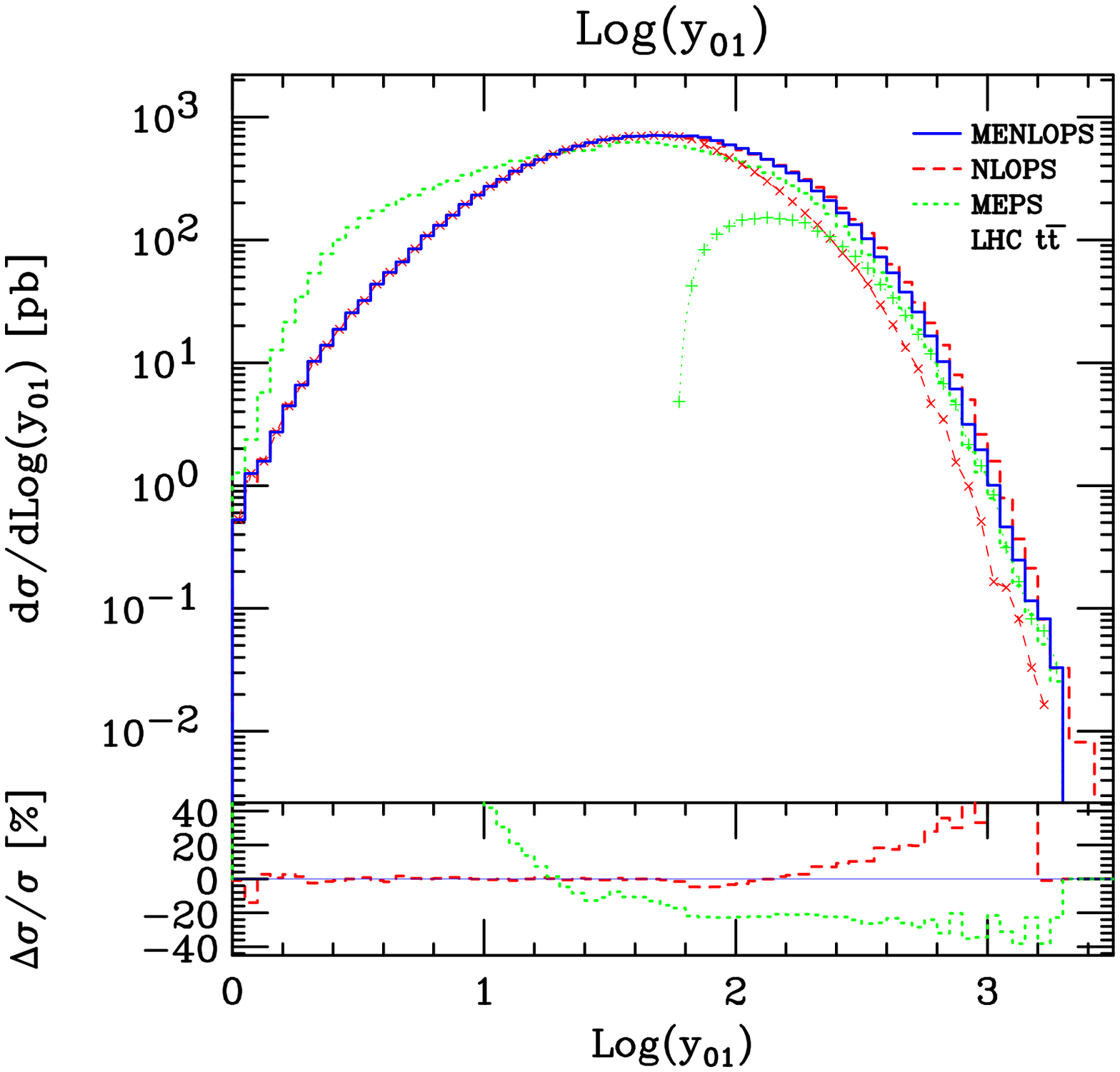}\hfill{}\includegraphics[width=0.4\textwidth,angle=90]{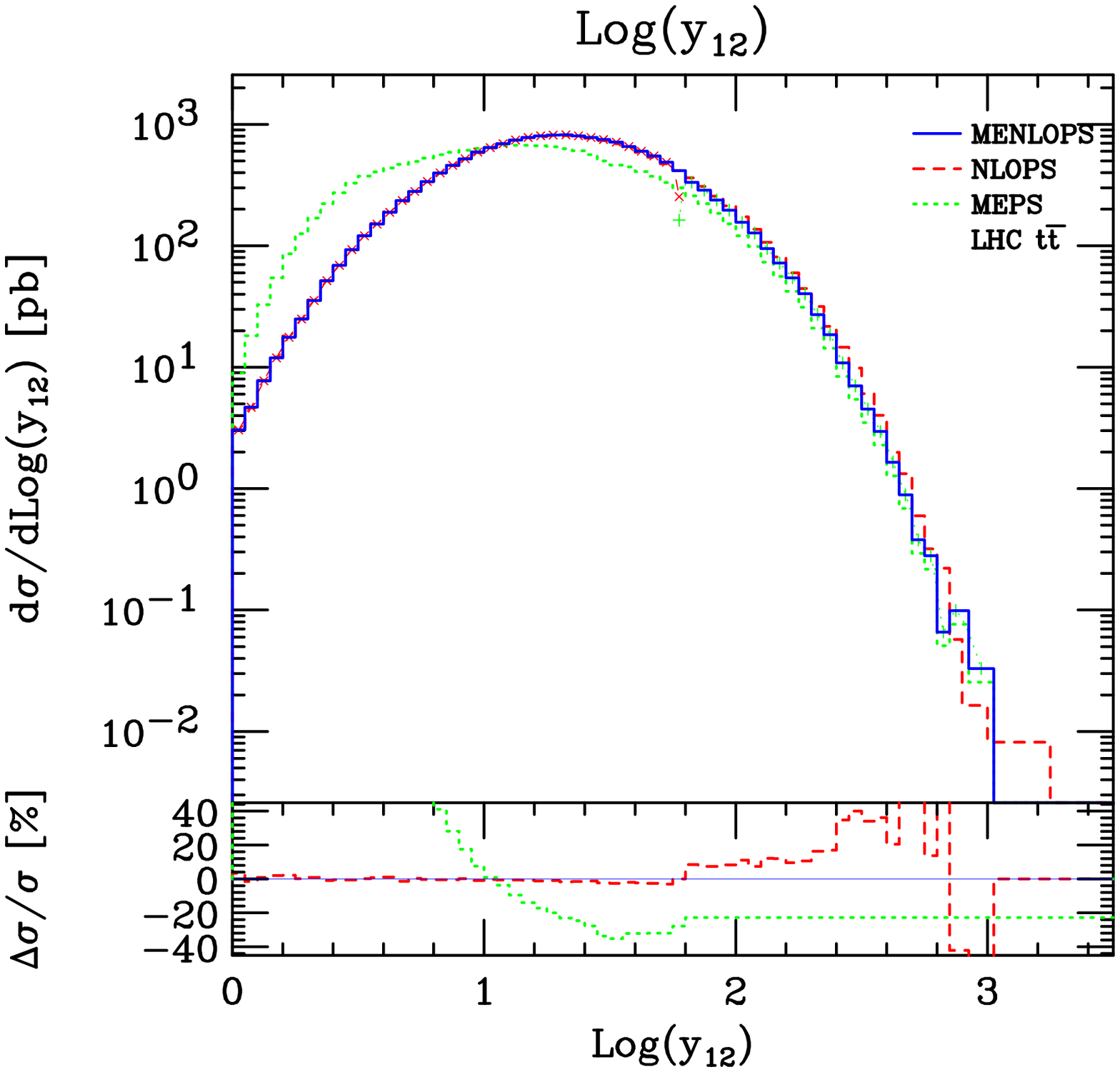} 
\par\end{centering}

\vspace{5mm}

\begin{centering}
\includegraphics[width=0.4\textwidth,angle=90]{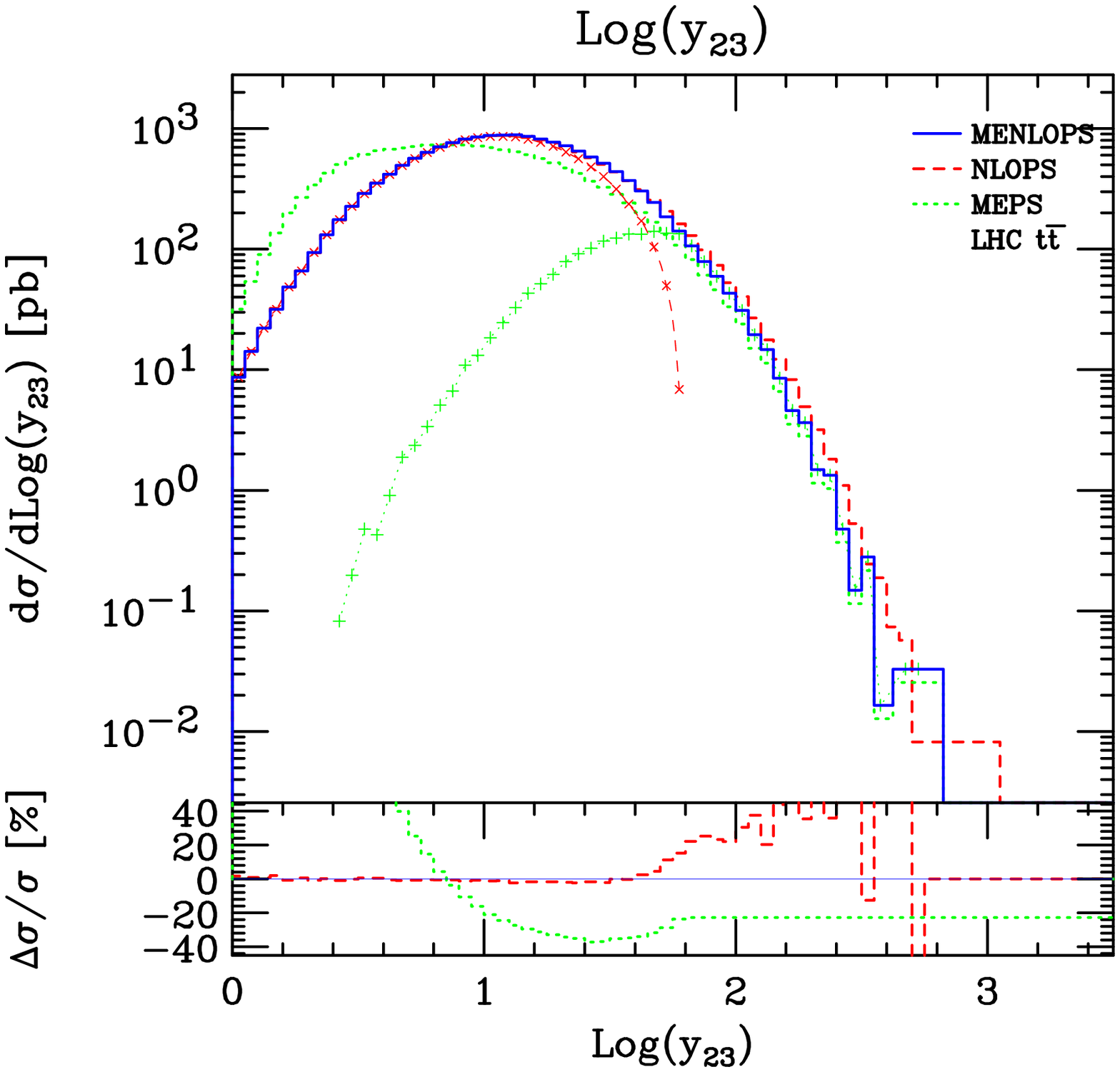}
\hfill{}\includegraphics[width=0.4\textwidth,angle=90]{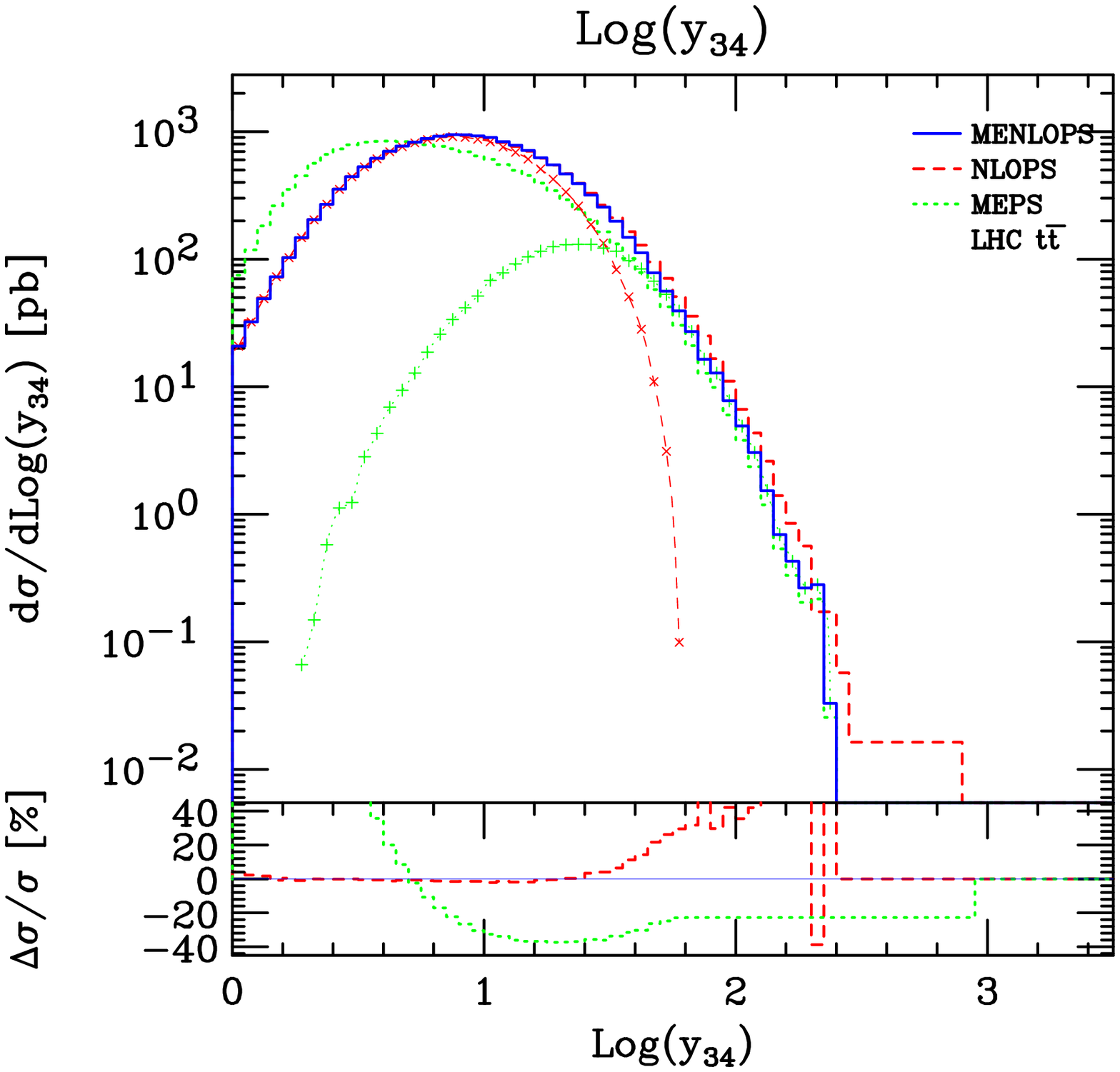} 
\par\end{centering}

\caption{Here we show the logarithm of the value of the jet clustering scale
$y_{nm}$ at which an $n$-jet event is resolved as an $m=n+1$-jet
event in each of the samples.}

\label{fig:tt_djrs} 
\end{figure}

We now end our analysis of top quark pair production by discussing
the differential jet rates displayed in Figure~\ref{fig:tt_djrs}.
We remind the reader that the quantity, $y_{nm}$, being plotted in
each histogram is the value of the clustering scale $\sqrt{d_{nm}}$,
at which an $n$ jet event becomes resolved as an $m=n+1$ jet event.
These distributions then probe directly the behavior of the \textsc{Meps}
and \textsc{Menlops} samples either side of their respective merging
scale boundaries. Recall that the merging scale in the former sample
was taken to be 30 GeV, while in the default \textsc{Menlops} combination
a value of 60 GeV was taken. In both cases, different types of simulation
populate either side of these unphysical boundaries, and so discontinuities
could be expected to appear in these jet rates.

In the \textsc{Meps} case the merging between the parton shower and
the matrix elements involves a phase space partition for every different
multiplicity. In the \textsc{Menlops} case all events with 0 or 1
jet are described by the one \textsc{Nlops} simulation, with the \textsc{Meps}
sample alone describing the rest. Hence, in the latter case all jet
rates should be free of discontinuities, with the exception, possibly,
of the $y_{12}$ jet rate, where there is an abrupt transition at 60
GeV from the \textsc{Meps} description to the \textsc{Nlops} one.

In all cases one can see that the predictions from the pure \textsc{Meps}
sample (dots) are smooth with no evidence of any merging scale
dependence. Likewise the pure \textsc{Nlops} results are also smooth,
which has to be the case, given that the pure \textsc{Nlops} sample
involves no phase space partitions. The two distributions exhibit
some differences in the soft region, the \textsc{Meps} sample favoring
more soft emission. This behavior was already noted in the discussion
of the $\mathrm{t}\bar{\mathrm{t}}$ pair transverse momentum spectrum
(Fig.\,\ref{fig:tt_pT_of_tt}), where an explanation based on the
scales used to evaluate the \textsc{PDF}s was given. 

We also see that the \textsc{Menlops} predictions (solid) are similarly
very smooth in all cases, in spite of the sharp transition from the
\textsc{Nlops} to the \textsc{Meps} description in the $y_{12}$ distribution
(their contributions to the \textsc{Menlops} sample are also shown). The
reason for this smooth transition is simply due to the fact that the
distributions constructed out of each class of events differ by sub-dominant
terms but are, theoretically,
the same at the level of large singular terms that are dominant around
the 60 GeV threshold.

\section{Conclusions\label{sec:Conclusions}}

At the beginning of this document we have reviewed the next-to-leading
order parton shower matching method \textsc{Powheg} and re-examined
matrix element-parton shower merging in the context of that formalism.
Our initial aim has been to determine an exact means by which \textsc{Nlops
}simulations, in particular, \textsc{Powheg}, may be combined with
those implementing \textsc{Meps }merging prescriptions, such as \textsc{MLM}
and \textsc{CKKW}, so as to retain full \textsc{NLO} accuracy for
inclusive observables and, at the same time, give an accurate description
of multi-jet final states. 

In Section~\ref{sec:Hardest-emission-xsecs}, making use of rather
modest assumptions concerning the behavior of an \textsc{Meps,} we
proposed a general expression for the corresponding hardest emission
cross section which, in the \textsc{Powheg} case, is key to achieving
\textsc{NLO} accuracy. We have been able to reformulate these expressions
in such a way as to identify how the \textsc{Meps} simulation must
be augmented, in order to make its hardest emission cross section
converge with that of \textsc{Powheg}. A key consideration in these
manipulations has been the enforcement of unitarity in the \textsc{Meps}
approach, for all possible configurations of the underlying Born event.
The conclusion of the theoretical analysis in Section~\ref{sec:Hardest-emission-xsecs}
was that \textsc{NLO} accuracy and an accurate model of multi-jet
radiation may be unified in a single simulation, by reweighting the
\textsc{Meps }events with a factor $\overline{B}\left(\Phi_{B}\right)/\overline{B}_{\mathrm{ME}}\left(\Phi_{B}\right)$:
the ratio of the \textsc{NLO} cross section, differential in the kinematics
of the underlying Born event, divided by the \textsc{Meps }simulation's
leading order approximation to it. Note that the latter is not identical
to the leading order cross section but exhibits differences at $\mathcal{O}\left(\alpha_{\mathrm{S}}\right)$. 

Despite the apparent simplicity of this conclusion, in general it is technically
very challenging, since the multi-dimensional weight factors must
be computed and stored prior to generating \textsc{Meps }events (both
of which are highly intensive numerical operations), which must then
be followed by a further reweighting/rejection procedure. For simple
processes, involving the production of just a single particle, the
effort needed to realize the method is likely to be reasonable, as
is the reduction in event generation efficiency. We leave this as the subject
of a future study.

Motivated by the observation that \textsc{Meps} merging schemes require,
in practice, the introduction of a phase space partition for each
jet multiplicity, beneath which radiation is described by the parton
shower approximation, we were led to consider the question of to what
extent it may be possible to achieve the same enhancements, by simpler
means, without modifications to the \emph{large}, \emph{mature}, \emph{existing},
body of simulations. It is clear that for events containing no additional
jets, defined according to the \textsc{Meps }merging scale, the \textsc{Nlops}
description is, categorically, always better than the \textsc{Meps}
one, while, for events with one additional jet the \textsc{Nlops}
is always at least as good (Sect.~\ref{sec:Combining-Powheg-and-Meps}).
With this in mind the preceding question becomes equivalent to the
question: do events with two or more jets, defined at the merging
scale, comprise more than a fraction $\alpha_{\mathrm{S}}$ of an
inclusive\textsc{ Meps} merged sample?

Based on this we formulated a method for combining \textsc{Meps }and\textsc{
Nlops }events into\textsc{ Menlops }samples according to a further
eponymous clustering scale. In this approach, if the \textsc{Menlops}
merging scale can be set equal to the \textsc{Meps} merging scale,
without including more than a fraction $\mathcal{O}\left(\alpha_{\mathrm{S}}\right)$
of the leading order, \textsc{Meps} events, the resulting event sample
is \textsc{NLO} accurate and the description of multi-jet final states
is exactly as in the \textsc{Meps} simulation. Requiring that the
\textsc{Meps} content not exceed a fraction $\mathcal{O}\left(\alpha_{\mathrm{S}}\right)$
means that it may not always be possible to lower the \textsc{Menlops}
merging scale to that used in a given \textsc{Meps} sample. Since
we always intend that the \textsc{Menlops} merging scale be restricted
such that the fraction of, technically leading order, \textsc{Meps}
events in the sample is less than $\mathcal{O}\left(\alpha_{\mathrm{S}}\right)$,
to avoid compromising \textsc{NLO} accuracy, this approach should
be viewed as a means of improving \textsc{Nlops} simulations in the
direction of \textsc{Meps} ones, as opposed to the opposite sense.

In Section~\ref{sec:Results} we carried out a detailed analysis
of \textsc{Menlops} samples for $\mathrm{W}^{-}$ and $\mathrm{t}\bar{\mathrm{t}}$
production. In the case of $\mathrm{t}\bar{\mathrm{t}}$ production
we analyzed a sample merging \textsc{Meps} and \textsc{Nlops} events
at a $k_{\perp}$ clustering scale of 60 GeV, comprised of 12.5\%
\textsc{Meps} events, 30 GeV above the merging scale recommended (and
used) to create the \textsc{Meps} sample with the P{\footnotesize YTHIA}
virtuality ordered parton shower%
\footnote{We remind the reader that the recommended \textsc{Meps }merging scale
in the case of the transverse momentum ordered shower was 100 GeV
\cite{Alwall:2008qv}.%
}. For $\mathrm{W}^{-}$ production the \textsc{Menlops} sample under
study consisted of 96\% \textsc{Nlops} events, with the \textsc{Meps}
and \textsc{Nlops} samples being merged at 25 GeV, only 5 GeV above
that recommended for the \textsc{Meps} sample. In view of this fact
the implementation of the exact method for \textsc{Meps-Nlops} merging
becomes an academic exercise in the case of $\mathrm{W}$ production.
However, also in the $\mathrm{t}\bar{\mathrm{t}}$ case the practical gain
from doing so is likely to be negligible.

In all cases, for inclusive quantities, the differences between the
pure \textsc{Nlops} and \textsc{Menlops} results were found to be
negligible $\mathcal{O}\left(1\%\right)$. On the other hand differences
of $\mathcal{O}\left(\alpha_{\mathrm{S}}\right)$ can be seen in,
for example, the $\mathrm{W}^{-}$ and $\mathrm{t}\bar{\mathrm{t}}$
rapidity distributions, when comparing the \textsc{Nlops} and \textsc{Menlops}
predictions to those from the \textsc{Meps} simulation. We attribute
these to the absence of \textsc{NLO} corrections in the latter. Semi-inclusive
observables, probing the distribution of the leading jet, exhibit
differences between the pure \textsc{Nlops }and\textsc{ Meps} results,
with the \textsc{Menlops} prediction, naturally, lying between the
two. These $\mathcal{O}\left(\alpha_{\mathrm{S}}\right)$ differences
are, however, expected, since all techniques here have only LO accuracy
for such quantities. By contrast, for observables directly sensitive
to the second hardest jet, the \textsc{Meps }and\textsc{ Menlops}
results become identical and reveal large corrections when compared
to those of the \textsc{Nlops }method. These corrections are particularly
acute in the case of $\mathrm{W}^{-}$ production.

In conclusion we wish to recommend the \textsc{Menlops} procedure
as a transparent and versatile means to pool existing Monte Carlo
resources together, in a way which encapsulates to a large extent
all of their best qualities.

\section{Acknowledgments\label{sec:Acknowledgments}}

Keith Hamilton would like to thank Simon de Visscher and Rikkert Frederix
for encouraging us to use \textsc{Madgraph}, with which our experience
was all very positive. We are also very grateful to the CP3, Louvain-la-Neuve,
especially Pavel Demin and Fabio Maltoni, for access to the excellent
computing facilities there.

\bibliographystyle{jhep}
\bibliography{menlops}
 
\end{document}